\documentclass[
preprint,
floatfix,
 superscriptaddress,amsmath,amssymb,
 aps,
pra,
]{revtex4-2}
\usepackage{placeins}
\usepackage{graphicx}
\usepackage{dcolumn}
\usepackage{bm}
\usepackage{xcolor}
\usepackage{textcomp, gensymb}
\usepackage{ulem}

\begin{document}
\title{Isotropic spin and inverse spin Hall effect in epitaxial (111)-oriented Pt/Co bilayers}
\author{Adrián Gudín}
\affiliation{IMDEA Nanociencia, 28049 Madrid, Spain.}

\author{Alberto Anadón}%
\email{alberto.anadon@univ-lorraine.fr, paolo.perna@imdea.org, juan-carlos.rojas-sanchez@univ-lorraine.fr}
\affiliation{Institut Jean Lamour, Université de Lorraine / CNRS, UMR7198, 54011 Nancy, France}

\author{Iciar Arnay}%
\affiliation{IMDEA Nanociencia, 28049 Madrid, Spain.}

\author{Rubén Guerrero}%
\affiliation{IMDEA Nanociencia, 28049 Madrid, Spain.}
\affiliation{Departamento de física aplicada, Universidad de Castilla la Mancha, 45071, Toledo, Spain}

\author{Julio Camarero}
\affiliation{IMDEA Nanociencia, 28049 Madrid, Spain.}
\affiliation{Departamento de Física de la Materia Condensada \& Departamento de Física Aplicada \& Instituto Nicolás Cabrera, Universidad Autónoma de Madrid, 28049 Madrid, Spain.}
\affiliation{IFIMAC, Universidad Autónoma de Madrid, 28049 Madrid, Spain.}

\author{Sebastien Petit-Watelot}%
\affiliation{Institut Jean Lamour, Université de Lorraine / CNRS, UMR7198, 54011 Nancy, France}

\author{Paolo Perna$^*$}%
\affiliation{IMDEA Nanociencia, 28049 Madrid, Spain.}

\author{Juan-Carlos Rojas-Sánchez$^*$}%
\affiliation{Institut Jean Lamour, Université de Lorraine / CNRS, UMR7198, 54011 Nancy, France}

\date{\today}

\begin{abstract}
The spin-to-charge current interconversion in bilayers composed of ferromagnetic and nonmagnetic layers with strong spin-orbit coupling has garnered considerable attention due to its exceptional potential in advancing spintronics devices for data storage and logic applications. Platinum (Pt) stands out as one of the most effective materials for generating spin current. While the spin conversion efficiency is isotropic in polycrystalline Pt samples, an ongoing debate persists regarding its dependence on the crystalline direction in single crystalline samples. In this study, we aim to comprehensively evaluate the in-plane anisotropy of spin-charge interconversion using an array of complementary Spin Hall and inverse Spin Hall techniques with both incoherent and coherent excitation. Specifically, we investigate the spin-to-charge interconversion in epitaxial, (111)-oriented, Co/Pt bilayers with low surface roughness, as resulted from x-ray experiments. By varying the thickness of the Pt layer, we gain insights into the spin-charge interconversion in epitaxial Pt and highlight the effects of the interfaces. Our results demonstrate an isotropic behavior within the limits of our detection uncertainty. This finding significantly enhances our understanding of spin conversion in one of the most relevant systems in spintronics and paves the way for future research in this field.
\end{abstract}

\maketitle


 \section{\label{sec:intro}Introduction}

During the last few years, emerging spin-orbitronics technology has been predominantly focused on the generation, manipulation and detection of spin currents \cite{Brataas2012}. Focusing on the suitable engineering of multilayer stacks composed of alternated magnetic/non-magnetic (FM/NM) \cite{Garello2013} metals it has been possible to enhance the spin-charge current interconversion.

The generation and detection of spin currents mainly occur in materials exhibiting substantial spin-orbit coupling (SOC), such as heavy metals (HM) like Ta  \cite{Liu2012} or Pt  \cite{Liu2011}. This large SOC enables the conversion of a pure transverse spin current into charge current due to the spin Hall effect (SHE) \cite{Hirsch1999,Sinova2015,Hoffmann2013,Slonczewski1996}. Similarly, we can detect spin currents through the inverse spin Hall effect  by  spin pumping experiments \cite{Costache2006,saitoh2006,Ando2011,rojas2014spin} or utilizing temperature gradients \cite{Uchida2008,Uchida2010,Anadon2022ThermalGa_0.6Fe_1.4O_3b,Anadon2016}. In brief, like electric currents are carried by moving charge, the spin currents occur due to the displacement of spin angular momentum.  In systems with SOC, these spin currents can be transferred to the magnetization of a magnetic film, allowing for the manipulation of its magnetization state. This effect is known as spin-orbit torque (SOT) \cite{Gambardella2011,Anadon2021Cu,Mangin2006} and it is the result of an interaction between itinerant electrons in a FM that are spin polarized and the magnetization. Additionally to SOT some correlative effects can arise due to the SHE such as the spin Hall magnetoresistance (SMR)  \cite{Nakayama2013,Homkar2021SpinGa0.6Fe1.4O3Bilayers} or the inverse Rashba-Edelstein Effect (IREE)  \cite{Sanchez2013}. The utilization of these effects opens the gate to new technological applications such as spin orbit torque magnetic random access memories \cite{MihaiMiron2010,Dieny2020}, spin Hall nano-oscillators  \cite{mazraati2016low}, magnetic sensors  \cite{xu2018ultrathin} and domain wall devices  \cite{ryu2013chiral,luo2020current}. 

The spin-charge conversion efficiency in FM/NM stacks (when the NM presents a significant SOC) is typically evaluated in terms of spin Hall angle ($\theta_{SH}$). The state of the art literature in the field focuses on sputtered polycrystalline samples, reporting large $\theta_{SH}$ in HMs such as $\beta$-Ta ($\sim$ -0.12) \cite{Liu2012}, $\beta$-W($\sim$ 0.45)  \cite{yu2016large}, Pt($\sim$ 0.05-0.1)  \cite{Liu2011a,rojas2014spin}, transition metal alloys like Fe-Pt($\sim$ 0.3) \cite{ou2018strong} and CuBi($\sim$ 
-0.24) \cite{niimi2012giant}. Pt is the most studied material in spin Hall experiments, exhibiting large $\theta_{SH}$, yet a substantial discrepancy exists among the reported values. This arises from the different scattering mechanism affecting the spins, which heavily depend on the structural properties of the films \cite{Sagasta2016}.  Thus, if we assume that the spin relaxation mechanism in Pt is influenced by its crystal structure \cite{ryu2016observation}, one may expect that the SOT induced by Pt does also depend on the crystallographic direction in which the current is applied and on the grain morphology. During the last few years, many research groups have experimentally studied the value of $\theta\textsubscript{SH}$ at different crystallographic orientations in epitaxial Pt in contact with Fe  \cite{guillemard2018,Keller2018DeterminationExperiments}, FeNi  \cite{ikebuchi2022crystal}, CoFeB  \cite{Thompson2020,PhysRevApplied.15.014055}, Co  \cite{Bai2021,choi2022thickness} or Py  \cite{Xiao2022}. However, the reported values continue to exhibit contradictory results regarding the isotropic or anisotropic nature of the effect in epitaxial Pt.  Since different FM materials and techniques have been used in these studies, it is crucial to highlight not only the properties of epitaxial Pt itself but also the properties of the FM material and the interface between the FM and Pt. The influence of crystalline orientation on spin conversion at the interface between the ferromagnetic material and Pt has been a subject of interest in prior research. Several studies have explored the impact of interfacial structure on spin transport, including the interface transparency due to spin memory loss or spin flip scattering at the interface \cite{Rojas-Sanchez2014}, the mismatch of the electronic bands \cite{Zhang2015role},  as well as studies focusing on the role of the crystal structure \cite{PhysRevLett.115.056601}. 
   

 In this work we aim to clarify the existing discrepancies in the literature on the spin conversion anisotropy in Co/Pt(111) epitaxial bilayers using three different and complementary techniques, namely spin pumping ferromagnetic  resonance (SP-FMR), spin-torque ferromagnetic resonance (ST-FMR) and thermo-spin measurements. Although the spin conversion in platinum is one of the most extensively investigated systems in the field of spintronics in recent years those techniques include the direct and inverse spin Hall effects, as well as both incoherent and coherent excitation. Each experiment has been done measuring the spin-conversion voltage along the same two orthogonal in-plane crystallographic directions as a function of the Pt thickness.  To ensure the reliability and reproducibility of our results, we use the same samples for all three experiments, i.e., the lithography process for the three techniques is done simultaneously in the same piece of sample. Notably, all these techniques exhibit high sensitivity to interfacial effects \cite{BRATAAS20041981,PhysRevB.66.224403,Jimenez-Cavero2021}, enabling us to account for them and obtain a comprehensive understanding of the spin conversion process.
 
 \section{\label{sec:methods}Methods}
 \subsection{\label{sec:meth:growth}Growth}

An epitaxial set of Pt-buffer samples were grown on commercial Al$_2$O$_3$(0001) single crystals. Al$_2$O$_3$ substrates were annealed in-situ at 670 K for 1 hour prior to the deposition inside the sputtering chamber with a base pressure of 3$\cdot$10$^{-8}$ mbar. Once the substrate has thermalized, epitaxial (111)-oriented Pt buffer with thicknesses ranging from 5 to 20 nm were deposited by DC sputtering at 670 K, at Ar partial pressure of 8 x10$^{-3}$ mbar with a deposition rate of 0.8 \r{A}/s measured \textit{in-situ} by a quartz microbalance \cite{Ajejas2018}. To confirm the epitaxial growth of the samples, the thickness and the crystallographic properties of fabricated Pt-buffers were assessed by XRD measurements. Once the structural quality of the Pt buffer was proven, a new set of samples with an additional 5 nm layer Co were grown at room temperature by DC sputtering with a lower rate (0.25 \r{A}/s) to avoid damage of the Pt surface. After this, we deposited a capping layer of Al by RF sputtering, to produce a naturally oxidized AlO$_x$ layer with a nominal thickness of 3.5 nm once transferred in air. The final stack composition was AlO$_x$(3.5 nm)/Co(5 nm)/Pt(t$\textsubscript{Pt}$)//Al$_2$O$_3$(0001). 

 \subsection{\label{sec:meth:structural}Structural characterization}

\begin{figure*}
\includegraphics[trim={0 1cm 0 0},width=\textwidth]{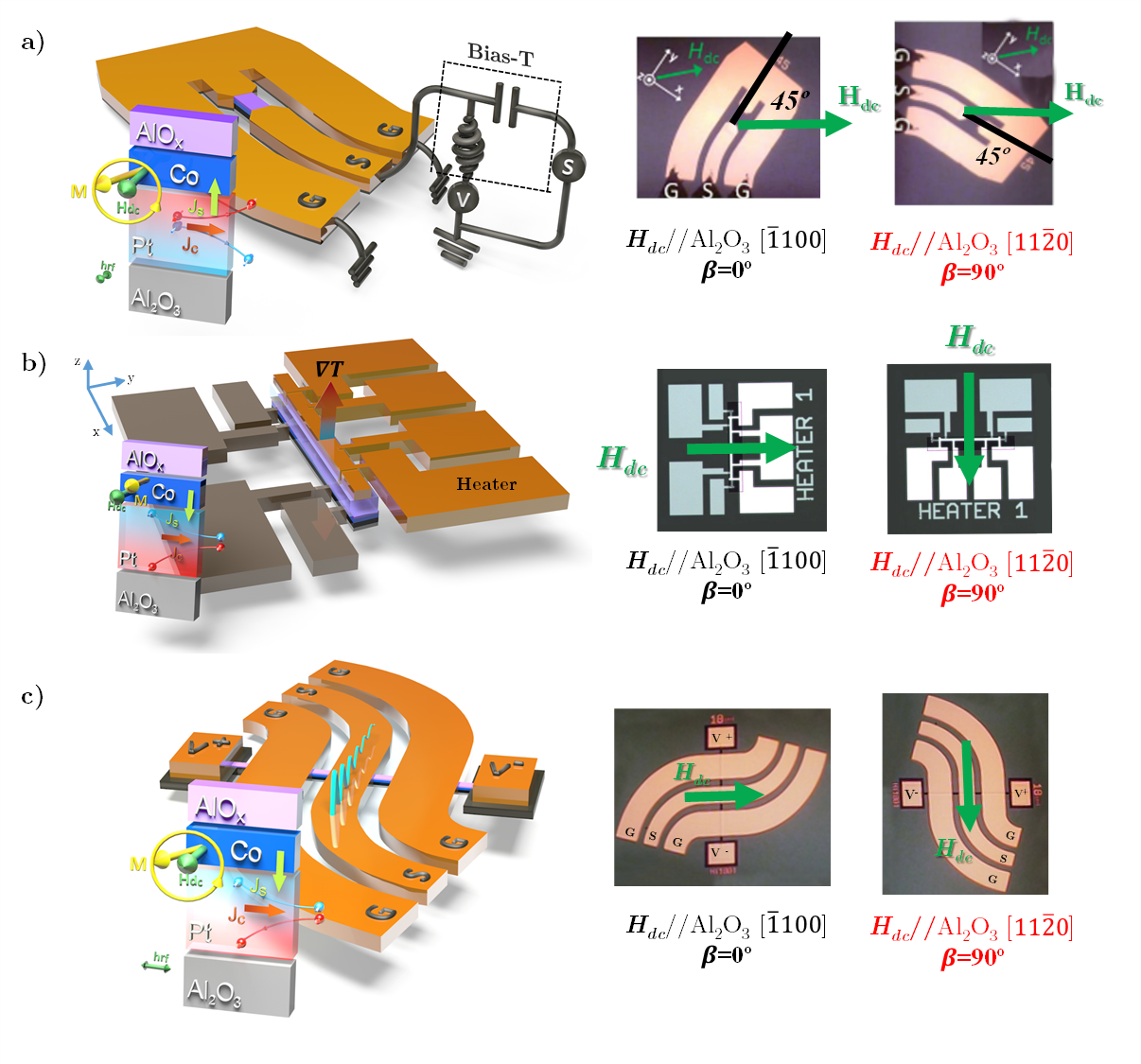}
\caption{\label{fig:schematics} \textbf{Schematics of the ST-FMR, Thermo-spin and SP-FMR devices.}  Schematic diagrams of  a) ST-FMR for the direct SHE, b) Thermo-spin, and c) SP-FMR devices for the inverse SHE in AlO$_x$(3.5 nm)/Co(5 nm)Pt(t$\textsubscript{Pt}$)// Al$_2$O$_3$(0001) stacks. Insets show the optical microscope images of the devices for the two crystallographic directions in which the H$\textsubscript{dc}$ is applied. The green arrows stand for the direction of the applied DC magnetic field. In the ST-FMR H\textsubscript{dc} is applied at 45\degree  \hspace{0.1cm}with respect to the RF electric current, while in thermo-spin and spin pumping it is applied perpendicularly to the measured voltage signal.}
\end{figure*}

Systematic structural characterization of Pt as well as Co/Pt stacks was performed using a commercial Rigaku SmartLab SE multipurpose x-ray diffractometer, equipped with a non-monochromatic Cu K$\alpha$ source ($\lambda$=1.54 \text{\normalfont\AA}). Low angle X-Ray Reflectivity (XRR) measurements were performed with an angular step of 0.004$^{o}$. Long range $\theta$-2$\theta$ diffraction scan was performed with an angular resolution of 0.015$^{o}$ and high resolution scan was recorded with a higher angular resolution of 0.001$^{o}$. Additional structural measurements were performed in final Co/Pt samples at the Spanish beamtime BM25-SpLine at The European Synchrotron (ESRF) in Grenoble, France. Grazing incidence measurements {($\alpha$=0.5$^{o}$)} were performed using a beam energy of 25 keV ($\lambda$=0.4954 \text{\normalfont\AA}).

The in-plane symmetry of the spin-charge interconversion efficiency has been evaluated along the two in-plane non-equivalent $\varGamma-K$ and $\varGamma-M$ directions of the Pt hexagonal lattice, which corresponds to  [$\bar{1}$100] and [11$\bar{2}$0] of the Al$_2$O$_3$ substrate. Such are the two main orthogonal crystallographic directions of Pt and any other direction in-plane would be a combination of these two in the reciprocal space. We define those  directions as $\beta$=0° for Al$_2$O$_3$[$\bar{1}$100] and $\beta$=90° for Al$_2$O$_3$[11$\bar{2}$0]. Those will define the spin direction in the injected spin current. Thus in the spin pumping and the thermo-spin it corresponds to the applied in-plane magnetic field $H\textsubscript{dc}$ as the magnetization fix yields the spin direction.


 \subsection{\label{sec:meth:transport}Device fabrication, electrical and thermal transport measurements}

The devices for spin pumping, spin-torque ferromagnetic resonance (ST-FMR), and thermo-spin measurements were prepared using conventional UV lithography.  The devices for the three different experiments were patterned simultaneously using the same samples starting for the Pt/Co/AlO\textsubscript{x} thin films. The substrate was kept in the same crystallographic direction with respect to the mask for all the samples in every step. The full stack was first patterned and subsequently ion milled, controlling the milled thickness by an ion mass spectrometer using a 4-wave IBE14L01-FA system. After that, in a second step, an insulating SiO$_2$ layer with a thickness of 200 nm was grown by RF sputtering using a Si target and Ar$^+$ and O$^{2-}$ plasma in a Kenositec KS400HR PVD. In a third lithography step, the contacts were patterned and evaporated using an evaporator PLASSYS MEB400S. The dimensions of the active bar are 10 $\times$ 600 $\mu$m for the spin pumping devices and 10 $\times$ 60 $\mu$m for the ST-FMR devices. Due to the small width of the bar, we do not expect significant artifacts from rectification effects in the spin pumping and ST-FMR signals.  \cite{Sanchez2013,Martin-Rio2022,Rojas-Sanchez2016} The geometry of the devices, including the thickness of the insulating SiO$_2$, the dimensions of the coplanar waveguide, the heater in the thermal devices and the lateral dimensions of the milled samples are the same in all the devices (of the same type) shown in this study to reliably compare the spin pumping and ST-FMR voltages. The dimensions of the active bar in the thermo-spin devices are 10 $\times$ 160 $\mu$m. 

Throughout this manuscript, we will refer to the crystallographic directions in which the external DC magnetic field is applied. In spin pumping and spin Seebeck, it is applied perpendicular to the direction where the ISHE voltage. However, in spin-torque FMR, the slab is at 45 degrees to the applied field as indicated in Fig.  \ref{fig:schematics}. 

Spin-torque ferromagnetic resonance is a technique that uses the spin-orbit torque effect to excite and detect the dynamics of the magnetization in a magnetic material. An RF current is applied to the non-magnetic metallic layer (Pt), generating a spin-polarized current that flows perpendicular to the plane of the magnetic layer. The resulting spin-transfer torque can drive the magnetization of the magnetic layer into resonance, which can be detected by measuring the voltage across the magnetic layer. In contrast, spin pumping FMR relies rather on the precession of the magnetization induced by an alternating magnetic field in the ferromagnetic material to generate a spin accumulation that is detected in the non-magnetic layer. 

We conducted experiments to measure ST- and SP-FMR using a probe station with an in-plane DC magnetic field (\textit{H}) up to 0.6 T generated by an electromagnet. The schematic of the ST- and SP-FMR devices used in the experiments are shown in figures \ref{fig:schematics}a) and \ref{fig:schematics}c) respectively. 
In the SP-FMR measurements, we inject a fixed RF frequency current (\textit{f}) of the order of GHz into the coplanar waveguide, producing an RF magnetic field (\textit{h}\textsubscript{RF}) on the sample. For certain frequencies the magnetization of Co may precess in resonance with the applied field, and by the spin-pumping effect, and thus generates a transverse spin current that is injected from the Co into the non-magnetic layer (Pt). 
This spin current is converted in the non-magnetic material into a charge accumulation by means of the inverse SHE (ISHE). We then measure the voltage (\textit{V}\textsubscript{SP}) produced by modulating the RF power injected into the coplanar waveguide and using a lock-in voltmeter that is matched to this modulation while sweeping the external \textit{H}. The power modulation is a sine function with a depth of 100\% and a modulation frequency of 433 Hz. The measured SP voltage peak exhibits a characteristic Lorentzian curve symmetric around the resonance field (\textit{H}\textsubscript{res}) when the system reaches the resonance condition, as shown in figure \ref{fig:schematics}c).
To determine the effective magnetization (\textit{M}\textsubscript{eff}) and the Gilbert damping ($\alpha$) of the ferromagnetic layer, the center (\textit{H}\textsubscript{res}) and width ($\Delta H$) of the Lorentzian peak in \textit{V}\textsubscript{SP} were analyzed as a function of frequency using a conventional method of fitting the voltage to a sum of a symmetric and an antisymmetric Lorentzian function \cite{Fache2020,Arango2022a,Damas2022,CespedesBerrocal2021CurrentInducedLayers}. The antisymmetric part of the signal was negligible in the SP measurements, and only the symmetric part was considered in the fit. The Kittel formula for an in-plane easy axis was then used to calculate \textit{M}\textsubscript{eff}. The damping was obtained by considering the linear dependence of $\Delta$H with frequency, where the Gilbert damping ($\alpha$) gives the slope and the intercept, $\Delta H_0$, is the frequency-independent inhomogeneous contribution \cite{arango2022spintocharge}.

By comparing the damping in the system with a reference AlO$_x$(3.5 nm)/Co(5 nm)//Al$_2$O$_3$(0001) sample (Co reference), we can compute the real part of the effective spin-mixing conductance, $g_{\uparrow\downarrow}^{eff}$, for Pt thickness greater than l$_{sf}$ following the standard spin pumping model\cite{Tserkovnyak2002a}:

\begin{equation}
  g_{\uparrow\downarrow}^{eff} = \frac{4\pi M_{s}t_{Co}}{g\mu_B}\left(\alpha_{Co/Pt}-\alpha_{Co}\right),
\end{equation}

where \textit{M}\textsubscript{s} is the saturation magnetization, \textit{t}\textsubscript{Co} the Co reference layer thickness, \textit{g} the Landé g-factor of the magnetic layer, and $\mu_B$ the Bohr magnetron.  $g_{\uparrow\downarrow}^{eff}$ can be obtained by subtracting the total damping of the Co/Pt bilayer ($\alpha\textsubscript{Co/Pt}$) from the damping of the reference Co thin film ($\alpha\textsubscript{Co}$=0.025(1)). In our sample, t$_{Pt}$ varies between 5 and 20 nm, which is significantly larger than the l$_{sf}$ in our Pt layer, around 2 nm, as we will see in section \ref{sec:transportSP}.

In the thermo-spin measurements we are sensing a combination of two different effects: SSE and ANE \cite{Anadon2016,Ramos2015}. They both arise from the thermal gradient-induced spin current. In SSE, a temperature gradient across a magnetic material generates a spin current that can be detected by a non-magnetic material via the inverse spin Hall effect. In the ANE, a temperature gradient across a conducting magnetic material with a spin-orbit coupling generates a transverse charge current that can be detected by measuring the voltage across the sample in the presence of a magnetic field. They both share the same measurement geometry.

Thermo-spin measurements are carried out in these devices using an electromagnet to apply an external in-plane magnetic field (\textit{H}) as shown in figure \ref{fig:SSE}b). A DC current is passed through the outer contacts of the heater and after 5 minutes of stabilization, the resistances of the sample and the heater are carefully monitored using I-V measurements to properly quantify the thermo-spin voltage and heating power. The thermo-spin voltage is monitored using a Keithley 2182a nano voltmeter. 

Pt resistivity ($\rho\textsubscript{Pt}$) along the two in-plane directions was measured in Hall bar devices of 20 $\mu$m  width and 80 $\mu$m length patterned using photolithography and Ar$^-$ion milling. This is shown in the appendix \ref{sec:app1}
 
 \section{Experimental results}
 \subsection{\label{sec:sctruc}Structural characterization}

Layer thickness was evaluated using x-ray reflectivity measurements, proving good agreement with nominal thickness and low roughness as shown in appendix \ref{sec:app1} for single Pt layers with different thickness.

\begin{figure*}
\includegraphics[width=\textwidth]{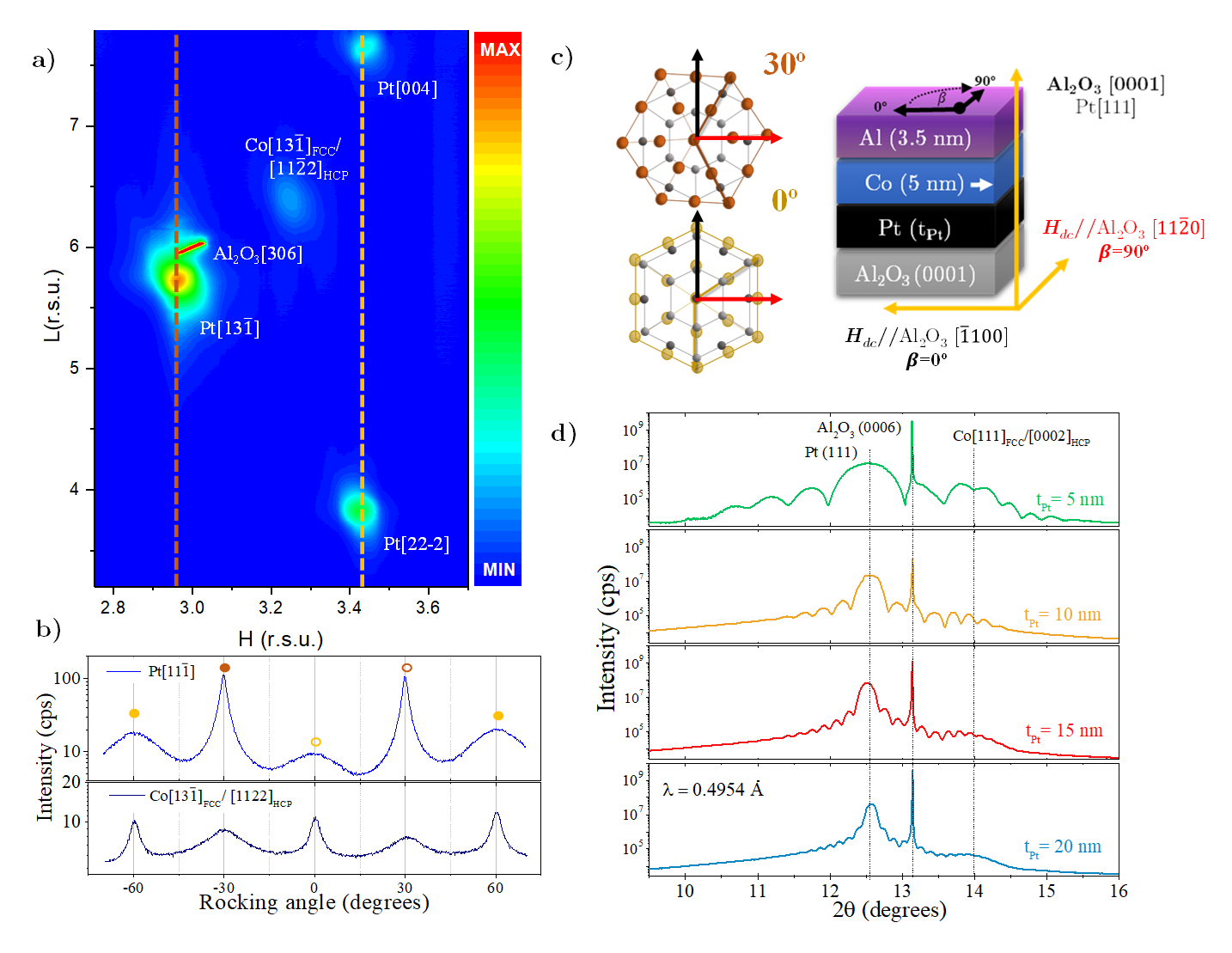}
\caption{\label{fig:XRD} \textbf{Structural characterization by X-ray diffaction.} a) Reciprocal space mapping (RSM) crossing Al$_2$O$_3$[306], Pt[13$\bar{1}$] and Co[13$\bar{1}$] reflections. The apparition of additional Pt[004] and Pt[22$\bar{2}$] is related to the presence of different rotational domains. b) Integration of diffracted intensity as a function of rocking angle for Pt [11$\bar{1}$] and Co[13$\bar{1}$]  reflections. Reflections at 0 and 30 degrees \textit{(bold circles)} show a 6-fold symmetry due to the presence of twin domains \textit{(open circles)}.} c) Schematics of the full stacks indicating crystallographic orientations for  $\beta$=0° and $\beta$=90° [$\bar{1}100$] and [11$\bar{2}$0], respectively. Besides, schematic view of Pt in plane preferential orientations on top of the Al$_2$O$_3$ c-plane. d) Out of plane symmetric $\theta$-2$\theta$ scans showing the Pt(111) and Co(111) reflections. Kiessig fringes appear around diffraction peaks due to low roughness interfaces.
\end{figure*} 

GIXRD measurements using a synchrotron source provide deep knowledge about Pt in-plane structural properties. Reciprocal space maps (figure \ref{fig:XRD}a) covering a large area of the reciprocal space confirms the presence of rotational domains with two main configurations. Most favorable configuration corresponds with a 30 degrees in plane rotation, so the mismatch between the Al$_{2}$O$_{3}$ and Pt lattices is reduced from 13.5$\%$ to an almost perfect coupling with less than 0.2 $\%$. Therefore diffracted intensity for Pt reflections related to this configuration is significantly larger than those related to an axis-on-axis growth, for instance, Pt[131] intensity compared to Pt[004] and Pt[22$\bar{2}$] in figure \ref{fig:XRD}a). Rocking scans around Pt reflections indicate the presence of a 6-fold symmetry instead of a 3-fold symmetry due to the presence of equivalent twin domains rotated by 180$^{o}$, which was previously observed in similar systems \cite{anadon2021b}. Co grows incommensurate axis-on-axis on Pt, showing equivalent rotational domains. Hexagonal close-pack (hcp) is the stable phase for Co at room temperature. Nevertheless, since the resemblance between the packing density \cite{BAKONYI20052509} and the energy \cite{Lizarraga2017} of both hcp and face-centered cubic (fcc) phases, the formation of fcc Co crystals is highly probable as a metastable phase. Co typically undergoes an initial fcc growth, especially when deposited on an epitaxial fcc Pt buffer, followed by a relaxation into the more stable hcp phase, although both usually coexist\cite{sewak2022temperature}. However, the similarity between both structures impedes a categorical distinction by means of XRD measurements, at least considering the particular reciprocal space regions proven in this work.

Figure \ref{fig:XRD}c) shows a schematic view of the final stacks indicating the relation between the Pt lattice and measurement directions, $\beta$=0° and $\beta$=90° in both main configurations. Notably, due to the high degree of symmetry of the Pt lattice and the presence of rotational domains, the $\beta$=0° direction in domains rotated by 30 degrees is equivalent to $\beta$=90° in the non-rotated domains. However, the relevant predominance of one over the other makes both directions clearly non-equivalent. Again, observation of Kiessig fringes around Bragg peaks in out of plane $\theta$-2$\theta$ scans indicates smooth interfaces even after Co deposition (figure \ref{fig:XRD}d).

 \subsection{\label{sec:Kerr}Magnetic characterization}

\begin{figure*}
\includegraphics[trim={0 1cm 1cm 0},width=\textwidth]{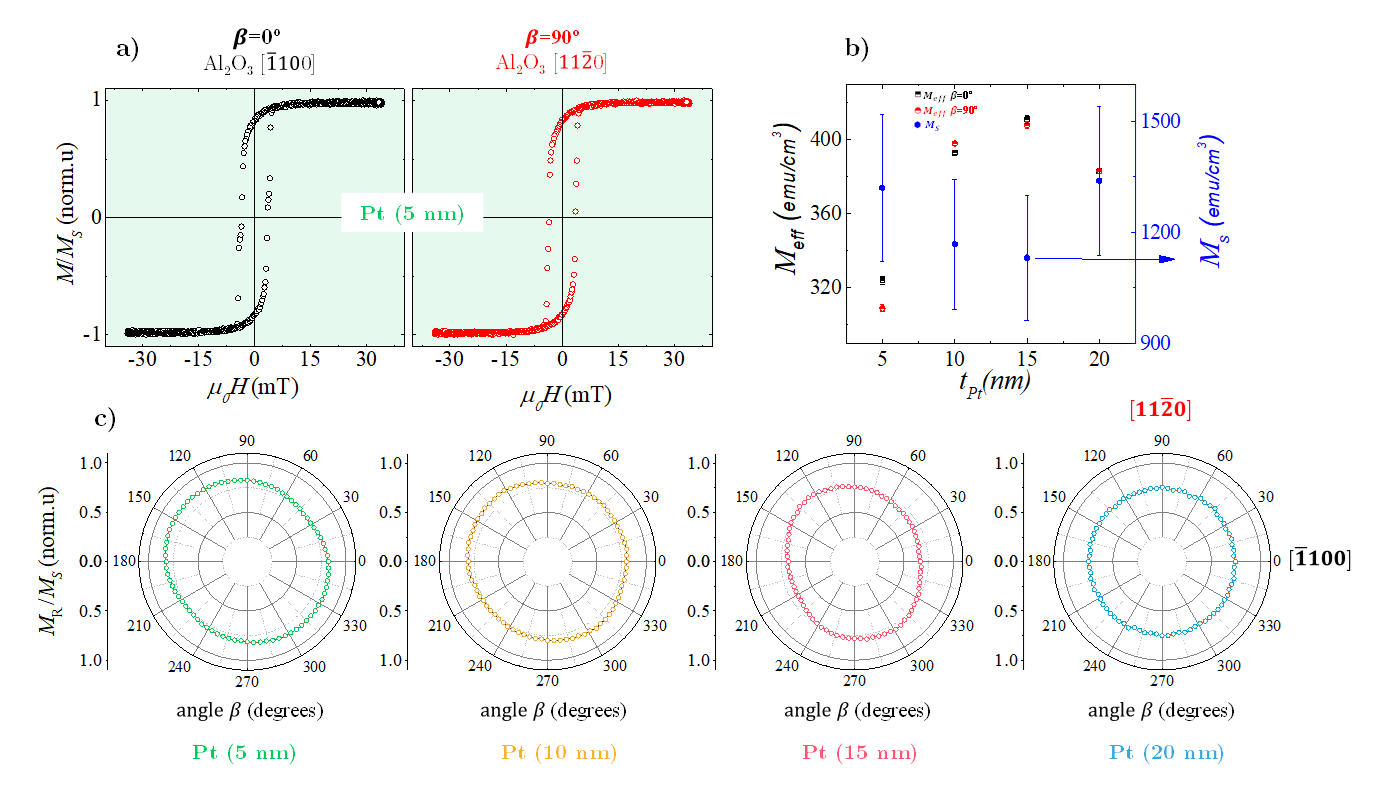}
\caption{\label{fig:KerrRot} \textbf{Magnetic characterization.}  a) Kerr magnetization component parallel to the applied in-plane magnetic field ($\mu_0$H) and normalized to the saturation value M$_S$ acquired in longitudinal geometry at room temperature for each Pt thickness measured along the two in-plane directions $\beta$=0° and $\beta=90$°. b) Evolution of the effective magnetization measured by ST-FMR and the saturation magnetization measured by VSM. c) Angular evolution of the remanenence magnetization for each Pt thickness (right panel). The weak in-plane magnetic anisotropy disappears as the Pt thickness increases.}
\end{figure*}

The magnetic properties of the Al/Co/Pt samples have been measured by VSM and in-plane MOKE (figure \ref{fig:KerrRot}) prior to the optical lithography processes.
In the polar plots of Figure \ref{fig:KerrRot}c) we can observe that there is small uniaxial magnetic anisotropy showing a deviation of M\textsubscript{R}/M\textsubscript{S} of 13\%, 10\%, 18\% and 5\%  for Pt thicknesses from 5 to 20 nm respectively. This may be due to different domains formation, defects, or unexpected changes in the growth conditions. Such anisotropy is much smaller than the difference between M\textsubscript{eff} and M\textsubscript{S} and the Kittel model for an uniaxial easy axis should still hold. Although we cannot elucidate the origin of this small anisotropy and considering that we are extracting the spin conversion parameters from measurements at magnetic fields much higher than this small anisotropy in-plane, we do not expect a significant difference in the extraction of these parameters for the different crystallographic directions, specially for the thicker samples.

Coercivity and remanence magnetization remain similar in all the cases, indicating that the Co layer presents similar magnetic properties regardless of the Pt thickness. An example is shown in figure \ref{fig:KerrRot}a) for the sample with t\textsubscript{Pt}=5 nm.

Figure \ref{fig:KerrRot}b), shows the evolution of \textit{M}\textsubscript{eff} as a function of the Pt thickness. The set of \textit{M}\textsubscript{eff} values found is relatively small, compared to the \textit{M}\textsubscript{S} measured for each thickness by VSM. However, these values are similar to those extracted from SP-FMR measurements. Such difference could arise from a non-negligible out-of-plane magnetization component as expected for hcp Co structure \cite{El-Tahawy2022}. Note that from our x-ray analysis we can not distinguish between the hcp and fcc structures of the Co layer.

The magnetic anisotropy with respect to M\textsubscript{eff} in a uniaxial system is given by:

\begin{equation}
    M_{eff}=M_S-\frac{2K_u}{\mu_0 M_S},
\end{equation}
where K\textsubscript{u} represents the uniaxial anisotropy constant. 

We obtain a uniaxial anisotropy ($2K_u/\mu_0M_S$) of 70$\pm$10 mT for the sample with 15 nm of Pt as an example. Values in the same range are found for the other samples showing a non-monotonic evolution with thickness. We find that these changes in the anisotropy found may be linked to the Co layer's texture and quality, influenced by possible changes in the Pt layer's properties. We verified the consistency of the Landé g-factor and find its value for all the samples within experimental error and the monotonic increase of the interface roughness shown in appendix A indicates that the interface is likey not the origin of the non-monotonic perpendicular component of the anisotropy. These findings suggest that the variations in anisotropy likely result from the non-trivial coexistence of fcc and hcp phases with varying proportions in each sample, as no other factors can fully explain the observed trend in figure \ref{fig:KerrRot}b). We cannot distinguish these two phases either from x-ray measurements as discussed before, nor from magnetization measurements, since the error in the estimation of M\textsubscript{S}  is larger than its difference of around 10\% \cite{BETANCOURTCANTERA20194995}

 \subsection{\label{sec:transport}Transport characterization}

The samples were deposited on 10x10 mm$^2$ Al$_2$O$_3$(0001) substrate being large enough to be patterned into four zones, each one dedicated to each transport measurement technique (ST-FMR, SSE and SP-FMR) with their specific devices arranged in different crystallographic directions as we have detailed previously in section \ref{sec:meth:transport}. We could anticipate similar results across the techniques due to the Onsager reciprocity of the spin Hall effect. 

However, although spin transport in both directions at an interface is expected to be equivalent dynamic processes, a discrepancy is found in the literature regarding the quantification of the efficiencies of interconversion or the spin-orbit torque in nominally identical systems analyzed with different techniques. For example, spin pumping for spin-charge and spin-torque for charge-spin yield different results, and even another widely used technique, the second harmonic, provides different results than spin-torque FMR having both the same physical origin. This variation in results could be attributed to the difficulty to quantify properly the different parameters in each experiment. For example, the exact distribution of the radiofrequency current in ST-FMR or the distribution and amplitude of the radiofrequency field in SP-FMR. Moreover, most of the models used to extract the efficiency rely on the macrospin aproximation for the magnetization dynamics. However, this detailed discussion is beyond the scope of our present study.

\subsubsection{\label{sec:transportST}Spin-torque FMR measurements}
\begin{figure*}
\includegraphics[trim={0.5cm 1cm 0.5cm 0.5cm},width=0.85\textwidth]{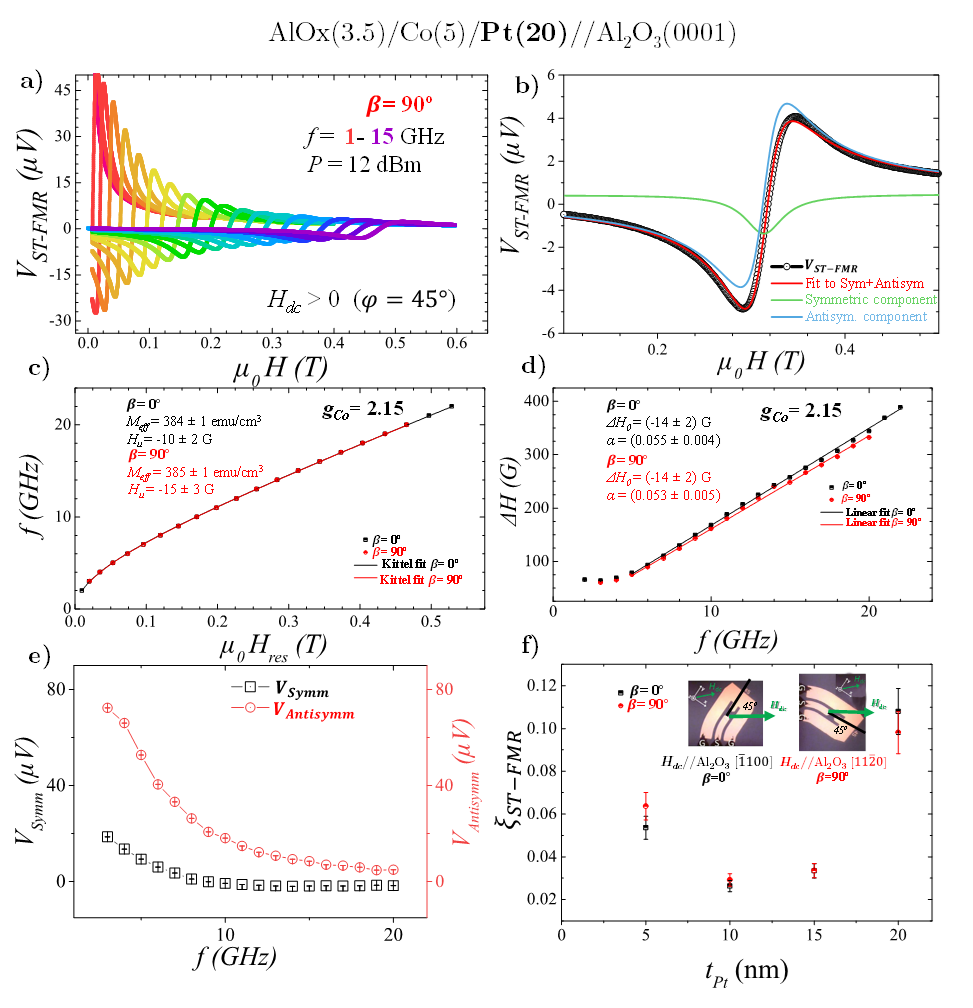}
\caption{\label{fig:STFMR} \textbf{Spin-torque FMR characterization.} a) Raw data of the ST-FMR scans for the case of t$\textsubscript{Pt}$ =20 nm. $H\textsubscript{dc}$ is applied parallel to Al$_2$O$_3$[11$\bar{2}$0], which has been defined as $\beta$=90°. b) Fit, symmetric and antisymmetric components of the V\textsubscript{ST-FMR} of the same sample at 10 GHz. c) Resonance frequency as a function of the resonance field and Kittel fits. Here we can identify the two orthogonal directions studied. d) Determination of Gilbert magnetic damping constant $\alpha$ along the two different crystallographic directions. The damping in both directions is similar within the experimental error. In graphs c and d symbols represent experimental measurements whereas solid lines correspond to fits to the Kittel relation and with a linear fit respectively. e) Evolution of the symmetric and antisymmetric components of V\textsubscript{ST-FMR} as a function of \textit{f}. f) Determination of the effective spin Hall angle from the ST-FMR measurement as a function of the Pt thickness for both crystallographic directions at 15 GHz.}
\end{figure*}

The measured ST voltage (\textit{V}\textsubscript{ST-FMR}) is shown in figure \ref{fig:STFMR}a) for the sample with t\textsubscript{Pt}=20 and it composed of a mixture of symmetric and antisymmetric Lorentzian functions \ref{fig:STFMR}b) around the resonance field $H\textsubscript{res}$. The symmetric Lorentzian amplitude is linked to the damping-like (DL) torque, while the antisymmetric Lorentzian is linked to the torques created by the Oersted field and the field-like (FL) torques, which are parallel \cite{Pai2015,guillemard2018}.  Generally, the antisymmetric component dominates, as it is the case in our experiments. From the estimation of the center and width of the resonance peaks we can estimate the effective magnetization (Fig. \ref{fig:STFMR}c)) and Gilbert damping constant (Fig. \ref{fig:STFMR}d)). We have measured V\textsubscript{ST-FMR} for different frequencies from 3 to 20 GHz for all the samples and extracted the \textit{M}\textsubscript{eff} and damping determination for every thickness in both crystallographic directions. We find typically similar values for both magnitudes regardless of the crystalline direction. Experimentally, for each given thickness of the Pt layer, we obtain that $g_{\uparrow\downarrow}^{eff}$ is always within the $\approx$10\% experimental error for the different crystallographic directions. We thus, cannot infer a dependence of the effective spin mixing conductance with the crystallographic direction. Comparing the different Pt thicknesses, we find a value for $g_{\uparrow\downarrow}^{eff}$ of 1.2$\pm$0.6$\cdot 10^{20}$ $m^{-2}$, where the deviation is probably related to small differences in the growth conditions. This has been previously shown in literature\cite{Mizukami_2011,rojas2014spin}.

We can also estimate from this the amplitude of both components as a function of the frequency, as depicted in figure \ref{fig:STFMR}e). At lower frequencies, the estimation of this components is not as reliable due to the proximity of the ${H}$=0, where the thermal effects appear. This is especially important in Co considering the large width of the peak compared to other metallic FM like Py. That is the reason for the symmetric component to be positive at lower frequencies and negative after 8 GHz. From the symmetric and antisymmetric components, we can calculate the so-called effective spin hall angle $\xi_{ST-FMR}$ \cite{Pai2015,PhysRevLett.116.127601,Damas2022}. It is important to notice that this parameter is not the same as $\theta_{SH}$, since it includes some approximations for its calculation \cite{PhysRevApplied.11.024039,PhysRevApplied.12.044074,guillemard2018} and has a mixed contribution from the field-like torque.  Although the power injected in the device is not the same for all frequencies, the effect of input power should be compensated as $\xi_{ST-FMR}$ is calculated as the ratio between the symmetric and antisymmetric components. 

Figure \ref{fig:STFMR}f) shows the $\xi_{ST-FMR}$ values extracted from the  comparison of the symmetric and antisymmetric parts of the ST-FMR peak at 15 GHz as \cite{Liu2011a,Pai2015,Damas2022}:
\begin{equation}
\xi_{ST-FMR}=\frac{V\textsubscript{sym}}{V\textsubscript{anti}}\frac{e}{\hbar}\mu_0M_st_{Co}t_{Pt}\sqrt{1+4\pi M_{eff}/H_{res}}
\end{equation} 

At first glance, we can observe that $\xi_{ST-FMR}$ presents lower values for the intermediate thicknesses, being higher for 5 nm and 20 nm. On the other hand, the difference between crystallographic directions is appreciable for 5 nm and 20 nm, within their error bars, while for the intermediate thicknesses, hardly any differences have been found. This observation is in accordance with a recent paper in Py/Pt with Pt in the [111] direction \cite{Xiao2022}, where they find an isotropic behaviour by ST-FMR using the same type of analysis. In addition, the estimated values do not reflect a clear trend. Nonetheless, all of them are within the range found in the literature \cite{Kondou2012,Ganguly2014,Xiao2022}.

This method may give rise to a wrong estimation of $\theta_{SHE}$ for the case of a non-negligible FL torque, which might arise due to the interfacial effects. Since the FL torque can also produce an antisymmetric Lorentzian line shape signal similar to $h_{rf}$, the value of $\xi_{ST-FMR}$ might be overestimated or underestimated from the method $V\textsubscript{sym}/V\textsubscript{anti}$  \cite{Damas2022,skinner2014, allen2015}. Regardless of this, it can provide relevant information about the directionality of the effect \cite{guillemard2018,PhysRevApplied.12.044074,PhysRevB.98.094407}.

\subsubsection{\label{sec:transportSSE}Thermo-spin measurements.}

\begin{figure*}
\includegraphics[trim={0.5cm 1cm 0cm 0.5cm},width=\textwidth]{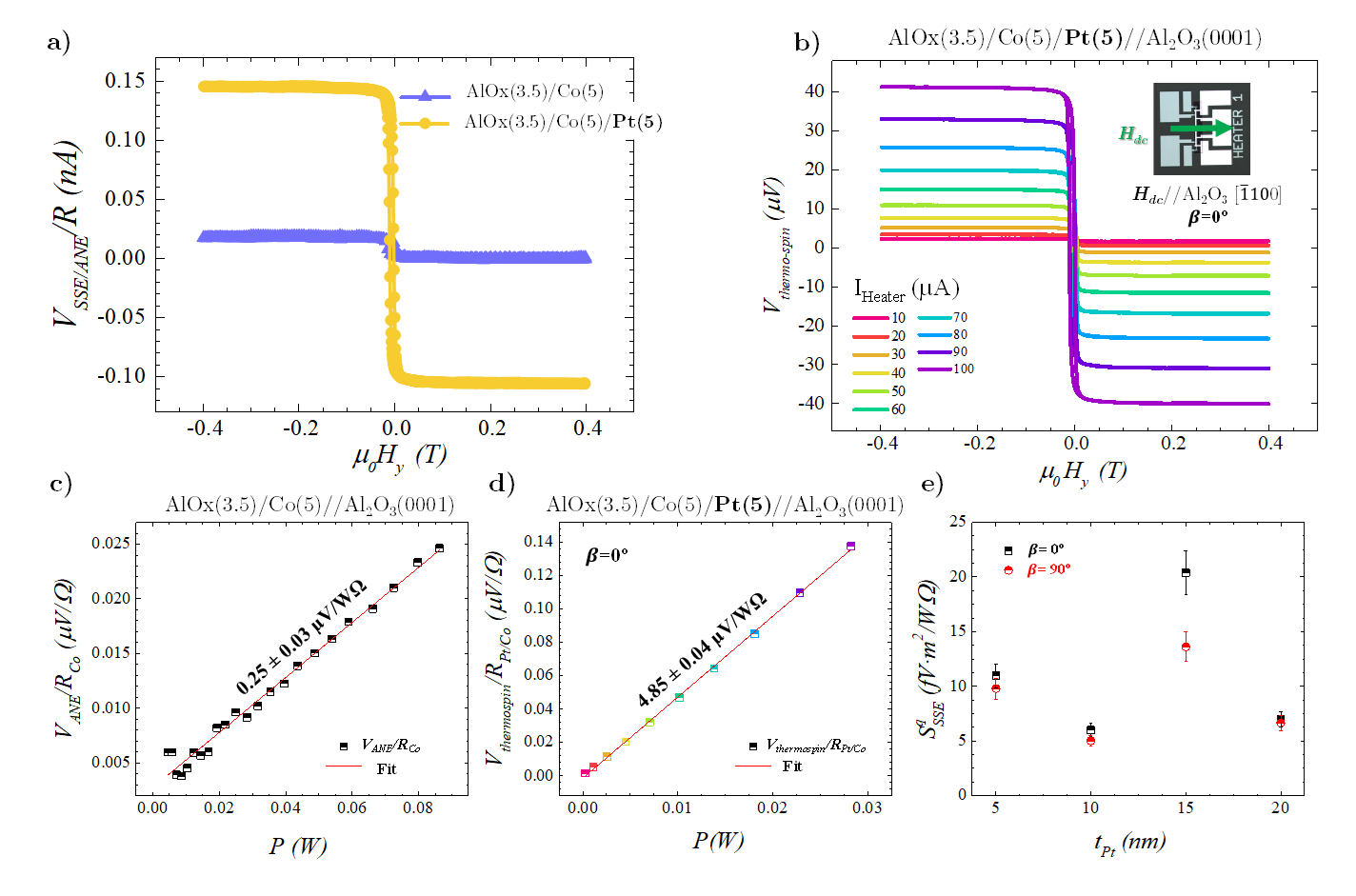}
\caption{\label{fig:SSE} \textbf{Thermo-spin measurements.} a) Thermo-spin voltage divided by the resistance of the device in AlO$_x$(3.5 nm)/Co(5 nm)/Pt(5 nm)//Al$_2$O$_3$(0001) (ANE+SSE) and AlO$_x$(3.5 nm)/Co(5 nm)//Al$_2$O$_3$(0001) (ANE only) at the same heater current of 100 mA. b) Raw data of the V$\textsubscript{thermo-spin}$ for different heater powers at AlO$_x$(3.5 nm)/Co(5 nm)/Pt(5 nm)//Al$_2$O$_3$(0001) when H$\textsubscript{dc}$ is applied parallel to Al$_2$O$_3$[$\bar{1}$100] ($\beta$=0°). c) V$\textsubscript{ANE}$ as a function of the heater power at the reference sample AlO$_x$(3.5 nm)/Co(5 nm)//Al$_2$O$_3$(0001), where the V$\textsubscript{SSE}$ does not appears due to the lack of HM layer. d) V$\textsubscript{thermo-spin}$ as a function of the heater power. e) SSE coefficient as a function of Pt thickness for the two crystallographic directions.}
\end{figure*}

While ST-FMR is a widely employed method for estimating spin conversion efficiency exploiting the direct SHE, we aim to explore additional options to gain a more comprehensive understanding studying also the inverse effect. Another commonly used method for this purpose is thermo-spin measurements, which differ from ST- and SP-FMR as they originate from incoherent thermal excitation. In our study, we conducted thermo-spin measurement at different heater currents ranging from 10 mA to 100 mA in the two different chosen crystallographic directions for all the Pt thickness. 

The thermo-spin voltage,  \textit{V}\textsubscript{thermo-spin}, presents an antisymmetric behavior with the magnetic field following the magnetic hysteresis loop of the Co layer as depicted in figures \ref{fig:SSE}a) and b). Some of the samples could present a small offset in the voltage loop that is due to a small misalignment between the heater and the generating spurious thermal gradients in other directions and is generally smaller than \textit{V}\textsubscript{thermo-spin}. The antisymmetric component corresponds to the sum of the voltage coming from the ISHE in Pt (i.e. the SSE) and the ANE from the Co layer. In order to estimate the \textit{V}\textsubscript{SSE}, we need to subtract the contribution from ANE $V^{contr}_{ANE}$ for each sample as  \cite{Ramos2013,Anadon2021} given by:
\begin{equation}
V^{contr}_{ANE}=(r/1+r)V\textsubscript{ANE},\label{eq:ANEcontr}
\end{equation}
 where V\textsubscript{ANE} is the ANE voltage of a single metallic FM layer, as commented above and $r=(\rho_{HM}/\rho_{FM})(t_{FM}/t_{HM})$, with $\rho_HM$ and $\rho_FM$ representing the Pt and Co resistivities and $t_{HM}$ and $t_{FM}$ representing their thickness, respectively.
 
In figure \ref{fig:SSE}, we show a comparison of \textit{V}\textsubscript{thermo-spin} divided the device resistance between a reference 5 nm Co layer without Pt and the sample with t\textsubscript{Pt}=5 nm. The SSE clearly dominates over the ANE. In figure \ref{fig:SSE}b), we show the evolution of \textit{V}\textsubscript{thermo-spin} as a function of the heater current for the same sample (t\textsubscript{Pt}=5 nm).

Figure \ref{fig:SSE}c) and d) shows the $V^{contr}_{ANE}$ and \textit{V}\textsubscript{thermo-spin} as a function of the heater power. In both cases, we represent the voltage difference between the positive and negative saturation fields, divided by two. To determine \textit{V}\textsubscript{SSE}, we subtracted the contribution of $V^{contr}{ANE}$ for each thickness of Pt using the previous equation. However, for a more accurate determination of the spin Seebeck effect coefficient, it is preferable to consider the heat flux instead of the thermal gradient  \cite{Sola2017,Anadon2022ThermalGa_0.6Fe_1.4O_3b}. Moreover, on-chip devices with lithographed heaters have also been utilized in this type of experiment, enhancing reproducibility between samples  \cite{Wu2015, luo2021}.

We can define the SSE coefficient as follows $S^{q}_{SSE}=V\textsubscript{SSE}/\phi_qR_{Pt}$   \cite{Anadon2022ThermalGa_0.6Fe_1.4O_3b}, where V\textsubscript{SSE} is the difference between \textit{V}\textsubscript{thermo-spin} and $V^{contr}_{ANE}$, $\phi_q$ is the heat flux through the FM considering that there are no thermal losses due to radiation, and $R_{Pt}$ is the sample resistance by 4-probes. The value of the SSE coefficient is represented in figure \ref{fig:SSE}e). 

The $S^{q}_{SSE}$ coefficient can be qualitatively compared to $\theta_{SH}$ obtained from SP-FMR and  $\xi_{ST-FMR}$ derived from ST-FMR, providing valuable insights into the efficiency of spin-charge interconversion.  Changes in the Co phase composition between samples are known to have a drastic effect in the anisotropic magnetoresistance\cite{ELTAHAWY2022169660} and, consequently, the ANE. This could account for the changes in the $S^{q}_{SSE}$ coefficient as a function of t$_{Pt}$, but information for different crystallographic directions for each sample is still reliable under this consideration. Thus, our observations reveal a predominantly isotropic behavior in the system.  Typically, thermo-spin measurements serve as a reliable technique to validate the generation of spin currents in insulating thick ferrimagnets, where the interfacial contribution constitutes a small fraction of the voltage \cite{Jimenez-Cavero2021a}.  In the case of  conducting ferromagnets like Co, there can be a contribution from the magnetic proximity effect. This contribution should not be significantly different from sample to sample in our study, since magnetic proximity normally does not depend on the thickness of Pt when above a few nm.

\subsubsection{\label{sec:transportSP}Spin pumping measurements}

Among the three techniques, spin pumping allows for the most careful quantification of the spin conversion. It arises from coherent excitation and is extremely sensitive to interfacial effects, since the spin current injected in the NM comes directly from the precessing electrons at the FM/NM interface \cite{Tserkovnyak2002a}. From the SP voltage (\textit{V}\textsubscript{SP-FMR}) measured, we can extract the charge current (\textit{I}\textsubscript{c}) generated in the systems just dividing it by the resistance of the FM/HM slab\cite{Fache2020,rojas2014spin}:

\begin{equation}
\label{eq:SPIc}
    I\textsubscript{C}=V\textsubscript{SP-FMR}/R=Wl_{sf}^{Pt}\theta_{SH}J_s^{eff}tanh(t_{Pt}/2l_{sf}^{Pt}),
\end{equation}
where $R=R\textsubscript{sheet}L/W$, $W= 10 \mu$m is the width of the device, $l_{sf}^{Pt}$ is the spin diffusion length and $\theta_{SH}$ the effective spin Hall angle of Pt. J\textsubscript{eff} is the effective spin current injected at the interface with the NM layer and is given by \cite{Fache2020,rojas2014spin}:

\begin{equation}
J_s^{eff}=\frac{eg_{\uparrow\downarrow}\gamma^2h_{RF}^2}{4\pi\alpha^2}\frac{\gamma\mu_0 M_{eff}+\sqrt{(\gamma\mu_0M_{eff})^2+4\omega^2}}{(\gamma\mu_0 M_{eff})^2+4\omega^2},
\end{equation}
where e is the electron charge, $\gamma$ the gyromagnetic ratio and $\omega=f/2\pi$.

\begin{figure*}
\includegraphics[width=\textwidth]{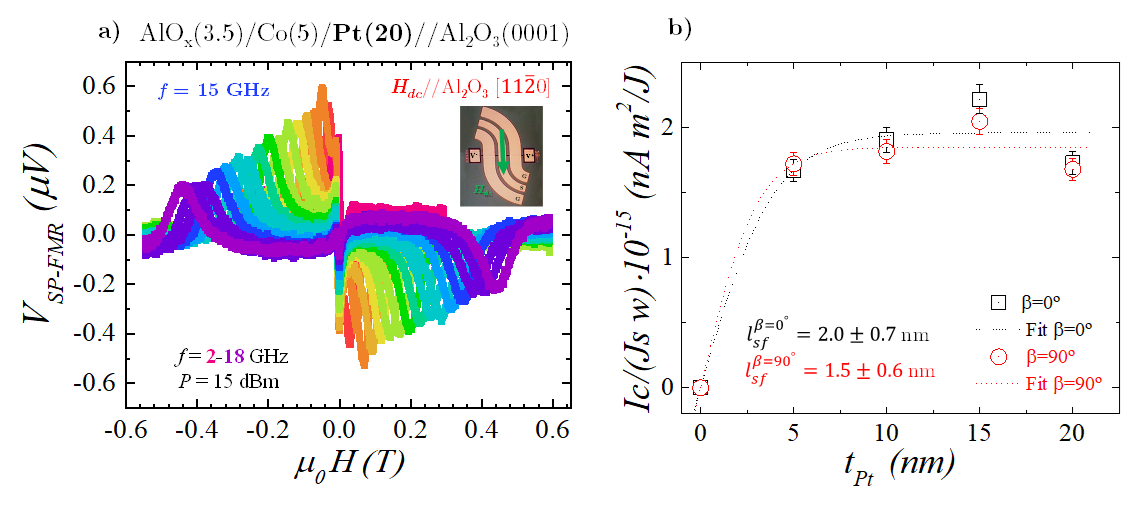}
\caption{\label{fig:SPFMR} \textbf{Spin-pumping FMR characterization.} a) Spin pumping voltage for AlO$_x$(3.5 nm)/Pt(20 nm)/Co(5 nm)// Al$_2$O$_3$(0001) at $\beta$=90° with f = 2-18 GHz at P = 15 dBm. b) Estimation of the spin diffusion length from spin pumping charge current divided by the spin current injected in the Pt and the width of the device as a function of the Pt thickness (t$\textsubscript{Pt}$) for 15 GHz at the two orthogonal crystallographic directions.}
\end{figure*}

As shown in figure \ref{fig:STFMR}d), the Gilbert damping and thus the spin mixing conductance do not depend on the crystallographic direction. As expected, this also holds for the spin pumping measurements. In Figure \ref{fig:SPFMR}b), we present the relationship between the spin pumping charge current, normalized by the injected spin current in Pt and the width of the device, and the Pt thickness in Co/Pt epitaxial bilayers for the two crystallographic directions under examination. By analyzing this behavior using equation \ref{eq:SPIc}, we extract valuable insights into the symmetry of both $\theta_{SH}$ and $l_{sf}$. The extracted values for $l_{sf}^{\beta=0\degree}= 2.0 \pm 0.7$ nm and $l_{sf}^{\beta=90\degree}= 1.5 \pm 0.6$ nm are consistent with experimental measurements reported in the literature using SP-FMR, which range from 0.5 nm \cite{boone2013} to 10 nm \cite{mosendz2010}. Our findings show no significant differences between the two orientations, as both values fall within the error range and are comparable to each other, albeit slightly lower than other epitaxial Pt \cite{Sagasta2016}. The product $\rho  l_{sf}$ typically remains below 0.61$f\Omega m^2$ (0.4$\pm$0.1 $f\Omega m^2$ for the 5-nm Pt sample), the theoretical limit when the Elliott-Yafet mechanism dominates over Dyakonov-Perel \cite{PhysRevLett.113.207202}. Although most experimental work supports the dominance of the Elliott-Yafet mechanism \cite{Sagasta2016,rojas2014spin,10.1063/1.4898593,boone2013}, recent studies have suggested a significant Dyakonov-Perel contribution at low Pt thicknesses \cite{PhysRevMaterials.5.064404}. Furthermore, the lower value of $l_{sf}$ implies a likely higher $\theta_{SH}$ since the product $l_{sf}\theta_{SH}$ is generally considered to be around 0.2 nm for Pt \cite{Sagasta2016,rojas2014spin}.

From this fit, where we have considered the individual values of $J_s^{eff}$ and $g_{\uparrow\downarrow}^{eff}$ for each Pt thickness, we directly obtain $\theta_{SH}^{\beta=0\degree}= 0.07 \pm 0.02$ and $\theta_{SH}^{\beta=90\degree}= 0.08 \pm 0.02$. These values correlate fairly well with previous results in the literature for the spin hall angle of Pt  \cite{guillemard2018,Zhang2015role,Obstbaum2014,PhysRevMaterials.5.064404} further supporting the isotropic nature of spin conversion in epitaxial (111)-oriented Pt within our error estimates. We find thus a $l_{sf}\theta_{SH}$ of 0.14$\pm$0.06 nm, in accordance with literature \cite{Sagasta2016,rojas2014spin}. From X-ray reflectivity measurements, we observe a slight increase in the roughness of the Pt layer as the Pt thickness increases, as shown in appendix A. Although this could potentially lead to changes in the transparency of the interface and the surface magnetic anisotropy, it's noteworthy that we do not observe a significant alteration in the effective spin mixing conductance ($g_{\uparrow\downarrow}^{eff}$) or a monotonic change of the effectiive magnetization for the samples with different Pt thicknesses, as we have discussed in section \ref{sec:Kerr} and \ref{sec:meth:transport}. This suggests that while there may be variations in interface roughness, they do not have a pronounced impact on the spin transport properties of the Pt/Co interface.

The observations in this study are coherent with  \cite{xiao2022crystalline} in the same crystallographic directions of Pt using the lineshape analisys in ST-FMR. They find similarly that the spin conversion is isotropic in (001)-oriented Pt and only find some anisotropy in (220)-oriented Pt (also in  \cite{PhysRevApplied.15.014055}). Similarly, in  \cite{guillemard2018} they have also shown isotropic behaviour in epitaxial and cubic Pt/Fe//MgO along in-plane directions for (001)-oriented Pt. Our work confirms within experimental error that the conversion between spin current to charge current, direct and inverse effects, in epitaxial (111)-oriented Pt is isotropic in the two in-plane directions studied on epitaxial Co/Pt.



 \section{Conclusions}
There are clear discrepancies in the literature regarding spin conversion in epitaxial Pt. We aim to clarify this observations in the case of the Co/Pt system, which is one of the most studied in spintronics. To do so, we have prepared a set of epitaxial samples consisting of AlO$_x$(3.5 nm)/Co(5 nm)/Pt(t$_\text{Pt}$)//Al$_2$O$_3$(0001) stacks with different thickness of the Pt buffer layer by DC sputtering. We confirm the epitaxial growth of the Pt buffer layers, their thickness and crystallographic properties using X-ray diffraction measurements. We characterized the Pt layers by resistivity measurements along the [$\bar{1}$100] and [11$\bar{2}$0] crystallographic directions of Al$_2$O$_3$(0001), finding similar results in both directions. 

The results reported in literature for the spin conversion in epitaxial Pt interfaced with different magnetic materials differ significantly. Thus, we have determined the spin Hall efficiency using three complementary experimental techniques: spin pumping ferromagnetic resonance, spin-torque ferromagnetic resonance, and thermo-spin measurements along two orthogonal non-equivalent in-plane crystallographic directions as a function of Pt thickness. These techniques cover the direct and inverse effects as well as coherent and incoherent excitation in the ferromagnet and are sensitive to interfacial effect. They all yield compatible results within experimental error, showing no significant anisotropy in the spin conversion behavior. Our findings provide important insights into the spin conversion properties of epitaxial Pt and contribute to a better understanding of spintronics in epitaxial systems.

\begin{acknowledgments}
AG and AA thank the funding by the IEEE Magnetics Society Educational Seed Funding.
This research is supported by the FLAG-ERA grant SOgraphMEM Project PCI2019-111867-2. IMDEA team acknowledges support by the Regional Government of Madrid through Project P2018/NMT-4321 (NANOMAGCOST-CM) and by the Spanish MICINN Projects PID2021-122980OB-C52 (ECLIPSE-ECoSOx), CNS2022-136143 (SPINCODE) and the 'Severo Ochoa' Programme for Centres of Excellence in R\&D (CEX2020-001039-S). This work was also supported by the French National Research Agency (ANR) through the ANR-19-CE24-0016-01 ‘Toptronic ANR’. Devices in the present study were patterned at Institut Jean Lamour's clean room facilities (MiNaLor). These facilities are partially funded by FEDER and Grand Est region through the RANGE project. The authors also acknowledge The European Synchrotron (ESRF) for provision of synchrotron radiation facilities and we would like to thank Dr. Juan Rubio-Zuazo and beamline staff for assistance and support in using beamline BM25-SpLine. We thank prof. E. Fullerton for fruitful discussions.
\end{acknowledgments}

\appendix
\section{X-ray reflectivity and resitivity of the Pt thin films for $\beta=0$ and 90\degree }
\label{sec:app1}

\begin{figure*}
\includegraphics[width=\textwidth]{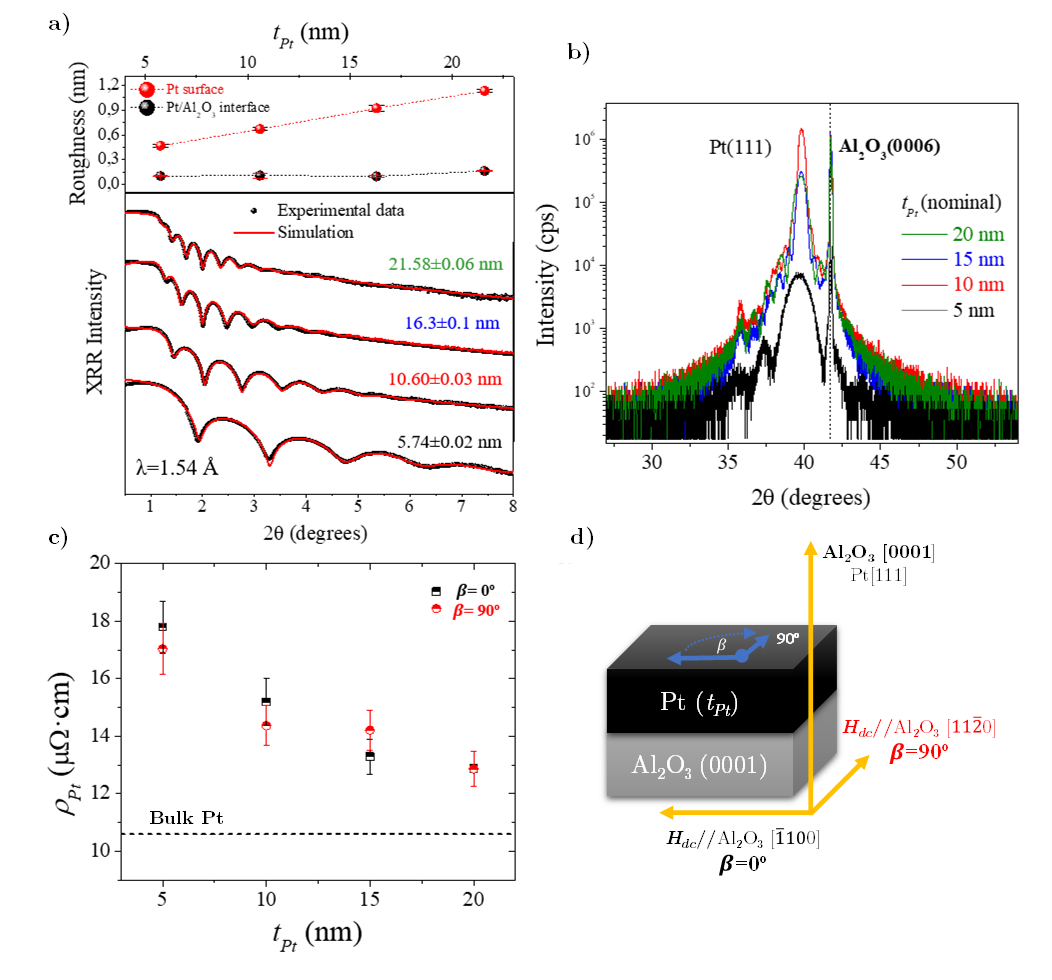}
\caption{\label{fig:XRR} \textbf{Pt/Co characterization by different techniques.}   a)  X-ray reflectivity (XRR) measurements for the different thickness of Pt together with thickness and roughness values obtained from simulation. b) X-ray diffraction (XRD) scans of Pt(t$\textsubscript{Pt}$)/Al$_2$O$_3$(0001)}. ($\otimes$) Indicates Pt(111) contribution from Cu K-$\beta$ radiation. c) Resistivity of the epitaxial Pt($\rho\textsubscript{Pt}$) for two different crystallographic directions as a function of the thickness of the Pt and its bulk resistivity \cite{Agustsson2008ElectricalSiO2}. d) Schematic of the epitaxial Pt control samples. 
\end{figure*}
Figure \ref{fig:XRR} shows Pt layer characterization prior to Co deposition. Layer thickness was evaluated using XRR, proving good agreement with nominal thickness and the deposition rate determined by quartz balance (figure \ref{fig:XRR}a). Fitting of the XRR patterns show an increase of Pt surface roughness with increasing thickness going from $0.47\pm0.02$ nm for thinnest nominal thickness ($5$ nm) to $1.13\pm0.02$ nm for the thickest nominal thickness (20 nm).  Out of plane $\theta$-2$\theta$ scans shown in figure \ref{fig:XRR}b) prove epitaxial growth of the Pt film with Pt[111] axis parallel to Al$_2$O$_3$[0006] for all chosen thicknesses. The apparition of Kiessig fringes, not only at low angle but also around Bragg peaks, indicate low surface and interface roughness. 

One of the possible origins of asymmetry in spin conversion could arise from different carrier scattering in the two different crystallographic directions $\alpha$ and $\beta$ in Pt. To account for this option, figure \ref{fig:XRR}c) shows Pt resistivity values measured in both $\beta$=0° and $\beta$=90°, corresponding to in plane orthogonal and non-equivalent crystallographic directions (figure \ref{fig:XRR}d). Although small variations are observed between $\beta$=0° and $\beta$=90° directions, the resistivities and the effective spin mixing conductance are within the error bars in all the cases, pointing to an isotropic behavior. 

\newpage

\bibliography{apssamp}

\begin{thebibliography}{89}%
\makeatletter
\providecommand \@ifxundefined [1]{%
 \@ifx{#1\undefined}
}%
\providecommand \@ifnum [1]{%
 \ifnum #1\expandafter \@firstoftwo
 \else \expandafter \@secondoftwo
 \fi
}%
\providecommand \@ifx [1]{%
 \ifx #1\expandafter \@firstoftwo
 \else \expandafter \@secondoftwo
 \fi
}%
\providecommand \natexlab [1]{#1}%
\providecommand \enquote  [1]{``#1''}%
\providecommand \bibnamefont  [1]{#1}%
\providecommand \bibfnamefont [1]{#1}%
\providecommand \citenamefont [1]{#1}%
\providecommand \href@noop [0]{\@secondoftwo}%
\providecommand \href [0]{\begingroup \@sanitize@url \@href}%
\providecommand \@href[1]{\@@startlink{#1}\@@href}%
\providecommand \@@href[1]{\endgroup#1\@@endlink}%
\providecommand \@sanitize@url [0]{\catcode `\\12\catcode `\$12\catcode `\&12\catcode `\#12\catcode `\^12\catcode `\_12\catcode `\%12\relax}%
\providecommand \@@startlink[1]{}%
\providecommand \@@endlink[0]{}%
\providecommand \url  [0]{\begingroup\@sanitize@url \@url }%
\providecommand \@url [1]{\endgroup\@href {#1}{\urlprefix }}%
\providecommand \urlprefix  [0]{URL }%
\providecommand \Eprint [0]{\href }%
\providecommand \doibase [0]{https://doi.org/}%
\providecommand \selectlanguage [0]{\@gobble}%
\providecommand \bibinfo  [0]{\@secondoftwo}%
\providecommand \bibfield  [0]{\@secondoftwo}%
\providecommand \translation [1]{[#1]}%
\providecommand \BibitemOpen [0]{}%
\providecommand \bibitemStop [0]{}%
\providecommand \bibitemNoStop [0]{.\EOS\space}%
\providecommand \EOS [0]{\spacefactor3000\relax}%
\providecommand \BibitemShut  [1]{\csname bibitem#1\endcsname}%
\let\auto@bib@innerbib\@empty
\bibitem [{\citenamefont {Brataas}\ \emph {et~al.}(2012)\citenamefont {Brataas}, \citenamefont {Kent},\ and\ \citenamefont {Ohno}}]{Brataas2012}%
  \BibitemOpen
  \bibfield  {author} {\bibinfo {author} {\bibfnamefont {A.}~\bibnamefont {Brataas}}, \bibinfo {author} {\bibfnamefont {A.~D.}\ \bibnamefont {Kent}},\ and\ \bibinfo {author} {\bibfnamefont {H.}~\bibnamefont {Ohno}},\ }\href {https://doi.org/10.1038/nmat3311} {\bibfield  {journal} {\bibinfo  {journal} {Nature Materials}\ }\textbf {\bibinfo {volume} {11}},\ \bibinfo {pages} {372} (\bibinfo {year} {2012})}\BibitemShut {NoStop}%
\bibitem [{\citenamefont {Garello}\ \emph {et~al.}(2013)\citenamefont {Garello}, \citenamefont {Miron}, \citenamefont {Avci}, \citenamefont {Freimuth}, \citenamefont {Mokrousov}, \citenamefont {Bl{\"{u}}gel}, \citenamefont {Auffret}, \citenamefont {Boulle}, \citenamefont {Gaudin},\ and\ \citenamefont {Gambardella}}]{Garello2013}%
  \BibitemOpen
  \bibfield  {author} {\bibinfo {author} {\bibfnamefont {K.}~\bibnamefont {Garello}}, \bibinfo {author} {\bibfnamefont {I.~M.}\ \bibnamefont {Miron}}, \bibinfo {author} {\bibfnamefont {C.~O.}\ \bibnamefont {Avci}}, \bibinfo {author} {\bibfnamefont {F.}~\bibnamefont {Freimuth}}, \bibinfo {author} {\bibfnamefont {Y.}~\bibnamefont {Mokrousov}}, \bibinfo {author} {\bibfnamefont {S.}~\bibnamefont {Bl{\"{u}}gel}}, \bibinfo {author} {\bibfnamefont {S.}~\bibnamefont {Auffret}}, \bibinfo {author} {\bibfnamefont {O.}~\bibnamefont {Boulle}}, \bibinfo {author} {\bibfnamefont {G.}~\bibnamefont {Gaudin}},\ and\ \bibinfo {author} {\bibfnamefont {P.}~\bibnamefont {Gambardella}},\ }\href {https://doi.org/10.1038/nnano.2013.145} {\bibfield  {journal} {\bibinfo  {journal} {Nature nanotechnology}\ }\textbf {\bibinfo {volume} {8}},\ \bibinfo {pages} {587} (\bibinfo {year} {2013})}\BibitemShut {NoStop}%
\bibitem [{\citenamefont {Liu}\ \emph {et~al.}(2012)\citenamefont {Liu}, \citenamefont {Pai}, \citenamefont {Li}, \citenamefont {Tseng}, \citenamefont {Ralph},\ and\ \citenamefont {Buhrman}}]{Liu2012}%
  \BibitemOpen
  \bibfield  {author} {\bibinfo {author} {\bibfnamefont {L.}~\bibnamefont {Liu}}, \bibinfo {author} {\bibfnamefont {C.-F.}\ \bibnamefont {Pai}}, \bibinfo {author} {\bibfnamefont {Y.}~\bibnamefont {Li}}, \bibinfo {author} {\bibfnamefont {H.~W.}\ \bibnamefont {Tseng}}, \bibinfo {author} {\bibfnamefont {D.~C.}\ \bibnamefont {Ralph}},\ and\ \bibinfo {author} {\bibfnamefont {R.~A.}\ \bibnamefont {Buhrman}},\ }\href {https://doi.org/10.1126/science.1218197} {\bibfield  {journal} {\bibinfo  {journal} {Science}\ }\textbf {\bibinfo {volume} {336}},\ \bibinfo {pages} {555} (\bibinfo {year} {2012})}\BibitemShut {NoStop}%
\bibitem [{\citenamefont {Liu}\ \emph {et~al.}(2011{\natexlab{a}})\citenamefont {Liu}, \citenamefont {Zhang}, \citenamefont {Carter},\ and\ \citenamefont {Xiao}}]{Liu2011}%
  \BibitemOpen
  \bibfield  {author} {\bibinfo {author} {\bibfnamefont {X.}~\bibnamefont {Liu}}, \bibinfo {author} {\bibfnamefont {W.}~\bibnamefont {Zhang}}, \bibinfo {author} {\bibfnamefont {M.~J.}\ \bibnamefont {Carter}},\ and\ \bibinfo {author} {\bibfnamefont {G.}~\bibnamefont {Xiao}},\ }\href {https://doi.org/10.1063/1.3615961} {\bibfield  {journal} {\bibinfo  {journal} {Journal of Applied Physics}\ }\textbf {\bibinfo {volume} {110}},\ \bibinfo {pages} {033910} (\bibinfo {year} {2011}{\natexlab{a}})}\BibitemShut {NoStop}%
\bibitem [{\citenamefont {Hirsch}(1999)}]{Hirsch1999}%
  \BibitemOpen
  \bibfield  {author} {\bibinfo {author} {\bibfnamefont {J.}~\bibnamefont {Hirsch}},\ }\href {https://doi.org/10.1103/PhysRevLett.83.1834} {\bibfield  {journal} {\bibinfo  {journal} {Physical Review Letters}\ }\textbf {\bibinfo {volume} {83}},\ \bibinfo {pages} {1834} (\bibinfo {year} {1999})}\BibitemShut {NoStop}%
\bibitem [{\citenamefont {Sinova}\ \emph {et~al.}(2015)\citenamefont {Sinova}, \citenamefont {Valenzuela}, \citenamefont {Wunderlich}, \citenamefont {Back},\ and\ \citenamefont {Jungwirth}}]{Sinova2015}%
  \BibitemOpen
  \bibfield  {author} {\bibinfo {author} {\bibfnamefont {J.}~\bibnamefont {Sinova}}, \bibinfo {author} {\bibfnamefont {S.~O.}\ \bibnamefont {Valenzuela}}, \bibinfo {author} {\bibfnamefont {J.}~\bibnamefont {Wunderlich}}, \bibinfo {author} {\bibfnamefont {C.~H.}\ \bibnamefont {Back}},\ and\ \bibinfo {author} {\bibfnamefont {T.}~\bibnamefont {Jungwirth}},\ }\href {https://doi.org/10.1103/RevModPhys.87.1213} {\bibfield  {journal} {\bibinfo  {journal} {Reviews of Modern Physics}\ }\textbf {\bibinfo {volume} {87}},\ \bibinfo {pages} {1213} (\bibinfo {year} {2015})}\BibitemShut {NoStop}%
\bibitem [{\citenamefont {Hoffmann}(2013)}]{Hoffmann2013}%
  \BibitemOpen
  \bibfield  {author} {\bibinfo {author} {\bibfnamefont {A.}~\bibnamefont {Hoffmann}},\ }\href {https://doi.org/10.1109/TMAG.2013.2262947} {\bibfield  {journal} {\bibinfo  {journal} {IEEE Transactions on Magnetics}\ }\textbf {\bibinfo {volume} {49}},\ \bibinfo {pages} {5172} (\bibinfo {year} {2013})}\BibitemShut {NoStop}%
\bibitem [{\citenamefont {Slonczewski}(1996)}]{Slonczewski1996}%
  \BibitemOpen
  \bibfield  {author} {\bibinfo {author} {\bibfnamefont {J.~C.}\ \bibnamefont {Slonczewski}},\ }\href@noop {} {\bibfield  {journal} {\bibinfo  {journal} {Journal of Magnetism and Magnetic Materials}\ }\textbf {\bibinfo {volume} {159}},\ \bibinfo {pages} {1} (\bibinfo {year} {1996})}\BibitemShut {NoStop}%
\bibitem [{\citenamefont {Costache}\ \emph {et~al.}(2006)\citenamefont {Costache}, \citenamefont {Sladkov}, \citenamefont {Watts}, \citenamefont {van~der Wal},\ and\ \citenamefont {van Wees}}]{Costache2006}%
  \BibitemOpen
  \bibfield  {author} {\bibinfo {author} {\bibfnamefont {M.}~\bibnamefont {Costache}}, \bibinfo {author} {\bibfnamefont {M.}~\bibnamefont {Sladkov}}, \bibinfo {author} {\bibfnamefont {S.}~\bibnamefont {Watts}}, \bibinfo {author} {\bibfnamefont {C.}~\bibnamefont {van~der Wal}},\ and\ \bibinfo {author} {\bibfnamefont {B.}~\bibnamefont {van Wees}},\ }\href {https://doi.org/10.1103/PhysRevLett.97.216603} {\bibfield  {journal} {\bibinfo  {journal} {Physical Review Letters}\ }\textbf {\bibinfo {volume} {97}},\ \bibinfo {pages} {216603} (\bibinfo {year} {2006})}\BibitemShut {NoStop}%
\bibitem [{\citenamefont {Saitoh}\ \emph {et~al.}(2006)\citenamefont {Saitoh}, \citenamefont {Ueda}, \citenamefont {Miyajima},\ and\ \citenamefont {Tatara}}]{saitoh2006}%
  \BibitemOpen
  \bibfield  {author} {\bibinfo {author} {\bibfnamefont {E.}~\bibnamefont {Saitoh}}, \bibinfo {author} {\bibfnamefont {M.}~\bibnamefont {Ueda}}, \bibinfo {author} {\bibfnamefont {H.}~\bibnamefont {Miyajima}},\ and\ \bibinfo {author} {\bibfnamefont {G.}~\bibnamefont {Tatara}},\ }\href {https://doi.org/10.1063/1.2199473} {\bibfield  {journal} {\bibinfo  {journal} {Applied Physics Letters}\ }\textbf {\bibinfo {volume} {88}},\ \bibinfo {pages} {182509} (\bibinfo {year} {2006})}\BibitemShut {NoStop}%
\bibitem [{\citenamefont {Ando}\ \emph {et~al.}(2011)\citenamefont {Ando}, \citenamefont {Takahashi}, \citenamefont {Ieda}, \citenamefont {Kajiwara}, \citenamefont {Nakayama}, \citenamefont {Yoshino}, \citenamefont {Harii}, \citenamefont {Fujikawa}, \citenamefont {Matsuo}, \citenamefont {Maekawa},\ and\ \citenamefont {Saitoh}}]{Ando2011}%
  \BibitemOpen
  \bibfield  {author} {\bibinfo {author} {\bibfnamefont {K.}~\bibnamefont {Ando}}, \bibinfo {author} {\bibfnamefont {S.}~\bibnamefont {Takahashi}}, \bibinfo {author} {\bibfnamefont {J.}~\bibnamefont {Ieda}}, \bibinfo {author} {\bibfnamefont {Y.}~\bibnamefont {Kajiwara}}, \bibinfo {author} {\bibfnamefont {H.}~\bibnamefont {Nakayama}}, \bibinfo {author} {\bibfnamefont {T.}~\bibnamefont {Yoshino}}, \bibinfo {author} {\bibfnamefont {K.}~\bibnamefont {Harii}}, \bibinfo {author} {\bibfnamefont {Y.}~\bibnamefont {Fujikawa}}, \bibinfo {author} {\bibfnamefont {M.}~\bibnamefont {Matsuo}}, \bibinfo {author} {\bibfnamefont {S.}~\bibnamefont {Maekawa}},\ and\ \bibinfo {author} {\bibfnamefont {E.}~\bibnamefont {Saitoh}},\ }\href {https://doi.org/10.1063/1.3587173} {\bibfield  {journal} {\bibinfo  {journal} {Journal of Applied Physics}\ }\textbf {\bibinfo {volume} {109}},\ \bibinfo {pages} {103913} (\bibinfo {year} {2011})}\BibitemShut {NoStop}%
\bibitem [{\citenamefont {Rojas-S{\'a}nchez}\ \emph {et~al.}(2014)\citenamefont {Rojas-S{\'a}nchez}, \citenamefont {Reyren}, \citenamefont {Laczkowski}, \citenamefont {Savero}, \citenamefont {Attan{\'e}}, \citenamefont {Deranlot}, \citenamefont {Jamet}, \citenamefont {George}, \citenamefont {Vila},\ and\ \citenamefont {Jaffr{\`e}s}}]{rojas2014spin}%
  \BibitemOpen
  \bibfield  {author} {\bibinfo {author} {\bibfnamefont {J.-C.}\ \bibnamefont {Rojas-S{\'a}nchez}}, \bibinfo {author} {\bibfnamefont {N.}~\bibnamefont {Reyren}}, \bibinfo {author} {\bibfnamefont {P.}~\bibnamefont {Laczkowski}}, \bibinfo {author} {\bibfnamefont {W.}~\bibnamefont {Savero}}, \bibinfo {author} {\bibfnamefont {J.-P.}\ \bibnamefont {Attan{\'e}}}, \bibinfo {author} {\bibfnamefont {C.}~\bibnamefont {Deranlot}}, \bibinfo {author} {\bibfnamefont {M.}~\bibnamefont {Jamet}}, \bibinfo {author} {\bibfnamefont {J.-M.}\ \bibnamefont {George}}, \bibinfo {author} {\bibfnamefont {L.}~\bibnamefont {Vila}},\ and\ \bibinfo {author} {\bibfnamefont {H.}~\bibnamefont {Jaffr{\`e}s}},\ }\href@noop {} {\bibfield  {journal} {\bibinfo  {journal} {Physical review letters}\ }\textbf {\bibinfo {volume} {112}},\ \bibinfo {pages} {106602} (\bibinfo {year} {2014})}\BibitemShut {NoStop}%
\bibitem [{\citenamefont {Uchida}\ \emph {et~al.}(2008)\citenamefont {Uchida}, \citenamefont {Takahashi}, \citenamefont {Harii}, \citenamefont {Ieda}, \citenamefont {Koshibae}, \citenamefont {Ando}, \citenamefont {Maekawa},\ and\ \citenamefont {Saitoh}}]{Uchida2008}%
  \BibitemOpen
  \bibfield  {author} {\bibinfo {author} {\bibfnamefont {K.}~\bibnamefont {Uchida}}, \bibinfo {author} {\bibfnamefont {S.}~\bibnamefont {Takahashi}}, \bibinfo {author} {\bibfnamefont {K.}~\bibnamefont {Harii}}, \bibinfo {author} {\bibfnamefont {J.}~\bibnamefont {Ieda}}, \bibinfo {author} {\bibfnamefont {W.}~\bibnamefont {Koshibae}}, \bibinfo {author} {\bibfnamefont {K.}~\bibnamefont {Ando}}, \bibinfo {author} {\bibfnamefont {S.}~\bibnamefont {Maekawa}},\ and\ \bibinfo {author} {\bibfnamefont {E.}~\bibnamefont {Saitoh}},\ }\href {https://doi.org/10.1038/nature07321} {\bibfield  {journal} {\bibinfo  {journal} {Nature}\ }\textbf {\bibinfo {volume} {455}},\ \bibinfo {pages} {778} (\bibinfo {year} {2008})}\BibitemShut {NoStop}%
\bibitem [{\citenamefont {Uchida}\ \emph {et~al.}(2010)\citenamefont {Uchida}, \citenamefont {Xiao}, \citenamefont {Adachi}, \citenamefont {Ohe}, \citenamefont {Takahashi}, \citenamefont {Ieda}, \citenamefont {Ota}, \citenamefont {Kajiwara}, \citenamefont {Umezawa}, \citenamefont {Kawai}, \citenamefont {Bauer}, \citenamefont {Maekawa},\ and\ \citenamefont {Saitoh}}]{Uchida2010}%
  \BibitemOpen
  \bibfield  {author} {\bibinfo {author} {\bibfnamefont {K.}~\bibnamefont {Uchida}}, \bibinfo {author} {\bibfnamefont {J.}~\bibnamefont {Xiao}}, \bibinfo {author} {\bibfnamefont {H.}~\bibnamefont {Adachi}}, \bibinfo {author} {\bibfnamefont {J.}~\bibnamefont {Ohe}}, \bibinfo {author} {\bibfnamefont {S.}~\bibnamefont {Takahashi}}, \bibinfo {author} {\bibfnamefont {J.}~\bibnamefont {Ieda}}, \bibinfo {author} {\bibfnamefont {T.}~\bibnamefont {Ota}}, \bibinfo {author} {\bibfnamefont {Y.}~\bibnamefont {Kajiwara}}, \bibinfo {author} {\bibfnamefont {H.}~\bibnamefont {Umezawa}}, \bibinfo {author} {\bibfnamefont {H.}~\bibnamefont {Kawai}}, \bibinfo {author} {\bibfnamefont {G.~E.}\ \bibnamefont {Bauer}}, \bibinfo {author} {\bibfnamefont {S.}~\bibnamefont {Maekawa}},\ and\ \bibinfo {author} {\bibfnamefont {E.}~\bibnamefont {Saitoh}},\ }\href {https://doi.org/10.1038/nmat2856} {\bibfield  {journal} {\bibinfo  {journal} {Nature Materials}\ }\textbf {\bibinfo {volume} {9}},\ \bibinfo {pages} {894} (\bibinfo {year}
  {2010})}\BibitemShut {NoStop}%
\bibitem [{\citenamefont {Anad{\'{o}}n}\ \emph {et~al.}(2022)\citenamefont {Anad{\'{o}}n}, \citenamefont {Martin}, \citenamefont {Homkar}, \citenamefont {Meunier}, \citenamefont {Verges}, \citenamefont {Damas}, \citenamefont {Alegre}, \citenamefont {Lefevre}, \citenamefont {Roulland}, \citenamefont {Dubs}, \citenamefont {Lindner}, \citenamefont {Pasquier}, \citenamefont {Copie}, \citenamefont {Dumesnil}, \citenamefont {Ramos}, \citenamefont {Preziosi}, \citenamefont {Petit-Watelot}, \citenamefont {Viart},\ and\ \citenamefont {Rojas-S{\'{a}}nchez}}]{Anadon2022ThermalGa_0.6Fe_1.4O_3b}%
  \BibitemOpen
  \bibfield  {author} {\bibinfo {author} {\bibfnamefont {A.}~\bibnamefont {Anad{\'{o}}n}}, \bibinfo {author} {\bibfnamefont {E.}~\bibnamefont {Martin}}, \bibinfo {author} {\bibfnamefont {S.}~\bibnamefont {Homkar}}, \bibinfo {author} {\bibfnamefont {B.}~\bibnamefont {Meunier}}, \bibinfo {author} {\bibfnamefont {M.}~\bibnamefont {Verges}}, \bibinfo {author} {\bibfnamefont {H.}~\bibnamefont {Damas}}, \bibinfo {author} {\bibfnamefont {J.}~\bibnamefont {Alegre}}, \bibinfo {author} {\bibfnamefont {C.}~\bibnamefont {Lefevre}}, \bibinfo {author} {\bibfnamefont {F.}~\bibnamefont {Roulland}}, \bibinfo {author} {\bibfnamefont {C.}~\bibnamefont {Dubs}}, \bibinfo {author} {\bibfnamefont {M.}~\bibnamefont {Lindner}}, \bibinfo {author} {\bibfnamefont {L.}~\bibnamefont {Pasquier}}, \bibinfo {author} {\bibfnamefont {O.}~\bibnamefont {Copie}}, \bibinfo {author} {\bibfnamefont {K.}~\bibnamefont {Dumesnil}}, \bibinfo {author} {\bibfnamefont {R.}~\bibnamefont {Ramos}}, \bibinfo {author} {\bibfnamefont {D.}~\bibnamefont
  {Preziosi}}, \bibinfo {author} {\bibfnamefont {S.}~\bibnamefont {Petit-Watelot}}, \bibinfo {author} {\bibfnamefont {N.}~\bibnamefont {Viart}},\ and\ \bibinfo {author} {\bibfnamefont {J.-C.}\ \bibnamefont {Rojas-S{\'{a}}nchez}},\ }\href {https://doi.org/10.1103/PhysRevApplied.18.054087} {\bibfield  {journal} {\bibinfo  {journal} {Physical Review Applied}\ }\textbf {\bibinfo {volume} {10}},\ \bibinfo {pages} {1} (\bibinfo {year} {2022})}\BibitemShut {NoStop}%
\bibitem [{\citenamefont {Anad{\'{o}}n}\ \emph {et~al.}(2016)\citenamefont {Anad{\'{o}}n}, \citenamefont {Ramos}, \citenamefont {Lucas}, \citenamefont {Algarabel}, \citenamefont {Morell{\'{o}}n}, \citenamefont {Ibarra},\ and\ \citenamefont {Aguirre}}]{Anadon2016}%
  \BibitemOpen
  \bibfield  {author} {\bibinfo {author} {\bibfnamefont {A.}~\bibnamefont {Anad{\'{o}}n}}, \bibinfo {author} {\bibfnamefont {R.}~\bibnamefont {Ramos}}, \bibinfo {author} {\bibfnamefont {I.}~\bibnamefont {Lucas}}, \bibinfo {author} {\bibfnamefont {P.~A.}\ \bibnamefont {Algarabel}}, \bibinfo {author} {\bibfnamefont {L.}~\bibnamefont {Morell{\'{o}}n}}, \bibinfo {author} {\bibfnamefont {M.~R.}\ \bibnamefont {Ibarra}},\ and\ \bibinfo {author} {\bibfnamefont {M.~H.}\ \bibnamefont {Aguirre}},\ }\href {https://doi.org/10.1063/1.4955031} {\bibfield  {journal} {\bibinfo  {journal} {Applied Physics Letters}\ }\textbf {\bibinfo {volume} {109}},\ \bibinfo {pages} {012404} (\bibinfo {year} {2016})}\BibitemShut {NoStop}%
\bibitem [{\citenamefont {Gambardella}\ and\ \citenamefont {Miron}(2011)}]{Gambardella2011}%
  \BibitemOpen
  \bibfield  {author} {\bibinfo {author} {\bibfnamefont {P.}~\bibnamefont {Gambardella}}\ and\ \bibinfo {author} {\bibfnamefont {I.~M.}\ \bibnamefont {Miron}},\ }\href {https://doi.org/10.1098/rsta.2010.0336} {\bibfield  {journal} {\bibinfo  {journal} {Philosophical Transactions of the Royal Society A: Mathematical, Physical and Engineering Sciences}\ }\textbf {\bibinfo {volume} {369}},\ \bibinfo {pages} {3175} (\bibinfo {year} {2011})}\BibitemShut {NoStop}%
\bibitem [{\citenamefont {Anad{\'{o}}n}\ \emph {et~al.}(2021{\natexlab{a}})\citenamefont {Anad{\'{o}}n}, \citenamefont {Guerrero}, \citenamefont {Jover-Galtier}, \citenamefont {Gud{\'{i}}n}, \citenamefont {D{\'{i}}ez~Toledano}, \citenamefont {Olleros-Rodr{\'{i}}guez}, \citenamefont {Miranda}, \citenamefont {Camarero},\ and\ \citenamefont {Perna}}]{Anadon2021Cu}%
  \BibitemOpen
  \bibfield  {author} {\bibinfo {author} {\bibfnamefont {A.}~\bibnamefont {Anad{\'{o}}n}}, \bibinfo {author} {\bibfnamefont {R.}~\bibnamefont {Guerrero}}, \bibinfo {author} {\bibfnamefont {J.~A.}\ \bibnamefont {Jover-Galtier}}, \bibinfo {author} {\bibfnamefont {A.}~\bibnamefont {Gud{\'{i}}n}}, \bibinfo {author} {\bibfnamefont {J.~M.}\ \bibnamefont {D{\'{i}}ez~Toledano}}, \bibinfo {author} {\bibfnamefont {P.}~\bibnamefont {Olleros-Rodr{\'{i}}guez}}, \bibinfo {author} {\bibfnamefont {R.}~\bibnamefont {Miranda}}, \bibinfo {author} {\bibfnamefont {J.}~\bibnamefont {Camarero}},\ and\ \bibinfo {author} {\bibfnamefont {P.}~\bibnamefont {Perna}},\ }\href {https://doi.org/10.1021/acsanm.0c02808} {\bibfield  {journal} {\bibinfo  {journal} {ACS Applied Nano Materials}\ }\textbf {\bibinfo {volume} {4}},\ \bibinfo {pages} {487} (\bibinfo {year} {2021}{\natexlab{a}})}\BibitemShut {NoStop}%
\bibitem [{\citenamefont {Mangin}\ \emph {et~al.}(2006)\citenamefont {Mangin}, \citenamefont {Ravelosona}, \citenamefont {Katine},\ and\ \citenamefont {Fullerton}}]{Mangin2006}%
  \BibitemOpen
  \bibfield  {author} {\bibinfo {author} {\bibfnamefont {S.}~\bibnamefont {Mangin}}, \bibinfo {author} {\bibfnamefont {D.}~\bibnamefont {Ravelosona}}, \bibinfo {author} {\bibfnamefont {J.~A.}\ \bibnamefont {Katine}},\ and\ \bibinfo {author} {\bibfnamefont {E.~E.}\ \bibnamefont {Fullerton}},\ }\href {https://doi.org/10.1109/INTMAG.2006.375414} {\bibfield  {journal} {\bibinfo  {journal} {INTERMAG 2006 - IEEE International Magnetics Conference}\ }\textbf {\bibinfo {volume} {5}},\ \bibinfo {pages} {5} (\bibinfo {year} {2006})}\BibitemShut {NoStop}%
\bibitem [{\citenamefont {Nakayama}\ \emph {et~al.}(2013)\citenamefont {Nakayama}, \citenamefont {Althammer}, \citenamefont {Chen}, \citenamefont {Uchida}, \citenamefont {Kajiwara}, \citenamefont {Kikuchi}, \citenamefont {Ohtani}, \citenamefont {Gepr{\"{a}}gs}, \citenamefont {Opel}, \citenamefont {Takahashi}, \citenamefont {Gross}, \citenamefont {Bauer}, \citenamefont {Goennenwein},\ and\ \citenamefont {Saitoh}}]{Nakayama2013}%
  \BibitemOpen
  \bibfield  {author} {\bibinfo {author} {\bibfnamefont {H.}~\bibnamefont {Nakayama}}, \bibinfo {author} {\bibfnamefont {M.}~\bibnamefont {Althammer}}, \bibinfo {author} {\bibfnamefont {Y.-T.}\ \bibnamefont {Chen}}, \bibinfo {author} {\bibfnamefont {K.}~\bibnamefont {Uchida}}, \bibinfo {author} {\bibfnamefont {Y.}~\bibnamefont {Kajiwara}}, \bibinfo {author} {\bibfnamefont {D.}~\bibnamefont {Kikuchi}}, \bibinfo {author} {\bibfnamefont {T.}~\bibnamefont {Ohtani}}, \bibinfo {author} {\bibfnamefont {S.}~\bibnamefont {Gepr{\"{a}}gs}}, \bibinfo {author} {\bibfnamefont {M.}~\bibnamefont {Opel}}, \bibinfo {author} {\bibfnamefont {S.}~\bibnamefont {Takahashi}}, \bibinfo {author} {\bibfnamefont {R.}~\bibnamefont {Gross}}, \bibinfo {author} {\bibfnamefont {G.~E.~W.}\ \bibnamefont {Bauer}}, \bibinfo {author} {\bibfnamefont {S.~T.~B.}\ \bibnamefont {Goennenwein}},\ and\ \bibinfo {author} {\bibfnamefont {E.}~\bibnamefont {Saitoh}},\ }\href {https://doi.org/10.1103/PhysRevLett.110.206601} {\bibfield  {journal} {\bibinfo
  {journal} {Physical Review Letters}\ }\textbf {\bibinfo {volume} {110}},\ \bibinfo {pages} {206601} (\bibinfo {year} {2013})}\BibitemShut {NoStop}%
\bibitem [{\citenamefont {Homkar}\ \emph {et~al.}(2021)\citenamefont {Homkar}, \citenamefont {Martin}, \citenamefont {Meunier}, \citenamefont {Anadon-Barcelona}, \citenamefont {Bouillet}, \citenamefont {Gorchon}, \citenamefont {Dumesnil}, \citenamefont {Lef{\`{e}}vre}, \citenamefont {Roulland}, \citenamefont {Copie}, \citenamefont {Preziosi}, \citenamefont {Petit-Watelot}, \citenamefont {Rojas-S{\'{a}}nchez},\ and\ \citenamefont {Viart}}]{Homkar2021SpinGa0.6Fe1.4O3Bilayers}%
  \BibitemOpen
  \bibfield  {author} {\bibinfo {author} {\bibfnamefont {S.}~\bibnamefont {Homkar}}, \bibinfo {author} {\bibfnamefont {E.}~\bibnamefont {Martin}}, \bibinfo {author} {\bibfnamefont {B.}~\bibnamefont {Meunier}}, \bibinfo {author} {\bibfnamefont {A.}~\bibnamefont {Anadon-Barcelona}}, \bibinfo {author} {\bibfnamefont {C.}~\bibnamefont {Bouillet}}, \bibinfo {author} {\bibfnamefont {J.}~\bibnamefont {Gorchon}}, \bibinfo {author} {\bibfnamefont {K.}~\bibnamefont {Dumesnil}}, \bibinfo {author} {\bibfnamefont {C.}~\bibnamefont {Lef{\`{e}}vre}}, \bibinfo {author} {\bibfnamefont {F.}~\bibnamefont {Roulland}}, \bibinfo {author} {\bibfnamefont {O.}~\bibnamefont {Copie}}, \bibinfo {author} {\bibfnamefont {D.}~\bibnamefont {Preziosi}}, \bibinfo {author} {\bibfnamefont {S.}~\bibnamefont {Petit-Watelot}}, \bibinfo {author} {\bibfnamefont {J.~C.}\ \bibnamefont {Rojas-S{\'{a}}nchez}},\ and\ \bibinfo {author} {\bibfnamefont {N.}~\bibnamefont {Viart}},\ }\href {https://doi.org/10.1021/acsaelm.1c00586} {\bibfield  {journal}
  {\bibinfo  {journal} {ACS Applied Electronic Materials}\ }\textbf {\bibinfo {volume} {3}},\ \bibinfo {pages} {4433} (\bibinfo {year} {2021})}\BibitemShut {NoStop}%
\bibitem [{\citenamefont {Rojas-S{\'{a}}nchez}\ \emph {et~al.}(2013)\citenamefont {Rojas-S{\'{a}}nchez}, \citenamefont {Vila}, \citenamefont {Desfonds}, \citenamefont {Gambarelli}, \citenamefont {Attan{\'{e}}}, \citenamefont {De~Teresa}, \citenamefont {Mag{\'{e}}n},\ and\ \citenamefont {Fert}}]{Sanchez2013}%
  \BibitemOpen
  \bibfield  {author} {\bibinfo {author} {\bibfnamefont {J.~C.}\ \bibnamefont {Rojas-S{\'{a}}nchez}}, \bibinfo {author} {\bibfnamefont {L.}~\bibnamefont {Vila}}, \bibinfo {author} {\bibfnamefont {G.}~\bibnamefont {Desfonds}}, \bibinfo {author} {\bibfnamefont {S.}~\bibnamefont {Gambarelli}}, \bibinfo {author} {\bibfnamefont {J.~P.}\ \bibnamefont {Attan{\'{e}}}}, \bibinfo {author} {\bibfnamefont {J.~M.}\ \bibnamefont {De~Teresa}}, \bibinfo {author} {\bibfnamefont {C.}~\bibnamefont {Mag{\'{e}}n}},\ and\ \bibinfo {author} {\bibfnamefont {A.}~\bibnamefont {Fert}},\ }\href {https://doi.org/10.1038/ncomms3944} {\bibfield  {journal} {\bibinfo  {journal} {Nature Communications}\ }\textbf {\bibinfo {volume} {4}},\ \bibinfo {pages} {2944} (\bibinfo {year} {2013})}\BibitemShut {NoStop}%
\bibitem [{\citenamefont {Mihai~Miron}\ \emph {et~al.}(2010)\citenamefont {Mihai~Miron}, \citenamefont {Gaudin}, \citenamefont {Auffret}, \citenamefont {Rodmacq}, \citenamefont {Schuhl}, \citenamefont {Pizzini}, \citenamefont {Vogel},\ and\ \citenamefont {Gambardella}}]{MihaiMiron2010}%
  \BibitemOpen
  \bibfield  {author} {\bibinfo {author} {\bibfnamefont {I.}~\bibnamefont {Mihai~Miron}}, \bibinfo {author} {\bibfnamefont {G.}~\bibnamefont {Gaudin}}, \bibinfo {author} {\bibfnamefont {S.}~\bibnamefont {Auffret}}, \bibinfo {author} {\bibfnamefont {B.}~\bibnamefont {Rodmacq}}, \bibinfo {author} {\bibfnamefont {A.}~\bibnamefont {Schuhl}}, \bibinfo {author} {\bibfnamefont {S.}~\bibnamefont {Pizzini}}, \bibinfo {author} {\bibfnamefont {J.}~\bibnamefont {Vogel}},\ and\ \bibinfo {author} {\bibfnamefont {P.}~\bibnamefont {Gambardella}},\ }\href {https://doi.org/10.1038/nmat2613} {\bibfield  {journal} {\bibinfo  {journal} {Nature Materials}\ }\textbf {\bibinfo {volume} {9}},\ \bibinfo {pages} {230} (\bibinfo {year} {2010})}\BibitemShut {NoStop}%
\bibitem [{\citenamefont {Dieny}\ \emph {et~al.}(2020)\citenamefont {Dieny}, \citenamefont {Prejbeanu}, \citenamefont {Garello}, \citenamefont {Gambardella}, \citenamefont {Freitas}, \citenamefont {Lehndorff}, \citenamefont {Raberg}, \citenamefont {Ebels}, \citenamefont {Demokritov}, \citenamefont {Akerman}, \citenamefont {Deac}, \citenamefont {Pirro}, \citenamefont {Adelmann}, \citenamefont {Anane}, \citenamefont {Chumak}, \citenamefont {Hirohata}, \citenamefont {Mangin}, \citenamefont {Valenzuela}, \citenamefont {Onba{\c{s}}lı}, \citenamefont {d’Aquino}, \citenamefont {Prenat}, \citenamefont {Finocchio}, \citenamefont {Lopez-Diaz}, \citenamefont {Chantrell}, \citenamefont {Chubykalo-Fesenko},\ and\ \citenamefont {Bortolotti}}]{Dieny2020}%
  \BibitemOpen
  \bibfield  {author} {\bibinfo {author} {\bibfnamefont {B.}~\bibnamefont {Dieny}}, \bibinfo {author} {\bibfnamefont {I.~L.}\ \bibnamefont {Prejbeanu}}, \bibinfo {author} {\bibfnamefont {K.}~\bibnamefont {Garello}}, \bibinfo {author} {\bibfnamefont {P.}~\bibnamefont {Gambardella}}, \bibinfo {author} {\bibfnamefont {P.}~\bibnamefont {Freitas}}, \bibinfo {author} {\bibfnamefont {R.}~\bibnamefont {Lehndorff}}, \bibinfo {author} {\bibfnamefont {W.}~\bibnamefont {Raberg}}, \bibinfo {author} {\bibfnamefont {U.}~\bibnamefont {Ebels}}, \bibinfo {author} {\bibfnamefont {S.~O.}\ \bibnamefont {Demokritov}}, \bibinfo {author} {\bibfnamefont {J.}~\bibnamefont {Akerman}}, \bibinfo {author} {\bibfnamefont {A.}~\bibnamefont {Deac}}, \bibinfo {author} {\bibfnamefont {P.}~\bibnamefont {Pirro}}, \bibinfo {author} {\bibfnamefont {C.}~\bibnamefont {Adelmann}}, \bibinfo {author} {\bibfnamefont {A.}~\bibnamefont {Anane}}, \bibinfo {author} {\bibfnamefont {A.~V.}\ \bibnamefont {Chumak}}, \bibinfo {author} {\bibfnamefont
  {A.}~\bibnamefont {Hirohata}}, \bibinfo {author} {\bibfnamefont {S.}~\bibnamefont {Mangin}}, \bibinfo {author} {\bibfnamefont {S.~O.}\ \bibnamefont {Valenzuela}}, \bibinfo {author} {\bibfnamefont {M.~C.}\ \bibnamefont {Onba{\c{s}}lı}}, \bibinfo {author} {\bibfnamefont {M.}~\bibnamefont {d’Aquino}}, \bibinfo {author} {\bibfnamefont {G.}~\bibnamefont {Prenat}}, \bibinfo {author} {\bibfnamefont {G.}~\bibnamefont {Finocchio}}, \bibinfo {author} {\bibfnamefont {L.}~\bibnamefont {Lopez-Diaz}}, \bibinfo {author} {\bibfnamefont {R.}~\bibnamefont {Chantrell}}, \bibinfo {author} {\bibfnamefont {O.}~\bibnamefont {Chubykalo-Fesenko}},\ and\ \bibinfo {author} {\bibfnamefont {P.}~\bibnamefont {Bortolotti}},\ }\href {https://doi.org/10.1038/s41928-020-0461-5} {\bibfield  {journal} {\bibinfo  {journal} {Nature Electronics}\ }\textbf {\bibinfo {volume} {3}},\ \bibinfo {pages} {446} (\bibinfo {year} {2020})}\BibitemShut {NoStop}%
\bibitem [{\citenamefont {Mazraati}\ \emph {et~al.}(2016)\citenamefont {Mazraati}, \citenamefont {Chung}, \citenamefont {Houshang}, \citenamefont {Dvornik}, \citenamefont {Piazza}, \citenamefont {Qejvanaj}, \citenamefont {Jiang}, \citenamefont {Le}, \citenamefont {Weissenrieder},\ and\ \citenamefont {{\AA}kerman}}]{mazraati2016low}%
  \BibitemOpen
  \bibfield  {author} {\bibinfo {author} {\bibfnamefont {H.}~\bibnamefont {Mazraati}}, \bibinfo {author} {\bibfnamefont {S.}~\bibnamefont {Chung}}, \bibinfo {author} {\bibfnamefont {A.}~\bibnamefont {Houshang}}, \bibinfo {author} {\bibfnamefont {M.}~\bibnamefont {Dvornik}}, \bibinfo {author} {\bibfnamefont {L.}~\bibnamefont {Piazza}}, \bibinfo {author} {\bibfnamefont {F.}~\bibnamefont {Qejvanaj}}, \bibinfo {author} {\bibfnamefont {S.}~\bibnamefont {Jiang}}, \bibinfo {author} {\bibfnamefont {T.~Q.}\ \bibnamefont {Le}}, \bibinfo {author} {\bibfnamefont {J.}~\bibnamefont {Weissenrieder}},\ and\ \bibinfo {author} {\bibfnamefont {J.}~\bibnamefont {{\AA}kerman}},\ }\href@noop {} {\bibfield  {journal} {\bibinfo  {journal} {Applied Physics Letters}\ }\textbf {\bibinfo {volume} {109}},\ \bibinfo {pages} {242402} (\bibinfo {year} {2016})}\BibitemShut {NoStop}%
\bibitem [{\citenamefont {Xu}\ \emph {et~al.}(2018)\citenamefont {Xu}, \citenamefont {Yang}, \citenamefont {Zhang}, \citenamefont {Luo},\ and\ \citenamefont {Wu}}]{xu2018ultrathin}%
  \BibitemOpen
  \bibfield  {author} {\bibinfo {author} {\bibfnamefont {Y.}~\bibnamefont {Xu}}, \bibinfo {author} {\bibfnamefont {Y.}~\bibnamefont {Yang}}, \bibinfo {author} {\bibfnamefont {M.}~\bibnamefont {Zhang}}, \bibinfo {author} {\bibfnamefont {Z.}~\bibnamefont {Luo}},\ and\ \bibinfo {author} {\bibfnamefont {Y.}~\bibnamefont {Wu}},\ }\href@noop {} {\bibfield  {journal} {\bibinfo  {journal} {Advanced Materials Technologies}\ }\textbf {\bibinfo {volume} {3}},\ \bibinfo {pages} {1800073} (\bibinfo {year} {2018})}\BibitemShut {NoStop}%
\bibitem [{\citenamefont {Ryu}\ \emph {et~al.}(2013)\citenamefont {Ryu}, \citenamefont {Thomas}, \citenamefont {Yang},\ and\ \citenamefont {Parkin}}]{ryu2013chiral}%
  \BibitemOpen
  \bibfield  {author} {\bibinfo {author} {\bibfnamefont {K.-S.}\ \bibnamefont {Ryu}}, \bibinfo {author} {\bibfnamefont {L.}~\bibnamefont {Thomas}}, \bibinfo {author} {\bibfnamefont {S.-H.}\ \bibnamefont {Yang}},\ and\ \bibinfo {author} {\bibfnamefont {S.}~\bibnamefont {Parkin}},\ }\href@noop {} {\bibfield  {journal} {\bibinfo  {journal} {Nature nanotechnology}\ }\textbf {\bibinfo {volume} {8}},\ \bibinfo {pages} {527} (\bibinfo {year} {2013})}\BibitemShut {NoStop}%
\bibitem [{\citenamefont {Luo}\ \emph {et~al.}(2020)\citenamefont {Luo}, \citenamefont {Hrabec}, \citenamefont {Dao}, \citenamefont {Sala}, \citenamefont {Finizio}, \citenamefont {Feng}, \citenamefont {Mayr}, \citenamefont {Raabe}, \citenamefont {Gambardella},\ and\ \citenamefont {Heyderman}}]{luo2020current}%
  \BibitemOpen
  \bibfield  {author} {\bibinfo {author} {\bibfnamefont {Z.}~\bibnamefont {Luo}}, \bibinfo {author} {\bibfnamefont {A.}~\bibnamefont {Hrabec}}, \bibinfo {author} {\bibfnamefont {T.~P.}\ \bibnamefont {Dao}}, \bibinfo {author} {\bibfnamefont {G.}~\bibnamefont {Sala}}, \bibinfo {author} {\bibfnamefont {S.}~\bibnamefont {Finizio}}, \bibinfo {author} {\bibfnamefont {J.}~\bibnamefont {Feng}}, \bibinfo {author} {\bibfnamefont {S.}~\bibnamefont {Mayr}}, \bibinfo {author} {\bibfnamefont {J.}~\bibnamefont {Raabe}}, \bibinfo {author} {\bibfnamefont {P.}~\bibnamefont {Gambardella}},\ and\ \bibinfo {author} {\bibfnamefont {L.~J.}\ \bibnamefont {Heyderman}},\ }\href@noop {} {\bibfield  {journal} {\bibinfo  {journal} {Nature}\ }\textbf {\bibinfo {volume} {579}},\ \bibinfo {pages} {214} (\bibinfo {year} {2020})}\BibitemShut {NoStop}%
\bibitem [{\citenamefont {Yu}\ \emph {et~al.}(2016)\citenamefont {Yu}, \citenamefont {Qiu}, \citenamefont {Legrand},\ and\ \citenamefont {Yang}}]{yu2016large}%
  \BibitemOpen
  \bibfield  {author} {\bibinfo {author} {\bibfnamefont {J.}~\bibnamefont {Yu}}, \bibinfo {author} {\bibfnamefont {X.}~\bibnamefont {Qiu}}, \bibinfo {author} {\bibfnamefont {W.}~\bibnamefont {Legrand}},\ and\ \bibinfo {author} {\bibfnamefont {H.}~\bibnamefont {Yang}},\ }\href@noop {} {\bibfield  {journal} {\bibinfo  {journal} {Applied Physics Letters}\ }\textbf {\bibinfo {volume} {109}},\ \bibinfo {pages} {042403} (\bibinfo {year} {2016})}\BibitemShut {NoStop}%
\bibitem [{\citenamefont {Liu}\ \emph {et~al.}(2011{\natexlab{b}})\citenamefont {Liu}, \citenamefont {Moriyama}, \citenamefont {Ralph},\ and\ \citenamefont {Buhrman}}]{Liu2011a}%
  \BibitemOpen
  \bibfield  {author} {\bibinfo {author} {\bibfnamefont {L.}~\bibnamefont {Liu}}, \bibinfo {author} {\bibfnamefont {T.}~\bibnamefont {Moriyama}}, \bibinfo {author} {\bibfnamefont {D.~C.}\ \bibnamefont {Ralph}},\ and\ \bibinfo {author} {\bibfnamefont {R.~A.}\ \bibnamefont {Buhrman}},\ }\bibfield  {journal} {\bibinfo  {journal} {Physical Review Letters}\ }\href {https://doi.org/10.1103/PhysRevLett.106.036601} {10.1103/PhysRevLett.106.036601} (\bibinfo {year} {2011}{\natexlab{b}})\BibitemShut {NoStop}%
\bibitem [{\citenamefont {Ou}\ \emph {et~al.}(2018)\citenamefont {Ou}, \citenamefont {Ralph},\ and\ \citenamefont {Buhrman}}]{ou2018strong}%
  \BibitemOpen
  \bibfield  {author} {\bibinfo {author} {\bibfnamefont {Y.}~\bibnamefont {Ou}}, \bibinfo {author} {\bibfnamefont {D.}~\bibnamefont {Ralph}},\ and\ \bibinfo {author} {\bibfnamefont {R.}~\bibnamefont {Buhrman}},\ }\href@noop {} {\bibfield  {journal} {\bibinfo  {journal} {Physical Review Letters}\ }\textbf {\bibinfo {volume} {120}},\ \bibinfo {pages} {097203} (\bibinfo {year} {2018})}\BibitemShut {NoStop}%
\bibitem [{\citenamefont {Niimi}\ \emph {et~al.}(2012)\citenamefont {Niimi}, \citenamefont {Kawanishi}, \citenamefont {Wei}, \citenamefont {Deranlot}, \citenamefont {Yang}, \citenamefont {Chshiev}, \citenamefont {Valet}, \citenamefont {Fert},\ and\ \citenamefont {Otani}}]{niimi2012giant}%
  \BibitemOpen
  \bibfield  {author} {\bibinfo {author} {\bibfnamefont {Y.}~\bibnamefont {Niimi}}, \bibinfo {author} {\bibfnamefont {Y.}~\bibnamefont {Kawanishi}}, \bibinfo {author} {\bibfnamefont {D.}~\bibnamefont {Wei}}, \bibinfo {author} {\bibfnamefont {C.}~\bibnamefont {Deranlot}}, \bibinfo {author} {\bibfnamefont {H.}~\bibnamefont {Yang}}, \bibinfo {author} {\bibfnamefont {M.}~\bibnamefont {Chshiev}}, \bibinfo {author} {\bibfnamefont {T.}~\bibnamefont {Valet}}, \bibinfo {author} {\bibfnamefont {A.}~\bibnamefont {Fert}},\ and\ \bibinfo {author} {\bibfnamefont {Y.}~\bibnamefont {Otani}},\ }\href@noop {} {\bibfield  {journal} {\bibinfo  {journal} {Physical review letters}\ }\textbf {\bibinfo {volume} {109}},\ \bibinfo {pages} {156602} (\bibinfo {year} {2012})}\BibitemShut {NoStop}%
\bibitem [{\citenamefont {Sagasta}\ \emph {et~al.}(2016)\citenamefont {Sagasta}, \citenamefont {Omori}, \citenamefont {Isasa}, \citenamefont {Gradhand}, \citenamefont {Hueso}, \citenamefont {Niimi}, \citenamefont {Otani},\ and\ \citenamefont {Casanova}}]{Sagasta2016}%
  \BibitemOpen
  \bibfield  {author} {\bibinfo {author} {\bibfnamefont {E.}~\bibnamefont {Sagasta}}, \bibinfo {author} {\bibfnamefont {Y.}~\bibnamefont {Omori}}, \bibinfo {author} {\bibfnamefont {M.}~\bibnamefont {Isasa}}, \bibinfo {author} {\bibfnamefont {M.}~\bibnamefont {Gradhand}}, \bibinfo {author} {\bibfnamefont {L.~E.}\ \bibnamefont {Hueso}}, \bibinfo {author} {\bibfnamefont {Y.}~\bibnamefont {Niimi}}, \bibinfo {author} {\bibfnamefont {Y.}~\bibnamefont {Otani}},\ and\ \bibinfo {author} {\bibfnamefont {F.}~\bibnamefont {Casanova}},\ }\href {https://doi.org/10.1103/PhysRevB.94.060412} {\bibfield  {journal} {\bibinfo  {journal} {Physical Review B}\ }\textbf {\bibinfo {volume} {94}},\ \bibinfo {pages} {1} (\bibinfo {year} {2016})}\BibitemShut {NoStop}%
\bibitem [{\citenamefont {Ryu}\ \emph {et~al.}(2016)\citenamefont {Ryu}, \citenamefont {Kohda},\ and\ \citenamefont {Nitta}}]{ryu2016observation}%
  \BibitemOpen
  \bibfield  {author} {\bibinfo {author} {\bibfnamefont {J.}~\bibnamefont {Ryu}}, \bibinfo {author} {\bibfnamefont {M.}~\bibnamefont {Kohda}},\ and\ \bibinfo {author} {\bibfnamefont {J.}~\bibnamefont {Nitta}},\ }\href@noop {} {\bibfield  {journal} {\bibinfo  {journal} {Physical review letters}\ }\textbf {\bibinfo {volume} {116}},\ \bibinfo {pages} {256802} (\bibinfo {year} {2016})}\BibitemShut {NoStop}%
\bibitem [{\citenamefont {Guillemard}\ \emph {et~al.}(2018)\citenamefont {Guillemard}, \citenamefont {Petit-Watelot}, \citenamefont {Andrieu},\ and\ \citenamefont {Rojas-S{\'{a}}nchez}}]{guillemard2018}%
  \BibitemOpen
  \bibfield  {author} {\bibinfo {author} {\bibfnamefont {C.}~\bibnamefont {Guillemard}}, \bibinfo {author} {\bibfnamefont {S.}~\bibnamefont {Petit-Watelot}}, \bibinfo {author} {\bibfnamefont {S.}~\bibnamefont {Andrieu}},\ and\ \bibinfo {author} {\bibfnamefont {J.-C.}\ \bibnamefont {Rojas-S{\'{a}}nchez}},\ }\href {https://doi.org/10.1063/1.5079236} {\bibfield  {journal} {\bibinfo  {journal} {Applied Physics Letters}\ }\textbf {\bibinfo {volume} {113}},\ \bibinfo {pages} {262404} (\bibinfo {year} {2018})}\BibitemShut {NoStop}%
\bibitem [{\citenamefont {Keller}\ \emph {et~al.}(2018)\citenamefont {Keller}, \citenamefont {Mihalceanu}, \citenamefont {Schweizer}, \citenamefont {Lang}, \citenamefont {Heinz}, \citenamefont {Geilen}, \citenamefont {Br{\"{a}}cher}, \citenamefont {Pirro}, \citenamefont {Meyer}, \citenamefont {Conca}, \citenamefont {Karfaridis}, \citenamefont {Vourlias}, \citenamefont {Kehagias}, \citenamefont {Hillebrands},\ and\ \citenamefont {Papaioannou}}]{Keller2018DeterminationExperiments}%
  \BibitemOpen
  \bibfield  {author} {\bibinfo {author} {\bibfnamefont {S.}~\bibnamefont {Keller}}, \bibinfo {author} {\bibfnamefont {L.}~\bibnamefont {Mihalceanu}}, \bibinfo {author} {\bibfnamefont {M.~R.}\ \bibnamefont {Schweizer}}, \bibinfo {author} {\bibfnamefont {P.}~\bibnamefont {Lang}}, \bibinfo {author} {\bibfnamefont {B.}~\bibnamefont {Heinz}}, \bibinfo {author} {\bibfnamefont {M.}~\bibnamefont {Geilen}}, \bibinfo {author} {\bibfnamefont {T.}~\bibnamefont {Br{\"{a}}cher}}, \bibinfo {author} {\bibfnamefont {P.}~\bibnamefont {Pirro}}, \bibinfo {author} {\bibfnamefont {T.}~\bibnamefont {Meyer}}, \bibinfo {author} {\bibfnamefont {A.}~\bibnamefont {Conca}}, \bibinfo {author} {\bibfnamefont {D.}~\bibnamefont {Karfaridis}}, \bibinfo {author} {\bibfnamefont {G.}~\bibnamefont {Vourlias}}, \bibinfo {author} {\bibfnamefont {T.}~\bibnamefont {Kehagias}}, \bibinfo {author} {\bibfnamefont {B.}~\bibnamefont {Hillebrands}},\ and\ \bibinfo {author} {\bibfnamefont {E.~T.}\ \bibnamefont {Papaioannou}},\ }\bibfield  {journal}
  {\bibinfo  {journal} {New Journal of Physics}\ }\textbf {\bibinfo {volume} {20}},\ \href {https://doi.org/10.1088/1367-2630/aabc46} {10.1088/1367-2630/aabc46} (\bibinfo {year} {2018})\BibitemShut {NoStop}%
\bibitem [{\citenamefont {Ikebuchi}\ \emph {et~al.}(2022)\citenamefont {Ikebuchi}, \citenamefont {Shiota}, \citenamefont {Ono}, \citenamefont {Nakamura},\ and\ \citenamefont {Moriyama}}]{ikebuchi2022crystal}%
  \BibitemOpen
  \bibfield  {author} {\bibinfo {author} {\bibfnamefont {T.}~\bibnamefont {Ikebuchi}}, \bibinfo {author} {\bibfnamefont {Y.}~\bibnamefont {Shiota}}, \bibinfo {author} {\bibfnamefont {T.}~\bibnamefont {Ono}}, \bibinfo {author} {\bibfnamefont {K.}~\bibnamefont {Nakamura}},\ and\ \bibinfo {author} {\bibfnamefont {T.}~\bibnamefont {Moriyama}},\ }\href@noop {} {\bibfield  {journal} {\bibinfo  {journal} {Applied Physics Letters}\ }\textbf {\bibinfo {volume} {120}},\ \bibinfo {pages} {072406} (\bibinfo {year} {2022})}\BibitemShut {NoStop}%
\bibitem [{\citenamefont {Thompson}\ \emph {et~al.}(2020)\citenamefont {Thompson}, \citenamefont {Ryu}, \citenamefont {Du}, \citenamefont {Karube}, \citenamefont {Kohda},\ and\ \citenamefont {Nitta}}]{Thompson2020}%
  \BibitemOpen
  \bibfield  {author} {\bibinfo {author} {\bibfnamefont {R.}~\bibnamefont {Thompson}}, \bibinfo {author} {\bibfnamefont {J.}~\bibnamefont {Ryu}}, \bibinfo {author} {\bibfnamefont {Y.}~\bibnamefont {Du}}, \bibinfo {author} {\bibfnamefont {S.}~\bibnamefont {Karube}}, \bibinfo {author} {\bibfnamefont {M.}~\bibnamefont {Kohda}},\ and\ \bibinfo {author} {\bibfnamefont {J.}~\bibnamefont {Nitta}},\ }\href {https://doi.org/10.1103/PhysRevB.101.214415} {\bibfield  {journal} {\bibinfo  {journal} {Physical Review B}\ }\textbf {\bibinfo {volume} {101}},\ \bibinfo {pages} {214415} (\bibinfo {year} {2020})}\BibitemShut {NoStop}%
\bibitem [{\citenamefont {Thompson}\ \emph {et~al.}(2021)\citenamefont {Thompson}, \citenamefont {Ryu}, \citenamefont {Choi}, \citenamefont {Karube}, \citenamefont {Kohda}, \citenamefont {Nitta},\ and\ \citenamefont {Park}}]{PhysRevApplied.15.014055}%
  \BibitemOpen
  \bibfield  {author} {\bibinfo {author} {\bibfnamefont {R.}~\bibnamefont {Thompson}}, \bibinfo {author} {\bibfnamefont {J.}~\bibnamefont {Ryu}}, \bibinfo {author} {\bibfnamefont {G.}~\bibnamefont {Choi}}, \bibinfo {author} {\bibfnamefont {S.}~\bibnamefont {Karube}}, \bibinfo {author} {\bibfnamefont {M.}~\bibnamefont {Kohda}}, \bibinfo {author} {\bibfnamefont {J.}~\bibnamefont {Nitta}},\ and\ \bibinfo {author} {\bibfnamefont {B.-G.}\ \bibnamefont {Park}},\ }\href {https://doi.org/10.1103/PhysRevApplied.15.014055} {\bibfield  {journal} {\bibinfo  {journal} {Phys. Rev. Appl.}\ }\textbf {\bibinfo {volume} {15}},\ \bibinfo {pages} {014055} (\bibinfo {year} {2021})}\BibitemShut {NoStop}%
\bibitem [{\citenamefont {Bai}\ \emph {et~al.}(2021)\citenamefont {Bai}, \citenamefont {Mao}, \citenamefont {Yun}, \citenamefont {Zhai}, \citenamefont {Chang}, \citenamefont {Zhang}, \citenamefont {Zhang}, \citenamefont {Zuo},\ and\ \citenamefont {Xi}}]{Bai2021}%
  \BibitemOpen
  \bibfield  {author} {\bibinfo {author} {\bibfnamefont {Q.}~\bibnamefont {Bai}}, \bibinfo {author} {\bibfnamefont {J.}~\bibnamefont {Mao}}, \bibinfo {author} {\bibfnamefont {J.}~\bibnamefont {Yun}}, \bibinfo {author} {\bibfnamefont {Y.}~\bibnamefont {Zhai}}, \bibinfo {author} {\bibfnamefont {M.}~\bibnamefont {Chang}}, \bibinfo {author} {\bibfnamefont {X.}~\bibnamefont {Zhang}}, \bibinfo {author} {\bibfnamefont {J.}~\bibnamefont {Zhang}}, \bibinfo {author} {\bibfnamefont {Y.}~\bibnamefont {Zuo}},\ and\ \bibinfo {author} {\bibfnamefont {L.}~\bibnamefont {Xi}},\ }\href {https://doi.org/10.1063/5.0024153} {\bibfield  {journal} {\bibinfo  {journal} {Applied Physics Letters}\ }\textbf {\bibinfo {volume} {118}},\ \bibinfo {pages} {132403} (\bibinfo {year} {2021})}\BibitemShut {NoStop}%
\bibitem [{\citenamefont {Choi}\ \emph {et~al.}(2022)\citenamefont {Choi}, \citenamefont {Ryu}, \citenamefont {Thompson}, \citenamefont {Choi}, \citenamefont {Jeong}, \citenamefont {Lee}, \citenamefont {Kang}, \citenamefont {Kohda}, \citenamefont {Nitta},\ and\ \citenamefont {Park}}]{choi2022thickness}%
  \BibitemOpen
  \bibfield  {author} {\bibinfo {author} {\bibfnamefont {G.}~\bibnamefont {Choi}}, \bibinfo {author} {\bibfnamefont {J.}~\bibnamefont {Ryu}}, \bibinfo {author} {\bibfnamefont {R.}~\bibnamefont {Thompson}}, \bibinfo {author} {\bibfnamefont {J.-G.}\ \bibnamefont {Choi}}, \bibinfo {author} {\bibfnamefont {J.}~\bibnamefont {Jeong}}, \bibinfo {author} {\bibfnamefont {S.}~\bibnamefont {Lee}}, \bibinfo {author} {\bibfnamefont {M.-G.}\ \bibnamefont {Kang}}, \bibinfo {author} {\bibfnamefont {M.}~\bibnamefont {Kohda}}, \bibinfo {author} {\bibfnamefont {J.}~\bibnamefont {Nitta}},\ and\ \bibinfo {author} {\bibfnamefont {B.-G.}\ \bibnamefont {Park}},\ }\href@noop {} {\bibfield  {journal} {\bibinfo  {journal} {APL Materials}\ }\textbf {\bibinfo {volume} {10}},\ \bibinfo {pages} {011105} (\bibinfo {year} {2022})}\BibitemShut {NoStop}%
\bibitem [{\citenamefont {Xiao}\ \emph {et~al.}(2022{\natexlab{a}})\citenamefont {Xiao}, \citenamefont {Wang},\ and\ \citenamefont {Fullerton}}]{Xiao2022}%
  \BibitemOpen
  \bibfield  {author} {\bibinfo {author} {\bibfnamefont {Y.}~\bibnamefont {Xiao}}, \bibinfo {author} {\bibfnamefont {H.}~\bibnamefont {Wang}},\ and\ \bibinfo {author} {\bibfnamefont {E.~E.}\ \bibnamefont {Fullerton}},\ }\href {https://doi.org/10.3389/fphy.2021.791736} {\bibfield  {journal} {\bibinfo  {journal} {Frontiers in Physics}\ }\textbf {\bibinfo {volume} {9}},\ \bibinfo {pages} {1} (\bibinfo {year} {2022}{\natexlab{a}})}\BibitemShut {NoStop}%
\bibitem [{\citenamefont {Rojas-S{\'{a}}nchez}\ \emph {et~al.}(2014)\citenamefont {Rojas-S{\'{a}}nchez}, \citenamefont {Reyren}, \citenamefont {Laczkowski}, \citenamefont {Savero}, \citenamefont {Attan{\'{e}}}, \citenamefont {Deranlot}, \citenamefont {Gambarelli}, \citenamefont {Jamet}, \citenamefont {George}, \citenamefont {Vila},\ and\ \citenamefont {Jaffr{\`{e}}s}}]{Rojas-Sanchez2014}%
  \BibitemOpen
  \bibfield  {author} {\bibinfo {author} {\bibfnamefont {J.-C.}\ \bibnamefont {Rojas-S{\'{a}}nchez}}, \bibinfo {author} {\bibfnamefont {N.}~\bibnamefont {Reyren}}, \bibinfo {author} {\bibfnamefont {P.}~\bibnamefont {Laczkowski}}, \bibinfo {author} {\bibfnamefont {W.}~\bibnamefont {Savero}}, \bibinfo {author} {\bibfnamefont {J.-P.}\ \bibnamefont {Attan{\'{e}}}}, \bibinfo {author} {\bibfnamefont {C.}~\bibnamefont {Deranlot}}, \bibinfo {author} {\bibfnamefont {S.}~\bibnamefont {Gambarelli}}, \bibinfo {author} {\bibfnamefont {M.}~\bibnamefont {Jamet}}, \bibinfo {author} {\bibfnamefont {J.-M.}\ \bibnamefont {George}}, \bibinfo {author} {\bibfnamefont {L.}~\bibnamefont {Vila}},\ and\ \bibinfo {author} {\bibfnamefont {H.}~\bibnamefont {Jaffr{\`{e}}s}},\ }\href {https://doi.org/10.1117/12.2059646} {\bibfield  {journal} {\bibinfo  {journal} {Spintronics VII}\ }\textbf {\bibinfo {volume} {9167}},\ \bibinfo {pages} {916729} (\bibinfo {year} {2014})}\BibitemShut {NoStop}%
\bibitem [{\citenamefont {Zhang}\ \emph {et~al.}(2015)\citenamefont {Zhang}, \citenamefont {Han}, \citenamefont {Jiang}, \citenamefont {Yang},\ and\ \citenamefont {SP~Parkin}}]{Zhang2015role}%
  \BibitemOpen
  \bibfield  {author} {\bibinfo {author} {\bibfnamefont {W.}~\bibnamefont {Zhang}}, \bibinfo {author} {\bibfnamefont {W.}~\bibnamefont {Han}}, \bibinfo {author} {\bibfnamefont {X.}~\bibnamefont {Jiang}}, \bibinfo {author} {\bibfnamefont {S.-H.}\ \bibnamefont {Yang}},\ and\ \bibinfo {author} {\bibfnamefont {S.}~\bibnamefont {SP~Parkin}},\ }\href@noop {} {\bibfield  {journal} {\bibinfo  {journal} {Nature physics}\ }\textbf {\bibinfo {volume} {11}},\ \bibinfo {pages} {496} (\bibinfo {year} {2015})}\BibitemShut {NoStop}%
\bibitem [{\citenamefont {Toka\ifmmode~\mbox{\c{c}}\else \c{c}\fi{}}\ \emph {et~al.}(2015)\citenamefont {Toka\ifmmode~\mbox{\c{c}}\else \c{c}\fi{}}, \citenamefont {Bunyaev}, \citenamefont {Kakazei}, \citenamefont {Schmool}, \citenamefont {Atkinson},\ and\ \citenamefont {Hindmarch}}]{PhysRevLett.115.056601}%
  \BibitemOpen
  \bibfield  {author} {\bibinfo {author} {\bibfnamefont {M.}~\bibnamefont {Toka\ifmmode~\mbox{\c{c}}\else \c{c}\fi{}}}, \bibinfo {author} {\bibfnamefont {S.~A.}\ \bibnamefont {Bunyaev}}, \bibinfo {author} {\bibfnamefont {G.~N.}\ \bibnamefont {Kakazei}}, \bibinfo {author} {\bibfnamefont {D.~S.}\ \bibnamefont {Schmool}}, \bibinfo {author} {\bibfnamefont {D.}~\bibnamefont {Atkinson}},\ and\ \bibinfo {author} {\bibfnamefont {A.~T.}\ \bibnamefont {Hindmarch}},\ }\href {https://doi.org/10.1103/PhysRevLett.115.056601} {\bibfield  {journal} {\bibinfo  {journal} {Phys. Rev. Lett.}\ }\textbf {\bibinfo {volume} {115}},\ \bibinfo {pages} {056601} (\bibinfo {year} {2015})}\BibitemShut {NoStop}%
\bibitem [{\citenamefont {Brataas}\ \emph {et~al.}(2004)\citenamefont {Brataas}, \citenamefont {Tserkovnyak},\ and\ \citenamefont {Bauer}}]{BRATAAS20041981}%
  \BibitemOpen
  \bibfield  {author} {\bibinfo {author} {\bibfnamefont {A.}~\bibnamefont {Brataas}}, \bibinfo {author} {\bibfnamefont {Y.}~\bibnamefont {Tserkovnyak}},\ and\ \bibinfo {author} {\bibfnamefont {G.~E.}\ \bibnamefont {Bauer}},\ }\href {https://doi.org/https://doi.org/10.1016/j.jmmm.2003.12.783} {\bibfield  {journal} {\bibinfo  {journal} {Journal of Magnetism and Magnetic Materials}\ }\textbf {\bibinfo {volume} {272-276}},\ \bibinfo {pages} {1981} (\bibinfo {year} {2004})},\ \bibinfo {note} {proceedings of the International Conference on Magnetism (ICM 2003)}\BibitemShut {NoStop}%
\bibitem [{\citenamefont {Tserkovnyak}\ \emph {et~al.}(2002{\natexlab{a}})\citenamefont {Tserkovnyak}, \citenamefont {Brataas},\ and\ \citenamefont {Bauer}}]{PhysRevB.66.224403}%
  \BibitemOpen
  \bibfield  {author} {\bibinfo {author} {\bibfnamefont {Y.}~\bibnamefont {Tserkovnyak}}, \bibinfo {author} {\bibfnamefont {A.}~\bibnamefont {Brataas}},\ and\ \bibinfo {author} {\bibfnamefont {G.~E.~W.}\ \bibnamefont {Bauer}},\ }\href {https://doi.org/10.1103/PhysRevB.66.224403} {\bibfield  {journal} {\bibinfo  {journal} {Phys. Rev. B}\ }\textbf {\bibinfo {volume} {66}},\ \bibinfo {pages} {224403} (\bibinfo {year} {2002}{\natexlab{a}})}\BibitemShut {NoStop}%
\bibitem [{\citenamefont {Jim{\'{e}}nez-Cavero}\ \emph {et~al.}(2021{\natexlab{a}})\citenamefont {Jim{\'{e}}nez-Cavero}, \citenamefont {Lucas}, \citenamefont {Bugallo}, \citenamefont {L{\'{o}}pez-Bueno}, \citenamefont {Ramos}, \citenamefont {Algarabel}, \citenamefont {Ibarra}, \citenamefont {Rivadulla},\ and\ \citenamefont {Morell{\'{o}}n}}]{Jimenez-Cavero2021}%
  \BibitemOpen
  \bibfield  {author} {\bibinfo {author} {\bibfnamefont {P.}~\bibnamefont {Jim{\'{e}}nez-Cavero}}, \bibinfo {author} {\bibfnamefont {I.}~\bibnamefont {Lucas}}, \bibinfo {author} {\bibfnamefont {D.}~\bibnamefont {Bugallo}}, \bibinfo {author} {\bibfnamefont {C.}~\bibnamefont {L{\'{o}}pez-Bueno}}, \bibinfo {author} {\bibfnamefont {R.}~\bibnamefont {Ramos}}, \bibinfo {author} {\bibfnamefont {P.~A.}\ \bibnamefont {Algarabel}}, \bibinfo {author} {\bibfnamefont {M.~R.}\ \bibnamefont {Ibarra}}, \bibinfo {author} {\bibfnamefont {F.}~\bibnamefont {Rivadulla}},\ and\ \bibinfo {author} {\bibfnamefont {L.}~\bibnamefont {Morell{\'{o}}n}},\ }\href {https://doi.org/10.1063/5.0038192} {\bibfield  {journal} {\bibinfo  {journal} {Applied Physics Letters}\ }\textbf {\bibinfo {volume} {118}},\ \bibinfo {pages} {092404} (\bibinfo {year} {2021}{\natexlab{a}})}\BibitemShut {NoStop}%
\bibitem [{\citenamefont {Ajejas}\ \emph {et~al.}(2018)\citenamefont {Ajejas}, \citenamefont {Gud{\'{i}}n}, \citenamefont {Guerrero}, \citenamefont {Anad{\'{o}}n~Barcelona}, \citenamefont {Diez}, \citenamefont {De~Melo~Costa}, \citenamefont {Olleros}, \citenamefont {Ni{\~{n}}o}, \citenamefont {Pizzini}, \citenamefont {Vogel}, \citenamefont {Valvidares}, \citenamefont {Gargiani}, \citenamefont {Cabero}, \citenamefont {Varela}, \citenamefont {Camarero}, \citenamefont {Miranda},\ and\ \citenamefont {Perna}}]{Ajejas2018}%
  \BibitemOpen
  \bibfield  {author} {\bibinfo {author} {\bibfnamefont {F.}~\bibnamefont {Ajejas}}, \bibinfo {author} {\bibfnamefont {A.}~\bibnamefont {Gud{\'{i}}n}}, \bibinfo {author} {\bibfnamefont {R.}~\bibnamefont {Guerrero}}, \bibinfo {author} {\bibfnamefont {A.}~\bibnamefont {Anad{\'{o}}n~Barcelona}}, \bibinfo {author} {\bibfnamefont {J.~M.}\ \bibnamefont {Diez}}, \bibinfo {author} {\bibfnamefont {L.}~\bibnamefont {De~Melo~Costa}}, \bibinfo {author} {\bibfnamefont {P.}~\bibnamefont {Olleros}}, \bibinfo {author} {\bibfnamefont {M.~A.}\ \bibnamefont {Ni{\~{n}}o}}, \bibinfo {author} {\bibfnamefont {S.}~\bibnamefont {Pizzini}}, \bibinfo {author} {\bibfnamefont {J.}~\bibnamefont {Vogel}}, \bibinfo {author} {\bibfnamefont {M.}~\bibnamefont {Valvidares}}, \bibinfo {author} {\bibfnamefont {P.}~\bibnamefont {Gargiani}}, \bibinfo {author} {\bibfnamefont {M.}~\bibnamefont {Cabero}}, \bibinfo {author} {\bibfnamefont {M.}~\bibnamefont {Varela}}, \bibinfo {author} {\bibfnamefont {J.}~\bibnamefont {Camarero}}, \bibinfo {author}
  {\bibfnamefont {R.}~\bibnamefont {Miranda}},\ and\ \bibinfo {author} {\bibfnamefont {P.}~\bibnamefont {Perna}},\ }\href {https://doi.org/10.1021/acs.nanolett.8b00878} {\bibfield  {journal} {\bibinfo  {journal} {Nano Letters}\ }\textbf {\bibinfo {volume} {18}},\ \bibinfo {pages} {5364} (\bibinfo {year} {2018})}\BibitemShut {NoStop}%
\bibitem [{\citenamefont {Martin-Rio}\ \emph {et~al.}(2022)\citenamefont {Martin-Rio}, \citenamefont {Frontera}, \citenamefont {Pomar}, \citenamefont {Balcells},\ and\ \citenamefont {Martinez}}]{Martin-Rio2022}%
  \BibitemOpen
  \bibfield  {author} {\bibinfo {author} {\bibfnamefont {S.}~\bibnamefont {Martin-Rio}}, \bibinfo {author} {\bibfnamefont {C.}~\bibnamefont {Frontera}}, \bibinfo {author} {\bibfnamefont {A.}~\bibnamefont {Pomar}}, \bibinfo {author} {\bibfnamefont {L.}~\bibnamefont {Balcells}},\ and\ \bibinfo {author} {\bibfnamefont {B.}~\bibnamefont {Martinez}},\ }\href {https://doi.org/10.1038/s41598-021-04319-z} {\bibfield  {journal} {\bibinfo  {journal} {Scientific Reports}\ }\textbf {\bibinfo {volume} {12}},\ \bibinfo {pages} {224} (\bibinfo {year} {2022})}\BibitemShut {NoStop}%
\bibitem [{\citenamefont {Rojas-S{\'{a}}nchez}\ \emph {et~al.}(2016)\citenamefont {Rojas-S{\'{a}}nchez}, \citenamefont {Oyarz{\'{u}}n}, \citenamefont {Fu}, \citenamefont {Marty}, \citenamefont {Vergnaud}, \citenamefont {Gambarelli}, \citenamefont {Vila}, \citenamefont {Jamet}, \citenamefont {Ohtsubo}, \citenamefont {Taleb-Ibrahimi}, \citenamefont {Le~F{\`{e}}vre}, \citenamefont {Bertran}, \citenamefont {Reyren}, \citenamefont {George},\ and\ \citenamefont {Fert}}]{Rojas-Sanchez2016}%
  \BibitemOpen
  \bibfield  {author} {\bibinfo {author} {\bibfnamefont {J.-C.}\ \bibnamefont {Rojas-S{\'{a}}nchez}}, \bibinfo {author} {\bibfnamefont {S.}~\bibnamefont {Oyarz{\'{u}}n}}, \bibinfo {author} {\bibfnamefont {Y.}~\bibnamefont {Fu}}, \bibinfo {author} {\bibfnamefont {A.}~\bibnamefont {Marty}}, \bibinfo {author} {\bibfnamefont {C.}~\bibnamefont {Vergnaud}}, \bibinfo {author} {\bibfnamefont {S.}~\bibnamefont {Gambarelli}}, \bibinfo {author} {\bibfnamefont {L.}~\bibnamefont {Vila}}, \bibinfo {author} {\bibfnamefont {M.}~\bibnamefont {Jamet}}, \bibinfo {author} {\bibfnamefont {Y.}~\bibnamefont {Ohtsubo}}, \bibinfo {author} {\bibfnamefont {A.}~\bibnamefont {Taleb-Ibrahimi}}, \bibinfo {author} {\bibfnamefont {P.}~\bibnamefont {Le~F{\`{e}}vre}}, \bibinfo {author} {\bibfnamefont {F.}~\bibnamefont {Bertran}}, \bibinfo {author} {\bibfnamefont {N.}~\bibnamefont {Reyren}}, \bibinfo {author} {\bibfnamefont {J.-M.}\ \bibnamefont {George}},\ and\ \bibinfo {author} {\bibfnamefont {A.}~\bibnamefont {Fert}},\ }\href
  {https://doi.org/10.1103/PhysRevLett.116.096602} {\bibfield  {journal} {\bibinfo  {journal} {Physical Review Letters}\ }\textbf {\bibinfo {volume} {116}},\ \bibinfo {pages} {096602} (\bibinfo {year} {2016})}\BibitemShut {NoStop}%
\bibitem [{\citenamefont {Fache}\ \emph {et~al.}(2020)\citenamefont {Fache}, \citenamefont {Rojas-Sanchez}, \citenamefont {Badie}, \citenamefont {Mangin},\ and\ \citenamefont {Petit-Watelot}}]{Fache2020}%
  \BibitemOpen
  \bibfield  {author} {\bibinfo {author} {\bibfnamefont {T.}~\bibnamefont {Fache}}, \bibinfo {author} {\bibfnamefont {J.~C.}\ \bibnamefont {Rojas-Sanchez}}, \bibinfo {author} {\bibfnamefont {L.}~\bibnamefont {Badie}}, \bibinfo {author} {\bibfnamefont {S.}~\bibnamefont {Mangin}},\ and\ \bibinfo {author} {\bibfnamefont {S.}~\bibnamefont {Petit-Watelot}},\ }\href {https://doi.org/10.1103/PhysRevB.102.064425} {\bibfield  {journal} {\bibinfo  {journal} {Physical Review B}\ }\textbf {\bibinfo {volume} {102}},\ \bibinfo {pages} {064425} (\bibinfo {year} {2020})}\BibitemShut {NoStop}%
\bibitem [{\citenamefont {Arango}\ \emph {et~al.}(2022{\natexlab{a}})\citenamefont {Arango}, \citenamefont {Anad{\'{o}}n}, \citenamefont {Novoa}, \citenamefont {Pham}, \citenamefont {Choi}, \citenamefont {Alegre}, \citenamefont {Badie}, \citenamefont {Chuvilin}, \citenamefont {Petit-Watelot}, \citenamefont {Hueso}, \citenamefont {Casanova},\ and\ \citenamefont {Rojas-S{\'{a}}nchez}}]{Arango2022a}%
  \BibitemOpen
  \bibfield  {author} {\bibinfo {author} {\bibfnamefont {I.~C.}\ \bibnamefont {Arango}}, \bibinfo {author} {\bibfnamefont {A.}~\bibnamefont {Anad{\'{o}}n}}, \bibinfo {author} {\bibfnamefont {S.}~\bibnamefont {Novoa}}, \bibinfo {author} {\bibfnamefont {V.~T.}\ \bibnamefont {Pham}}, \bibinfo {author} {\bibfnamefont {W.~Y.}\ \bibnamefont {Choi}}, \bibinfo {author} {\bibfnamefont {J.}~\bibnamefont {Alegre}}, \bibinfo {author} {\bibfnamefont {L.}~\bibnamefont {Badie}}, \bibinfo {author} {\bibfnamefont {A.}~\bibnamefont {Chuvilin}}, \bibinfo {author} {\bibfnamefont {S.}~\bibnamefont {Petit-Watelot}}, \bibinfo {author} {\bibfnamefont {L.~E.}\ \bibnamefont {Hueso}}, \bibinfo {author} {\bibfnamefont {F.}~\bibnamefont {Casanova}},\ and\ \bibinfo {author} {\bibfnamefont {J.-C.}\ \bibnamefont {Rojas-S{\'{a}}nchez}},\ }\href {http://arxiv.org/abs/2212.12697} {\bibfield  {journal} {\bibinfo  {journal} {arXiv preprint arXiv: 212.12697}\ ,\ \bibinfo {pages} {1}} (\bibinfo {year} {2022}{\natexlab{a}})}\BibitemShut {NoStop}%
\bibitem [{\citenamefont {Damas}\ \emph {et~al.}(2022)\citenamefont {Damas}, \citenamefont {Anadon}, \citenamefont {C{\'{e}}spedes-Berrocal}, \citenamefont {Alegre-Saenz}, \citenamefont {Bello}, \citenamefont {Arriola-C{\'{o}}rdova}, \citenamefont {Migot}, \citenamefont {Ghanbaja}, \citenamefont {Copie}, \citenamefont {Hehn}, \citenamefont {Cros}, \citenamefont {Petit-Watelot},\ and\ \citenamefont {Rojas-S{\'{a}}nchez}}]{Damas2022}%
  \BibitemOpen
  \bibfield  {author} {\bibinfo {author} {\bibfnamefont {H.}~\bibnamefont {Damas}}, \bibinfo {author} {\bibfnamefont {A.}~\bibnamefont {Anadon}}, \bibinfo {author} {\bibfnamefont {D.}~\bibnamefont {C{\'{e}}spedes-Berrocal}}, \bibinfo {author} {\bibfnamefont {J.}~\bibnamefont {Alegre-Saenz}}, \bibinfo {author} {\bibfnamefont {J.-L.}\ \bibnamefont {Bello}}, \bibinfo {author} {\bibfnamefont {A.}~\bibnamefont {Arriola-C{\'{o}}rdova}}, \bibinfo {author} {\bibfnamefont {S.}~\bibnamefont {Migot}}, \bibinfo {author} {\bibfnamefont {J.}~\bibnamefont {Ghanbaja}}, \bibinfo {author} {\bibfnamefont {O.}~\bibnamefont {Copie}}, \bibinfo {author} {\bibfnamefont {M.}~\bibnamefont {Hehn}}, \bibinfo {author} {\bibfnamefont {V.}~\bibnamefont {Cros}}, \bibinfo {author} {\bibfnamefont {S.}~\bibnamefont {Petit-Watelot}},\ and\ \bibinfo {author} {\bibfnamefont {J.-C.}\ \bibnamefont {Rojas-S{\'{a}}nchez}},\ }\href {https://doi.org/10.1002/pssr.202200035} {\bibfield  {journal} {\bibinfo  {journal} {physica status solidi (RRL) –
  Rapid Research Letters}\ }\textbf {\bibinfo {volume} {16}},\ \bibinfo {pages} {2200035} (\bibinfo {year} {2022})}\BibitemShut {NoStop}%
\bibitem [{\citenamefont {C{\'{e}}spedes‐Berrocal}\ \emph {et~al.}(2021)\citenamefont {C{\'{e}}spedes‐Berrocal}, \citenamefont {Damas}, \citenamefont {Petit‐Watelot}, \citenamefont {Maccariello}, \citenamefont {Tang}, \citenamefont {Arriola‐C{\'{o}}rdova}, \citenamefont {Vallobra}, \citenamefont {Xu}, \citenamefont {Bello}, \citenamefont {Martin}, \citenamefont {Migot}, \citenamefont {Ghanbaja}, \citenamefont {Zhang}, \citenamefont {Hehn}, \citenamefont {Mangin}, \citenamefont {Panagopoulos}, \citenamefont {Cros}, \citenamefont {Fert},\ and\ \citenamefont {Rojas‐S{\'{a}}nchez}}]{CespedesBerrocal2021CurrentInducedLayers}%
  \BibitemOpen
  \bibfield  {author} {\bibinfo {author} {\bibfnamefont {D.}~\bibnamefont {C{\'{e}}spedes‐Berrocal}}, \bibinfo {author} {\bibfnamefont {H.}~\bibnamefont {Damas}}, \bibinfo {author} {\bibfnamefont {S.}~\bibnamefont {Petit‐Watelot}}, \bibinfo {author} {\bibfnamefont {D.}~\bibnamefont {Maccariello}}, \bibinfo {author} {\bibfnamefont {P.}~\bibnamefont {Tang}}, \bibinfo {author} {\bibfnamefont {A.}~\bibnamefont {Arriola‐C{\'{o}}rdova}}, \bibinfo {author} {\bibfnamefont {P.}~\bibnamefont {Vallobra}}, \bibinfo {author} {\bibfnamefont {Y.}~\bibnamefont {Xu}}, \bibinfo {author} {\bibfnamefont {J.}~\bibnamefont {Bello}}, \bibinfo {author} {\bibfnamefont {E.}~\bibnamefont {Martin}}, \bibinfo {author} {\bibfnamefont {S.}~\bibnamefont {Migot}}, \bibinfo {author} {\bibfnamefont {J.}~\bibnamefont {Ghanbaja}}, \bibinfo {author} {\bibfnamefont {S.}~\bibnamefont {Zhang}}, \bibinfo {author} {\bibfnamefont {M.}~\bibnamefont {Hehn}}, \bibinfo {author} {\bibfnamefont {S.}~\bibnamefont {Mangin}}, \bibinfo {author}
  {\bibfnamefont {C.}~\bibnamefont {Panagopoulos}}, \bibinfo {author} {\bibfnamefont {V.}~\bibnamefont {Cros}}, \bibinfo {author} {\bibfnamefont {A.}~\bibnamefont {Fert}},\ and\ \bibinfo {author} {\bibfnamefont {J.}~\bibnamefont {Rojas‐S{\'{a}}nchez}},\ }\href {https://doi.org/10.1002/adma.202007047} {\bibfield  {journal} {\bibinfo  {journal} {Advanced Materials}\ }\textbf {\bibinfo {volume} {33}},\ \bibinfo {pages} {2007047} (\bibinfo {year} {2021})}\BibitemShut {NoStop}%
\bibitem [{\citenamefont {Arango}\ \emph {et~al.}(2022{\natexlab{b}})\citenamefont {Arango}, \citenamefont {Anadón}, \citenamefont {Novoa}, \citenamefont {Pham}, \citenamefont {Choi}, \citenamefont {Alegre}, \citenamefont {Badie}, \citenamefont {Chuvilin}, \citenamefont {Petit-Watelot}, \citenamefont {Hueso}, \citenamefont {Casanova},\ and\ \citenamefont {Rojas-Sánchez}}]{arango2022spintocharge}%
  \BibitemOpen
  \bibfield  {author} {\bibinfo {author} {\bibfnamefont {I.~C.}\ \bibnamefont {Arango}}, \bibinfo {author} {\bibfnamefont {A.}~\bibnamefont {Anadón}}, \bibinfo {author} {\bibfnamefont {S.}~\bibnamefont {Novoa}}, \bibinfo {author} {\bibfnamefont {V.~T.}\ \bibnamefont {Pham}}, \bibinfo {author} {\bibfnamefont {W.~Y.}\ \bibnamefont {Choi}}, \bibinfo {author} {\bibfnamefont {J.}~\bibnamefont {Alegre}}, \bibinfo {author} {\bibfnamefont {L.}~\bibnamefont {Badie}}, \bibinfo {author} {\bibfnamefont {A.}~\bibnamefont {Chuvilin}}, \bibinfo {author} {\bibfnamefont {S.}~\bibnamefont {Petit-Watelot}}, \bibinfo {author} {\bibfnamefont {L.~E.}\ \bibnamefont {Hueso}}, \bibinfo {author} {\bibfnamefont {F.}~\bibnamefont {Casanova}},\ and\ \bibinfo {author} {\bibfnamefont {J.-C.}\ \bibnamefont {Rojas-Sánchez}},\ }\href@noop {} {\bibinfo {title} {Spin-to-charge conversion by spin pumping in sputtered polycrystalline bi$_x$se$_{1-x}$}} (\bibinfo {year} {2022}{\natexlab{b}}),\ \Eprint {https://arxiv.org/abs/2212.12697}
  {arXiv:2212.12697 [cond-mat.mtrl-sci]} \BibitemShut {NoStop}%
\bibitem [{\citenamefont {Tserkovnyak}\ \emph {et~al.}(2002{\natexlab{b}})\citenamefont {Tserkovnyak}, \citenamefont {Brataas},\ and\ \citenamefont {Bauer}}]{Tserkovnyak2002a}%
  \BibitemOpen
  \bibfield  {author} {\bibinfo {author} {\bibfnamefont {Y.}~\bibnamefont {Tserkovnyak}}, \bibinfo {author} {\bibfnamefont {A.}~\bibnamefont {Brataas}},\ and\ \bibinfo {author} {\bibfnamefont {G.~E.}\ \bibnamefont {Bauer}},\ }\href {https://doi.org/10.1103/PhysRevB.66.224403} {\bibfield  {journal} {\bibinfo  {journal} {Physical Review B - Condensed Matter and Materials Physics}\ }\textbf {\bibinfo {volume} {66}},\ \bibinfo {pages} {1} (\bibinfo {year} {2002}{\natexlab{b}})}\BibitemShut {NoStop}%
\bibitem [{\citenamefont {Ramos}\ \emph {et~al.}(2015)\citenamefont {Ramos}, \citenamefont {Kikkawa}, \citenamefont {Aguirre}, \citenamefont {Lucas}, \citenamefont {Anad{\'{o}}n}, \citenamefont {Oyake}, \citenamefont {Uchida}, \citenamefont {Adachi}, \citenamefont {Shiomi}, \citenamefont {Algarabel}, \citenamefont {Morell{\'{o}}n}, \citenamefont {Maekawa}, \citenamefont {Saitoh},\ and\ \citenamefont {Ibarra}}]{Ramos2015}%
  \BibitemOpen
  \bibfield  {author} {\bibinfo {author} {\bibfnamefont {R.}~\bibnamefont {Ramos}}, \bibinfo {author} {\bibfnamefont {T.}~\bibnamefont {Kikkawa}}, \bibinfo {author} {\bibfnamefont {M.~H.}\ \bibnamefont {Aguirre}}, \bibinfo {author} {\bibfnamefont {I.}~\bibnamefont {Lucas}}, \bibinfo {author} {\bibfnamefont {A.}~\bibnamefont {Anad{\'{o}}n}}, \bibinfo {author} {\bibfnamefont {T.}~\bibnamefont {Oyake}}, \bibinfo {author} {\bibfnamefont {K.}~\bibnamefont {Uchida}}, \bibinfo {author} {\bibfnamefont {H.}~\bibnamefont {Adachi}}, \bibinfo {author} {\bibfnamefont {J.}~\bibnamefont {Shiomi}}, \bibinfo {author} {\bibfnamefont {P.~A.}\ \bibnamefont {Algarabel}}, \bibinfo {author} {\bibfnamefont {L.}~\bibnamefont {Morell{\'{o}}n}}, \bibinfo {author} {\bibfnamefont {S.}~\bibnamefont {Maekawa}}, \bibinfo {author} {\bibfnamefont {E.}~\bibnamefont {Saitoh}},\ and\ \bibinfo {author} {\bibfnamefont {M.~R.}\ \bibnamefont {Ibarra}},\ }\href {https://doi.org/10.1103/PhysRevB.92.220407} {\bibfield  {journal} {\bibinfo  {journal}
  {Physical Review B}\ }\textbf {\bibinfo {volume} {92}},\ \bibinfo {pages} {220407} (\bibinfo {year} {2015})}\BibitemShut {NoStop}%
\bibitem [{\citenamefont {Anadón}\ \emph {et~al.}(2021)\citenamefont {Anadón}, \citenamefont {Gudín}, \citenamefont {Guerrero}, \citenamefont {Arnay}, \citenamefont {Guedeja-Marron}, \citenamefont {Jiménez-Cavero}, \citenamefont {Díez~Toledano}, \citenamefont {Ajejas}, \citenamefont {Varela}, \citenamefont {Petit-Watelot}, \citenamefont {Lucas}, \citenamefont {Morellón}, \citenamefont {Algarabel}, \citenamefont {Ibarra}, \citenamefont {Miranda}, \citenamefont {Camarero}, \citenamefont {Rojas-Sánchez},\ and\ \citenamefont {Perna}}]{anadon2021b}%
  \BibitemOpen
  \bibfield  {author} {\bibinfo {author} {\bibfnamefont {A.}~\bibnamefont {Anadón}}, \bibinfo {author} {\bibfnamefont {A.}~\bibnamefont {Gudín}}, \bibinfo {author} {\bibfnamefont {R.~G.}\ \bibnamefont {Guerrero}}, \bibinfo {author} {\bibfnamefont {I.}~\bibnamefont {Arnay}}, \bibinfo {author} {\bibfnamefont {A.}~\bibnamefont {Guedeja-Marron}}, \bibinfo {author} {\bibfnamefont {P.}~\bibnamefont {Jiménez-Cavero}}, \bibinfo {author} {\bibfnamefont {J.~M.}\ \bibnamefont {Díez~Toledano}}, \bibinfo {author} {\bibfnamefont {F.}~\bibnamefont {Ajejas}}, \bibinfo {author} {\bibfnamefont {M.}~\bibnamefont {Varela}}, \bibinfo {author} {\bibfnamefont {S.}~\bibnamefont {Petit-Watelot}}, \bibinfo {author} {\bibfnamefont {I.}~\bibnamefont {Lucas}}, \bibinfo {author} {\bibfnamefont {L.}~\bibnamefont {Morellón}}, \bibinfo {author} {\bibfnamefont {P.~A.}\ \bibnamefont {Algarabel}}, \bibinfo {author} {\bibfnamefont {M.~R.}\ \bibnamefont {Ibarra}}, \bibinfo {author} {\bibfnamefont {R.}~\bibnamefont {Miranda}}, \bibinfo
  {author} {\bibfnamefont {J.}~\bibnamefont {Camarero}}, \bibinfo {author} {\bibfnamefont {J.~C.}\ \bibnamefont {Rojas-Sánchez}},\ and\ \bibinfo {author} {\bibfnamefont {P.}~\bibnamefont {Perna}},\ }\href@noop {} {\bibfield  {journal} {\bibinfo  {journal} {APL Materials}\ }\textbf {\bibinfo {volume} {9}},\ \bibinfo {pages} {061113} (\bibinfo {year} {2021})}\BibitemShut {NoStop}%
\bibitem [{\citenamefont {Bakonyi}(2005)}]{BAKONYI20052509}%
  \BibitemOpen
  \bibfield  {author} {\bibinfo {author} {\bibfnamefont {I.}~\bibnamefont {Bakonyi}},\ }\href {https://doi.org/https://doi.org/10.1016/j.actamat.2005.02.016} {\bibfield  {journal} {\bibinfo  {journal} {Acta Materialia}\ }\textbf {\bibinfo {volume} {53}},\ \bibinfo {pages} {2509} (\bibinfo {year} {2005})}\BibitemShut {NoStop}%
\bibitem [{\citenamefont {Lizárraga}\ \emph {et~al.}(2017)\citenamefont {Lizárraga}, \citenamefont {Pan}, \citenamefont {Bergqvist}, \citenamefont {Holmström}, \citenamefont {Gercsi},\ and\ \citenamefont {Vitos}}]{Lizarraga2017}%
  \BibitemOpen
  \bibfield  {author} {\bibinfo {author} {\bibfnamefont {R.}~\bibnamefont {Lizárraga}}, \bibinfo {author} {\bibfnamefont {F.}~\bibnamefont {Pan}}, \bibinfo {author} {\bibfnamefont {L.}~\bibnamefont {Bergqvist}}, \bibinfo {author} {\bibfnamefont {E.}~\bibnamefont {Holmström}}, \bibinfo {author} {\bibfnamefont {Z.}~\bibnamefont {Gercsi}},\ and\ \bibinfo {author} {\bibfnamefont {L.}~\bibnamefont {Vitos}},\ }\href {https://doi.org/10.1038/s41598-017-03877-5} {\bibfield  {journal} {\bibinfo  {journal} {Scientific Reports}\ }\textbf {\bibinfo {volume} {7}},\ \bibinfo {pages} {3778} (\bibinfo {year} {2017})}\BibitemShut {NoStop}%
\bibitem [{\citenamefont {Sewak}\ \emph {et~al.}(2022)\citenamefont {Sewak}, \citenamefont {Dey},\ and\ \citenamefont {Toprek}}]{sewak2022temperature}%
  \BibitemOpen
  \bibfield  {author} {\bibinfo {author} {\bibfnamefont {R.}~\bibnamefont {Sewak}}, \bibinfo {author} {\bibfnamefont {C.~C.}\ \bibnamefont {Dey}},\ and\ \bibinfo {author} {\bibfnamefont {D.}~\bibnamefont {Toprek}},\ }\href@noop {} {\bibfield  {journal} {\bibinfo  {journal} {Scientific Reports}\ }\textbf {\bibinfo {volume} {12}},\ \bibinfo {pages} {10054} (\bibinfo {year} {2022})}\BibitemShut {NoStop}%
\bibitem [{\citenamefont {El-Tahawy}\ \emph {et~al.}(2022{\natexlab{a}})\citenamefont {El-Tahawy}, \citenamefont {P{\'{e}}ter}, \citenamefont {Kiss}, \citenamefont {Gubicza}, \citenamefont {Czig{\'{a}}ny}, \citenamefont {Moln{\'{a}}r},\ and\ \citenamefont {Bakonyi}}]{El-Tahawy2022}%
  \BibitemOpen
  \bibfield  {author} {\bibinfo {author} {\bibfnamefont {M.}~\bibnamefont {El-Tahawy}}, \bibinfo {author} {\bibfnamefont {L.}~\bibnamefont {P{\'{e}}ter}}, \bibinfo {author} {\bibfnamefont {L.~F.}\ \bibnamefont {Kiss}}, \bibinfo {author} {\bibfnamefont {J.}~\bibnamefont {Gubicza}}, \bibinfo {author} {\bibfnamefont {Z.}~\bibnamefont {Czig{\'{a}}ny}}, \bibinfo {author} {\bibfnamefont {G.}~\bibnamefont {Moln{\'{a}}r}},\ and\ \bibinfo {author} {\bibfnamefont {I.}~\bibnamefont {Bakonyi}},\ }\bibfield  {journal} {\bibinfo  {journal} {Journal of Magnetism and Magnetic Materials}\ }\textbf {\bibinfo {volume} {560}},\ \href {https://doi.org/10.1016/j.jmmm.2022.169660} {10.1016/j.jmmm.2022.169660} (\bibinfo {year} {2022}{\natexlab{a}})\BibitemShut {NoStop}%
\bibitem [{\citenamefont {Betancourt-Cantera}\ \emph {et~al.}(2019)\citenamefont {Betancourt-Cantera}, \citenamefont {{Sánchez-De Jesús}}, \citenamefont {Bolarín-Miró}, \citenamefont {Torres-Villaseñor},\ and\ \citenamefont {Betancourt-Cantera}}]{BETANCOURTCANTERA20194995}%
  \BibitemOpen
  \bibfield  {author} {\bibinfo {author} {\bibfnamefont {J.}~\bibnamefont {Betancourt-Cantera}}, \bibinfo {author} {\bibfnamefont {F.}~\bibnamefont {{Sánchez-De Jesús}}}, \bibinfo {author} {\bibfnamefont {A.}~\bibnamefont {Bolarín-Miró}}, \bibinfo {author} {\bibfnamefont {G.}~\bibnamefont {Torres-Villaseñor}},\ and\ \bibinfo {author} {\bibfnamefont {L.}~\bibnamefont {Betancourt-Cantera}},\ }\href {https://doi.org/https://doi.org/10.1016/j.jmrt.2019.07.048} {\bibfield  {journal} {\bibinfo  {journal} {Journal of Materials Research and Technology}\ }\textbf {\bibinfo {volume} {8}},\ \bibinfo {pages} {4995} (\bibinfo {year} {2019})}\BibitemShut {NoStop}%
\bibitem [{\citenamefont {Pai}\ \emph {et~al.}(2015)\citenamefont {Pai}, \citenamefont {Ou}, \citenamefont {Vilela-Le{\~{a}}o}, \citenamefont {Ralph},\ and\ \citenamefont {Buhrman}}]{Pai2015}%
  \BibitemOpen
  \bibfield  {author} {\bibinfo {author} {\bibfnamefont {C.~F.}\ \bibnamefont {Pai}}, \bibinfo {author} {\bibfnamefont {Y.}~\bibnamefont {Ou}}, \bibinfo {author} {\bibfnamefont {L.~H.}\ \bibnamefont {Vilela-Le{\~{a}}o}}, \bibinfo {author} {\bibfnamefont {D.~C.}\ \bibnamefont {Ralph}},\ and\ \bibinfo {author} {\bibfnamefont {R.~A.}\ \bibnamefont {Buhrman}},\ }\bibfield  {journal} {\bibinfo  {journal} {Physical Review B - Condensed Matter and Materials Physics}\ }\textbf {\bibinfo {volume} {92}},\ \href {https://doi.org/10.1103/PhysRevB.92.064426} {10.1103/PhysRevB.92.064426} (\bibinfo {year} {2015})\BibitemShut {NoStop}%
\bibitem [{\citenamefont {Mizukami}\ \emph {et~al.}(2011)\citenamefont {Mizukami}, \citenamefont {Kubota}, \citenamefont {Zhang}, \citenamefont {Naganuma}, \citenamefont {Oogane}, \citenamefont {Ando},\ and\ \citenamefont {Miyazaki}}]{Mizukami_2011}%
  \BibitemOpen
  \bibfield  {author} {\bibinfo {author} {\bibfnamefont {S.}~\bibnamefont {Mizukami}}, \bibinfo {author} {\bibfnamefont {T.}~\bibnamefont {Kubota}}, \bibinfo {author} {\bibfnamefont {X.}~\bibnamefont {Zhang}}, \bibinfo {author} {\bibfnamefont {H.}~\bibnamefont {Naganuma}}, \bibinfo {author} {\bibfnamefont {M.}~\bibnamefont {Oogane}}, \bibinfo {author} {\bibfnamefont {Y.}~\bibnamefont {Ando}},\ and\ \bibinfo {author} {\bibfnamefont {T.}~\bibnamefont {Miyazaki}},\ }\href {https://doi.org/10.1143/JJAP.50.103003} {\bibfield  {journal} {\bibinfo  {journal} {Japanese Journal of Applied Physics}\ }\textbf {\bibinfo {volume} {50}},\ \bibinfo {pages} {103003} (\bibinfo {year} {2011})}\BibitemShut {NoStop}%
\bibitem [{\citenamefont {Nahas}\ \emph {et~al.}(2016)\citenamefont {Nahas}, \citenamefont {Prokhorenko}, \citenamefont {Kornev},\ and\ \citenamefont {Bellaiche}}]{PhysRevLett.116.127601}%
  \BibitemOpen
  \bibfield  {author} {\bibinfo {author} {\bibfnamefont {Y.}~\bibnamefont {Nahas}}, \bibinfo {author} {\bibfnamefont {S.}~\bibnamefont {Prokhorenko}}, \bibinfo {author} {\bibfnamefont {I.}~\bibnamefont {Kornev}},\ and\ \bibinfo {author} {\bibfnamefont {L.}~\bibnamefont {Bellaiche}},\ }\href {https://doi.org/10.1103/PhysRevLett.116.127601} {\bibfield  {journal} {\bibinfo  {journal} {Phys. Rev. Lett.}\ }\textbf {\bibinfo {volume} {116}},\ \bibinfo {pages} {127601} (\bibinfo {year} {2016})}\BibitemShut {NoStop}%
\bibitem [{\citenamefont {Skowronski}\ \emph {et~al.}(2019)\citenamefont {Skowronski}, \citenamefont {Karwacki}, \citenamefont {Zietek}, \citenamefont {Kanak}, \citenamefont {Lazarski}, \citenamefont {Grochot}, \citenamefont {Stobiecki}, \citenamefont {Kuswik}, \citenamefont {Stobiecki},\ and\ \citenamefont {Barnas}}]{PhysRevApplied.11.024039}%
  \BibitemOpen
  \bibfield  {author} {\bibinfo {author} {\bibfnamefont {W.}~\bibnamefont {Skowronski}}, \bibinfo {author} {\bibfnamefont {L.}~\bibnamefont {Karwacki}}, \bibinfo {author} {\bibfnamefont {S.}~\bibnamefont {Zietek}}, \bibinfo {author} {\bibfnamefont {J.}~\bibnamefont {Kanak}}, \bibinfo {author} {\bibfnamefont {S.}~\bibnamefont {Lazarski}}, \bibinfo {author} {\bibfnamefont {K.}~\bibnamefont {Grochot}}, \bibinfo {author} {\bibfnamefont {T.}~\bibnamefont {Stobiecki}}, \bibinfo {author} {\bibfnamefont {P.}~\bibnamefont {Kuswik}}, \bibinfo {author} {\bibfnamefont {F.}~\bibnamefont {Stobiecki}},\ and\ \bibinfo {author} {\bibfnamefont {J.}~\bibnamefont {Barnas}},\ }\href@noop {} {\bibfield  {journal} {\bibinfo  {journal} {Phys Rev Applied}\ }\textbf {\bibinfo {volume} {11}},\ \bibinfo {pages} {024039} (\bibinfo {year} {2019})}\BibitemShut {NoStop}%
\bibitem [{\citenamefont {Liu}\ \emph {et~al.}(2019)\citenamefont {Liu}, \citenamefont {Fache}, \citenamefont {Cespedes-Berrocal}, \citenamefont {Zhang}, \citenamefont {Petit-Watelot}, \citenamefont {Mangin}, \citenamefont {Xu},\ and\ \citenamefont {Rojas-S\'anchez}}]{PhysRevApplied.12.044074}%
  \BibitemOpen
  \bibfield  {author} {\bibinfo {author} {\bibfnamefont {E.}~\bibnamefont {Liu}}, \bibinfo {author} {\bibfnamefont {T.}~\bibnamefont {Fache}}, \bibinfo {author} {\bibfnamefont {D.}~\bibnamefont {Cespedes-Berrocal}}, \bibinfo {author} {\bibfnamefont {Z.}~\bibnamefont {Zhang}}, \bibinfo {author} {\bibfnamefont {S.}~\bibnamefont {Petit-Watelot}}, \bibinfo {author} {\bibfnamefont {S.}~\bibnamefont {Mangin}}, \bibinfo {author} {\bibfnamefont {F.}~\bibnamefont {Xu}},\ and\ \bibinfo {author} {\bibfnamefont {J.-C.}\ \bibnamefont {Rojas-S\'anchez}},\ }\href {https://doi.org/10.1103/PhysRevApplied.12.044074} {\bibfield  {journal} {\bibinfo  {journal} {Phys. Rev. Appl.}\ }\textbf {\bibinfo {volume} {12}},\ \bibinfo {pages} {044074} (\bibinfo {year} {2019})}\BibitemShut {NoStop}%
\bibitem [{\citenamefont {Kondou}\ \emph {et~al.}(2012)\citenamefont {Kondou}, \citenamefont {Sukegawa}, \citenamefont {Mitani}, \citenamefont {Tsukagoshi},\ and\ \citenamefont {Kasai}}]{Kondou2012}%
  \BibitemOpen
  \bibfield  {author} {\bibinfo {author} {\bibfnamefont {K.}~\bibnamefont {Kondou}}, \bibinfo {author} {\bibfnamefont {H.}~\bibnamefont {Sukegawa}}, \bibinfo {author} {\bibfnamefont {S.}~\bibnamefont {Mitani}}, \bibinfo {author} {\bibfnamefont {K.}~\bibnamefont {Tsukagoshi}},\ and\ \bibinfo {author} {\bibfnamefont {S.}~\bibnamefont {Kasai}},\ }\href@noop {} {\bibfield  {journal} {\bibinfo  {journal} {Applied Physics Express}\ }\textbf {\bibinfo {volume} {5}},\ \bibinfo {pages} {073002} (\bibinfo {year} {2012})}\BibitemShut {NoStop}%
\bibitem [{\citenamefont {Ganguly}\ \emph {et~al.}(2014)\citenamefont {Ganguly}, \citenamefont {Kondou}, \citenamefont {Sukegawa}, \citenamefont {Mitani}, \citenamefont {Kasai}, \citenamefont {Niimi}, \citenamefont {Otani},\ and\ \citenamefont {Barman}}]{Ganguly2014}%
  \BibitemOpen
  \bibfield  {author} {\bibinfo {author} {\bibfnamefont {A.}~\bibnamefont {Ganguly}}, \bibinfo {author} {\bibfnamefont {K.}~\bibnamefont {Kondou}}, \bibinfo {author} {\bibfnamefont {H.}~\bibnamefont {Sukegawa}}, \bibinfo {author} {\bibfnamefont {S.}~\bibnamefont {Mitani}}, \bibinfo {author} {\bibfnamefont {S.}~\bibnamefont {Kasai}}, \bibinfo {author} {\bibfnamefont {Y.}~\bibnamefont {Niimi}}, \bibinfo {author} {\bibfnamefont {Y.}~\bibnamefont {Otani}},\ and\ \bibinfo {author} {\bibfnamefont {A.}~\bibnamefont {Barman}},\ }\href@noop {} {\bibfield  {journal} {\bibinfo  {journal} {Applied Physics Letters}\ }\textbf {\bibinfo {volume} {104}},\ \bibinfo {pages} {072405} (\bibinfo {year} {2014})}\BibitemShut {NoStop}%
\bibitem [{\citenamefont {Skinner}\ \emph {et~al.}(2014)\citenamefont {Skinner}, \citenamefont {Wang}, \citenamefont {Hindmarch}, \citenamefont {Rushforth}, \citenamefont {Irvine}, \citenamefont {Heiss}, \citenamefont {Kurebayashi},\ and\ \citenamefont {Ferguson}}]{skinner2014}%
  \BibitemOpen
  \bibfield  {author} {\bibinfo {author} {\bibfnamefont {T.}~\bibnamefont {Skinner}}, \bibinfo {author} {\bibfnamefont {M.}~\bibnamefont {Wang}}, \bibinfo {author} {\bibfnamefont {A.}~\bibnamefont {Hindmarch}}, \bibinfo {author} {\bibfnamefont {A.}~\bibnamefont {Rushforth}}, \bibinfo {author} {\bibfnamefont {A.}~\bibnamefont {Irvine}}, \bibinfo {author} {\bibfnamefont {D.}~\bibnamefont {Heiss}}, \bibinfo {author} {\bibfnamefont {H.}~\bibnamefont {Kurebayashi}},\ and\ \bibinfo {author} {\bibfnamefont {A.}~\bibnamefont {Ferguson}},\ }\href@noop {} {\bibfield  {journal} {\bibinfo  {journal} {Applied Physics Letters}\ }\textbf {\bibinfo {volume} {104}},\ \bibinfo {pages} {062401} (\bibinfo {year} {2014})}\BibitemShut {NoStop}%
\bibitem [{\citenamefont {Allen}\ \emph {et~al.}(2015)\citenamefont {Allen}, \citenamefont {Manipatruni}, \citenamefont {Nikonov}, \citenamefont {Doczy},\ and\ \citenamefont {Young}}]{allen2015}%
  \BibitemOpen
  \bibfield  {author} {\bibinfo {author} {\bibfnamefont {G.}~\bibnamefont {Allen}}, \bibinfo {author} {\bibfnamefont {S.}~\bibnamefont {Manipatruni}}, \bibinfo {author} {\bibfnamefont {D.~E.}\ \bibnamefont {Nikonov}}, \bibinfo {author} {\bibfnamefont {M.}~\bibnamefont {Doczy}},\ and\ \bibinfo {author} {\bibfnamefont {I.~A.}\ \bibnamefont {Young}},\ }\href@noop {} {\bibfield  {journal} {\bibinfo  {journal} {Physical Review B}\ }\textbf {\bibinfo {volume} {91}},\ \bibinfo {pages} {144412} (\bibinfo {year} {2015})}\BibitemShut {NoStop}%
\bibitem [{\citenamefont {Saglam}\ \emph {et~al.}(2018)\citenamefont {Saglam}, \citenamefont {Rojas-Sanchez}, \citenamefont {Petit}, \citenamefont {Hehn}, \citenamefont {Zhang}, \citenamefont {Pearson}, \citenamefont {Mangin},\ and\ \citenamefont {Hoffmann}}]{PhysRevB.98.094407}%
  \BibitemOpen
  \bibfield  {author} {\bibinfo {author} {\bibfnamefont {H.}~\bibnamefont {Saglam}}, \bibinfo {author} {\bibfnamefont {J.~C.}\ \bibnamefont {Rojas-Sanchez}}, \bibinfo {author} {\bibfnamefont {S.}~\bibnamefont {Petit}}, \bibinfo {author} {\bibfnamefont {M.}~\bibnamefont {Hehn}}, \bibinfo {author} {\bibfnamefont {W.}~\bibnamefont {Zhang}}, \bibinfo {author} {\bibfnamefont {J.~E.}\ \bibnamefont {Pearson}}, \bibinfo {author} {\bibfnamefont {S.}~\bibnamefont {Mangin}},\ and\ \bibinfo {author} {\bibfnamefont {A.}~\bibnamefont {Hoffmann}},\ }\href {https://doi.org/10.1103/PhysRevB.98.094407} {\bibfield  {journal} {\bibinfo  {journal} {Phys. Rev. B}\ }\textbf {\bibinfo {volume} {98}},\ \bibinfo {pages} {094407} (\bibinfo {year} {2018})}\BibitemShut {NoStop}%
\bibitem [{\citenamefont {Ramos}\ \emph {et~al.}(2013)\citenamefont {Ramos}, \citenamefont {Kikkawa}, \citenamefont {Uchida}, \citenamefont {Adachi}, \citenamefont {Lucas}, \citenamefont {Aguirre}, \citenamefont {Algarabel}, \citenamefont {Morell{\'{o}}n}, \citenamefont {Maekawa}, \citenamefont {Saitoh},\ and\ \citenamefont {Ibarra}}]{Ramos2013}%
  \BibitemOpen
  \bibfield  {author} {\bibinfo {author} {\bibfnamefont {R.}~\bibnamefont {Ramos}}, \bibinfo {author} {\bibfnamefont {T.}~\bibnamefont {Kikkawa}}, \bibinfo {author} {\bibfnamefont {K.}~\bibnamefont {Uchida}}, \bibinfo {author} {\bibfnamefont {H.}~\bibnamefont {Adachi}}, \bibinfo {author} {\bibfnamefont {I.}~\bibnamefont {Lucas}}, \bibinfo {author} {\bibfnamefont {M.~H.}\ \bibnamefont {Aguirre}}, \bibinfo {author} {\bibfnamefont {P.}~\bibnamefont {Algarabel}}, \bibinfo {author} {\bibfnamefont {L.}~\bibnamefont {Morell{\'{o}}n}}, \bibinfo {author} {\bibfnamefont {S.}~\bibnamefont {Maekawa}}, \bibinfo {author} {\bibfnamefont {E.}~\bibnamefont {Saitoh}},\ and\ \bibinfo {author} {\bibfnamefont {M.~R.}\ \bibnamefont {Ibarra}},\ }\href {https://doi.org/10.1063/1.4793486} {\bibfield  {journal} {\bibinfo  {journal} {Applied Physics Letters}\ }\textbf {\bibinfo {volume} {102}},\ \bibinfo {pages} {072413} (\bibinfo {year} {2013})}\BibitemShut {NoStop}%
\bibitem [{\citenamefont {Anad{\'{o}}n}\ \emph {et~al.}(2021{\natexlab{b}})\citenamefont {Anad{\'{o}}n}, \citenamefont {Gud{\'{i}}n}, \citenamefont {Guerrero}, \citenamefont {Arnay}, \citenamefont {Guedeja-Marron}, \citenamefont {Jim{\'{e}}nez-Cavero}, \citenamefont {D{\'{i}}ez~Toledano}, \citenamefont {Ajejas}, \citenamefont {Varela}, \citenamefont {Petit-Watelot}, \citenamefont {Lucas}, \citenamefont {Morell{\'{o}}n}, \citenamefont {Algarabel}, \citenamefont {Ibarra}, \citenamefont {Miranda}, \citenamefont {Camarero}, \citenamefont {Rojas-S{\'{a}}nchez},\ and\ \citenamefont {Perna}}]{Anadon2021}%
  \BibitemOpen
  \bibfield  {author} {\bibinfo {author} {\bibfnamefont {A.}~\bibnamefont {Anad{\'{o}}n}}, \bibinfo {author} {\bibfnamefont {A.}~\bibnamefont {Gud{\'{i}}n}}, \bibinfo {author} {\bibfnamefont {R.}~\bibnamefont {Guerrero}}, \bibinfo {author} {\bibfnamefont {I.}~\bibnamefont {Arnay}}, \bibinfo {author} {\bibfnamefont {A.}~\bibnamefont {Guedeja-Marron}}, \bibinfo {author} {\bibfnamefont {P.}~\bibnamefont {Jim{\'{e}}nez-Cavero}}, \bibinfo {author} {\bibfnamefont {J.~M.}\ \bibnamefont {D{\'{i}}ez~Toledano}}, \bibinfo {author} {\bibfnamefont {F.}~\bibnamefont {Ajejas}}, \bibinfo {author} {\bibfnamefont {M.}~\bibnamefont {Varela}}, \bibinfo {author} {\bibfnamefont {S.}~\bibnamefont {Petit-Watelot}}, \bibinfo {author} {\bibfnamefont {I.}~\bibnamefont {Lucas}}, \bibinfo {author} {\bibfnamefont {L.}~\bibnamefont {Morell{\'{o}}n}}, \bibinfo {author} {\bibfnamefont {P.~A.}\ \bibnamefont {Algarabel}}, \bibinfo {author} {\bibfnamefont {M.~R.}\ \bibnamefont {Ibarra}}, \bibinfo {author} {\bibfnamefont {R.}~\bibnamefont
  {Miranda}}, \bibinfo {author} {\bibfnamefont {J.}~\bibnamefont {Camarero}}, \bibinfo {author} {\bibfnamefont {J.~C.}\ \bibnamefont {Rojas-S{\'{a}}nchez}},\ and\ \bibinfo {author} {\bibfnamefont {P.}~\bibnamefont {Perna}},\ }\href {https://doi.org/10.1063/5.0048612} {\bibfield  {journal} {\bibinfo  {journal} {APL Materials}\ }\textbf {\bibinfo {volume} {9}},\ \bibinfo {pages} {061113} (\bibinfo {year} {2021}{\natexlab{b}})}\BibitemShut {NoStop}%
\bibitem [{\citenamefont {Sola}\ \emph {et~al.}(2017)\citenamefont {Sola}, \citenamefont {Bougiatioti}, \citenamefont {Kuepferling}, \citenamefont {Meier}, \citenamefont {Reiss}, \citenamefont {Pasquale}, \citenamefont {Kuschel},\ and\ \citenamefont {Basso}}]{Sola2017}%
  \BibitemOpen
  \bibfield  {author} {\bibinfo {author} {\bibfnamefont {A.}~\bibnamefont {Sola}}, \bibinfo {author} {\bibfnamefont {P.}~\bibnamefont {Bougiatioti}}, \bibinfo {author} {\bibfnamefont {M.}~\bibnamefont {Kuepferling}}, \bibinfo {author} {\bibfnamefont {D.}~\bibnamefont {Meier}}, \bibinfo {author} {\bibfnamefont {G.}~\bibnamefont {Reiss}}, \bibinfo {author} {\bibfnamefont {M.}~\bibnamefont {Pasquale}}, \bibinfo {author} {\bibfnamefont {T.}~\bibnamefont {Kuschel}},\ and\ \bibinfo {author} {\bibfnamefont {V.}~\bibnamefont {Basso}},\ }\href {https://doi.org/10.1038/srep46752} {\bibfield  {journal} {\bibinfo  {journal} {Scientific Reports}\ }\textbf {\bibinfo {volume} {7}},\ \bibinfo {pages} {46752} (\bibinfo {year} {2017})}\BibitemShut {NoStop}%
\bibitem [{\citenamefont {Wu}\ \emph {et~al.}(2015)\citenamefont {Wu}, \citenamefont {Fradin}, \citenamefont {Hoffman}, \citenamefont {Hoffmann},\ and\ \citenamefont {Bhattacharya}}]{Wu2015}%
  \BibitemOpen
  \bibfield  {author} {\bibinfo {author} {\bibfnamefont {S.~M.}\ \bibnamefont {Wu}}, \bibinfo {author} {\bibfnamefont {F.~Y.}\ \bibnamefont {Fradin}}, \bibinfo {author} {\bibfnamefont {J.}~\bibnamefont {Hoffman}}, \bibinfo {author} {\bibfnamefont {A.}~\bibnamefont {Hoffmann}},\ and\ \bibinfo {author} {\bibfnamefont {A.}~\bibnamefont {Bhattacharya}},\ }\href {https://doi.org/10.1063/1.4916188} {\bibfield  {journal} {\bibinfo  {journal} {Journal of Applied Physics}\ }\textbf {\bibinfo {volume} {117}},\ \bibinfo {pages} {13} (\bibinfo {year} {2015})}\BibitemShut {NoStop}%
\bibitem [{\citenamefont {Luo}\ \emph {et~al.}(2021)\citenamefont {Luo}, \citenamefont {Liu}, \citenamefont {Saglam}, \citenamefont {Li}, \citenamefont {Zhang}, \citenamefont {Zhang}, \citenamefont {Pearson}, \citenamefont {Fisher}, \citenamefont {Zhou}, \citenamefont {Bhattacharya} \emph {et~al.}}]{luo2021}%
  \BibitemOpen
  \bibfield  {author} {\bibinfo {author} {\bibfnamefont {Y.}~\bibnamefont {Luo}}, \bibinfo {author} {\bibfnamefont {C.}~\bibnamefont {Liu}}, \bibinfo {author} {\bibfnamefont {H.}~\bibnamefont {Saglam}}, \bibinfo {author} {\bibfnamefont {Y.}~\bibnamefont {Li}}, \bibinfo {author} {\bibfnamefont {W.}~\bibnamefont {Zhang}}, \bibinfo {author} {\bibfnamefont {S.~S.-L.}\ \bibnamefont {Zhang}}, \bibinfo {author} {\bibfnamefont {J.~E.}\ \bibnamefont {Pearson}}, \bibinfo {author} {\bibfnamefont {B.}~\bibnamefont {Fisher}}, \bibinfo {author} {\bibfnamefont {T.}~\bibnamefont {Zhou}}, \bibinfo {author} {\bibfnamefont {A.}~\bibnamefont {Bhattacharya}}, \emph {et~al.},\ }\href@noop {} {\bibfield  {journal} {\bibinfo  {journal} {Physical Review B}\ }\textbf {\bibinfo {volume} {103}},\ \bibinfo {pages} {L020401} (\bibinfo {year} {2021})}\BibitemShut {NoStop}%
\bibitem [{\citenamefont {El-Tahawy}\ \emph {et~al.}(2022{\natexlab{b}})\citenamefont {El-Tahawy}, \citenamefont {Péter}, \citenamefont {Kiss}, \citenamefont {Gubicza}, \citenamefont {Czigány}, \citenamefont {Molnár},\ and\ \citenamefont {Bakonyi}}]{ELTAHAWY2022169660}%
  \BibitemOpen
  \bibfield  {author} {\bibinfo {author} {\bibfnamefont {M.}~\bibnamefont {El-Tahawy}}, \bibinfo {author} {\bibfnamefont {L.}~\bibnamefont {Péter}}, \bibinfo {author} {\bibfnamefont {L.}~\bibnamefont {Kiss}}, \bibinfo {author} {\bibfnamefont {J.}~\bibnamefont {Gubicza}}, \bibinfo {author} {\bibfnamefont {Z.}~\bibnamefont {Czigány}}, \bibinfo {author} {\bibfnamefont {G.}~\bibnamefont {Molnár}},\ and\ \bibinfo {author} {\bibfnamefont {I.}~\bibnamefont {Bakonyi}},\ }\href {https://doi.org/https://doi.org/10.1016/j.jmmm.2022.169660} {\bibfield  {journal} {\bibinfo  {journal} {Journal of Magnetism and Magnetic Materials}\ }\textbf {\bibinfo {volume} {560}},\ \bibinfo {pages} {169660} (\bibinfo {year} {2022}{\natexlab{b}})}\BibitemShut {NoStop}%
\bibitem [{\citenamefont {Jim{\'{e}}nez-Cavero}\ \emph {et~al.}(2021{\natexlab{b}})\citenamefont {Jim{\'{e}}nez-Cavero}, \citenamefont {Lucas}, \citenamefont {Bugallo}, \citenamefont {L{\'{o}}pez-Bueno}, \citenamefont {Ramos}, \citenamefont {Algarabel}, \citenamefont {Ibarra}, \citenamefont {Rivadulla},\ and\ \citenamefont {Morell{\'{o}}n}}]{Jimenez-Cavero2021a}%
  \BibitemOpen
  \bibfield  {author} {\bibinfo {author} {\bibfnamefont {P.}~\bibnamefont {Jim{\'{e}}nez-Cavero}}, \bibinfo {author} {\bibfnamefont {I.}~\bibnamefont {Lucas}}, \bibinfo {author} {\bibfnamefont {D.}~\bibnamefont {Bugallo}}, \bibinfo {author} {\bibfnamefont {C.}~\bibnamefont {L{\'{o}}pez-Bueno}}, \bibinfo {author} {\bibfnamefont {R.}~\bibnamefont {Ramos}}, \bibinfo {author} {\bibfnamefont {P.~A.}\ \bibnamefont {Algarabel}}, \bibinfo {author} {\bibfnamefont {M.~R.}\ \bibnamefont {Ibarra}}, \bibinfo {author} {\bibfnamefont {F.}~\bibnamefont {Rivadulla}},\ and\ \bibinfo {author} {\bibfnamefont {L.}~\bibnamefont {Morell{\'{o}}n}},\ }\bibfield  {journal} {\bibinfo  {journal} {Applied Physics Letters}\ }\textbf {\bibinfo {volume} {118}},\ \href {https://doi.org/10.1063/5.0038192} {10.1063/5.0038192} (\bibinfo {year} {2021}{\natexlab{b}})\BibitemShut {NoStop}%
\bibitem [{\citenamefont {Boone}\ \emph {et~al.}(2013)\citenamefont {Boone}, \citenamefont {Nembach}, \citenamefont {Shaw},\ and\ \citenamefont {Silva}}]{boone2013}%
  \BibitemOpen
  \bibfield  {author} {\bibinfo {author} {\bibfnamefont {C.~T.}\ \bibnamefont {Boone}}, \bibinfo {author} {\bibfnamefont {H.~T.}\ \bibnamefont {Nembach}}, \bibinfo {author} {\bibfnamefont {J.~M.}\ \bibnamefont {Shaw}},\ and\ \bibinfo {author} {\bibfnamefont {T.~J.}\ \bibnamefont {Silva}},\ }\href@noop {} {\bibfield  {journal} {\bibinfo  {journal} {Journal of Applied Physics}\ }\textbf {\bibinfo {volume} {113}},\ \bibinfo {pages} {153906} (\bibinfo {year} {2013})}\BibitemShut {NoStop}%
\bibitem [{\citenamefont {Mosendz}\ \emph {et~al.}(2010)\citenamefont {Mosendz}, \citenamefont {Pearson}, \citenamefont {Fradin}, \citenamefont {Bader},\ and\ \citenamefont {Hoffmann}}]{mosendz2010}%
  \BibitemOpen
  \bibfield  {author} {\bibinfo {author} {\bibfnamefont {O.}~\bibnamefont {Mosendz}}, \bibinfo {author} {\bibfnamefont {J.~E.}\ \bibnamefont {Pearson}}, \bibinfo {author} {\bibfnamefont {F.~Y.}\ \bibnamefont {Fradin}}, \bibinfo {author} {\bibfnamefont {S.~D.}\ \bibnamefont {Bader}},\ and\ \bibinfo {author} {\bibfnamefont {A.}~\bibnamefont {Hoffmann}},\ }\href {https://doi.org/10.1063/1.3280378} {\bibfield  {journal} {\bibinfo  {journal} {Applied Physics Letters}\ }\textbf {\bibinfo {volume} {96}},\ \bibinfo {pages} {1} (\bibinfo {year} {2010})}\BibitemShut {NoStop}%
\bibitem [{\citenamefont {Liu}\ \emph {et~al.}(2014)\citenamefont {Liu}, \citenamefont {Yuan}, \citenamefont {Wesselink}, \citenamefont {Starikov},\ and\ \citenamefont {Kelly}}]{PhysRevLett.113.207202}%
  \BibitemOpen
  \bibfield  {author} {\bibinfo {author} {\bibfnamefont {Y.}~\bibnamefont {Liu}}, \bibinfo {author} {\bibfnamefont {Z.}~\bibnamefont {Yuan}}, \bibinfo {author} {\bibfnamefont {R.~J.~H.}\ \bibnamefont {Wesselink}}, \bibinfo {author} {\bibfnamefont {A.~A.}\ \bibnamefont {Starikov}},\ and\ \bibinfo {author} {\bibfnamefont {P.~J.}\ \bibnamefont {Kelly}},\ }\href {https://doi.org/10.1103/PhysRevLett.113.207202} {\bibfield  {journal} {\bibinfo  {journal} {Phys. Rev. Lett.}\ }\textbf {\bibinfo {volume} {113}},\ \bibinfo {pages} {207202} (\bibinfo {year} {2014})}\BibitemShut {NoStop}%
\bibitem [{\citenamefont {Wang}\ \emph {et~al.}(2014)\citenamefont {Wang}, \citenamefont {Deorani}, \citenamefont {Qiu}, \citenamefont {Kwon},\ and\ \citenamefont {Yang}}]{10.1063/1.4898593}%
  \BibitemOpen
  \bibfield  {author} {\bibinfo {author} {\bibfnamefont {Y.}~\bibnamefont {Wang}}, \bibinfo {author} {\bibfnamefont {P.}~\bibnamefont {Deorani}}, \bibinfo {author} {\bibfnamefont {X.}~\bibnamefont {Qiu}}, \bibinfo {author} {\bibfnamefont {J.~H.}\ \bibnamefont {Kwon}},\ and\ \bibinfo {author} {\bibfnamefont {H.}~\bibnamefont {Yang}},\ }\href {https://doi.org/10.1063/1.4898593} {\bibfield  {journal} {\bibinfo  {journal} {Applied Physics Letters}\ }\textbf {\bibinfo {volume} {105}},\ \bibinfo {pages} {152412} (\bibinfo {year} {2014})},\ \Eprint {https://arxiv.org/abs/https://pubs.aip.org/aip/apl/article-pdf/doi/10.1063/1.4898593/14092999/152412\_1\_online.pdf} {https://pubs.aip.org/aip/apl/article-pdf/doi/10.1063/1.4898593/14092999/152412\_1\_online.pdf} \BibitemShut {NoStop}%
\bibitem [{\citenamefont {Li}\ \emph {et~al.}(2021)\citenamefont {Li}, \citenamefont {Riddiford}, \citenamefont {Bi}, \citenamefont {Wisser}, \citenamefont {Sun}, \citenamefont {Vailionis}, \citenamefont {Veit}, \citenamefont {Altman}, \citenamefont {Li}, \citenamefont {DC}, \citenamefont {Wang}, \citenamefont {Suzuki},\ and\ \citenamefont {Emori}}]{PhysRevMaterials.5.064404}%
  \BibitemOpen
  \bibfield  {author} {\bibinfo {author} {\bibfnamefont {P.}~\bibnamefont {Li}}, \bibinfo {author} {\bibfnamefont {L.~J.}\ \bibnamefont {Riddiford}}, \bibinfo {author} {\bibfnamefont {C.}~\bibnamefont {Bi}}, \bibinfo {author} {\bibfnamefont {J.~J.}\ \bibnamefont {Wisser}}, \bibinfo {author} {\bibfnamefont {X.-Q.}\ \bibnamefont {Sun}}, \bibinfo {author} {\bibfnamefont {A.}~\bibnamefont {Vailionis}}, \bibinfo {author} {\bibfnamefont {M.~J.}\ \bibnamefont {Veit}}, \bibinfo {author} {\bibfnamefont {A.}~\bibnamefont {Altman}}, \bibinfo {author} {\bibfnamefont {X.}~\bibnamefont {Li}}, \bibinfo {author} {\bibfnamefont {M.}~\bibnamefont {DC}}, \bibinfo {author} {\bibfnamefont {S.~X.}\ \bibnamefont {Wang}}, \bibinfo {author} {\bibfnamefont {Y.}~\bibnamefont {Suzuki}},\ and\ \bibinfo {author} {\bibfnamefont {S.}~\bibnamefont {Emori}},\ }\href {https://doi.org/10.1103/PhysRevMaterials.5.064404} {\bibfield  {journal} {\bibinfo  {journal} {Phys. Rev. Mater.}\ }\textbf {\bibinfo {volume} {5}},\ \bibinfo {pages} {064404}
  (\bibinfo {year} {2021})}\BibitemShut {NoStop}%
\bibitem [{\citenamefont {Obstbaum}\ \emph {et~al.}(2014)\citenamefont {Obstbaum}, \citenamefont {Hartinger}, \citenamefont {Bauer}, \citenamefont {Meier}, \citenamefont {Swientek}, \citenamefont {Back},\ and\ \citenamefont {Woltersdorf}}]{Obstbaum2014}%
  \BibitemOpen
  \bibfield  {author} {\bibinfo {author} {\bibfnamefont {M.}~\bibnamefont {Obstbaum}}, \bibinfo {author} {\bibfnamefont {M.}~\bibnamefont {Hartinger}}, \bibinfo {author} {\bibfnamefont {H.}~\bibnamefont {Bauer}}, \bibinfo {author} {\bibfnamefont {T.}~\bibnamefont {Meier}}, \bibinfo {author} {\bibfnamefont {F.}~\bibnamefont {Swientek}}, \bibinfo {author} {\bibfnamefont {C.}~\bibnamefont {Back}},\ and\ \bibinfo {author} {\bibfnamefont {G.}~\bibnamefont {Woltersdorf}},\ }\href@noop {} {\bibfield  {journal} {\bibinfo  {journal} {Physical Review B}\ }\textbf {\bibinfo {volume} {89}},\ \bibinfo {pages} {060407} (\bibinfo {year} {2014})}\BibitemShut {NoStop}%
\bibitem [{\citenamefont {Xiao}\ \emph {et~al.}(2022{\natexlab{b}})\citenamefont {Xiao}, \citenamefont {Wang},\ and\ \citenamefont {Fullerton}}]{xiao2022crystalline}%
  \BibitemOpen
  \bibfield  {author} {\bibinfo {author} {\bibfnamefont {Y.}~\bibnamefont {Xiao}}, \bibinfo {author} {\bibfnamefont {H.}~\bibnamefont {Wang}},\ and\ \bibinfo {author} {\bibfnamefont {E.~E.}\ \bibnamefont {Fullerton}},\ }\href@noop {} {\bibfield  {journal} {\bibinfo  {journal} {Frontiers in Physics}\ }\textbf {\bibinfo {volume} {9}},\ \bibinfo {pages} {785} (\bibinfo {year} {2022}{\natexlab{b}})}\BibitemShut {NoStop}%
\bibitem [{\citenamefont {Agustsson}\ \emph {et~al.}(2008)\citenamefont {Agustsson}, \citenamefont {Arnalds}, \citenamefont {Ingason}, \citenamefont {Gylfason}, \citenamefont {Johnsen}, \citenamefont {Olafsson},\ and\ \citenamefont {Gudmundsson}}]{Agustsson2008ElectricalSiO2}%
  \BibitemOpen
  \bibfield  {author} {\bibinfo {author} {\bibfnamefont {J.~S.}\ \bibnamefont {Agustsson}}, \bibinfo {author} {\bibfnamefont {U.~B.}\ \bibnamefont {Arnalds}}, \bibinfo {author} {\bibfnamefont {A.~S.}\ \bibnamefont {Ingason}}, \bibinfo {author} {\bibfnamefont {K.~B.}\ \bibnamefont {Gylfason}}, \bibinfo {author} {\bibfnamefont {K.}~\bibnamefont {Johnsen}}, \bibinfo {author} {\bibfnamefont {S.}~\bibnamefont {Olafsson}},\ and\ \bibinfo {author} {\bibfnamefont {J.~T.}\ \bibnamefont {Gudmundsson}},\ }\bibfield  {journal} {\bibinfo  {journal} {Journal of Physics: Conference Series}\ }\textbf {\bibinfo {volume} {100}},\ \href {https://doi.org/10.1088/1742-6596/100/8/082006} {10.1088/1742-6596/100/8/082006} (\bibinfo {year} {2008})\BibitemShut {NoStop}%
\end{thebibliography}%

\end{document}